\documentclass[graybox]{svmult}

\usepackage{mathptmx}       % selects Times Roman as basic font
\usepackage{helvet}         % selects Helvetica as sans-serif font
\usepackage{courier}        % selects Courier as typewriter font
\usepackage{type1cm}        % activate if the above 3 fonts are
\usepackage{subfigure}
\usepackage{xspace}
\usepackage{siunitx}
\usepackage{amsmath}

\usepackage{makeidx}         % allows index generation
\usepackage{graphicx}        % standard LaTeX graphics tool
\usepackage{multicol}        % used for the two-column index
\usepackage[bottom]{footmisc}% places footnotes at page bottom
\usepackage{soul}            % for high-lighting of text

\usepackage[square,numbers,compress]{natbib}

\usepackage{tabularx}
\usepackage{amssymb}

\usepackage{hyperref}        %for hyperlinks
\hypersetup{colorlinks=true,allcolors=blue}

\makeindex             % used for the subject index
\newcommand{\hess}{H.E.S.S.\xspace}
\newcommand{\code}[1]{\texttt{#1}}
\newcommand{\impct}{\code{ImPACT}\xspace}
\def\flat{\emph{Fermi}-LAT\xspace}
\def\fermi{\emph{Fermi}\xspace}
\def\fermitools{\code{fermitools}\xspace}
\makeatletter
\def\NAT@def@citea{\def\@citea{\NAT@separator}}%
\makeatother

\begin{document}
%\tableofcontents{}
\title*{Analysis Methods for Gamma-ray Astronomy}
\author{Denys Malyshev, Lars Mohrmann}
\institute{
  Denys Malyshev \at Institut f\"ur Astronomie und Astrophysik T\"ubingen,
                     Eberhard Karls Universit\"at T\"ubingen,\\ Sand-1, D-72076, T\"ubingen, Germany\\
                     \email{denys.malyshev@astro.uni-tuebingen.de}
  \and
  Lars Mohrmann \at Max-Planck-Institut f\"ur Kernphysik,
                    Saupfercheckweg 1, D-69117 Heidelberg, Germany\\
                    \email{lars.mohrmann@mpi-hd.mpg.de}
}

\maketitle

\vspace{-2cm}
\abstract{
The launch of the \fermi satellite in 2008, with its Large Area Telescope (LAT) on board, has opened a new era for the study of gamma-ray sources at GeV (\SI{e9}{eV}) energies.
Similarly, the commissioning of the third generation of imaging atmospheric Cherenkov telescopes (IACTs) -- \hess, MAGIC, and VERITAS -- in the mid-2000's has firmly established the field of TeV (\SI{e12}{eV}) gamma-ray astronomy.
Together, these instruments have revolutionised our understanding of the high-energy gamma-ray sky, and they continue to provide access to it over more than six decades in energy.
In recent years, the ground-level particle detector arrays HAWC, Tibet, and LHAASO have opened a new window to gamma rays of the highest energies, beyond \SI{100}{TeV}.
Soon, next-generation facilities such as CTA and SWGO will provide even better sensitivity, thus promising a bright future for the field.
In this chapter, we provide a brief overview of methods commonly employed for the analysis of gamma-ray data, focusing on those used for \flat and IACT observations.
We describe the standard data formats, explain event reconstruction and selection algorithms, and cover in detail high-level analysis approaches for imaging and extraction of spectra, including aperture photometry as well as advanced likelihood techniques.
}

\section*{Keywords}
Analysis methods; Gamma-ray astronomy; \flat; IACT; \hess; MAGIC; VERITAS; CTA; Aperture photometry; Likelihood analysis.

\section*{Introduction}
In the last 15--20 years, progress in gamma-ray astronomy has been dominated by two kinds of instruments: space-based telescopes and imaging atmospheric Cherenkov telescopes (IACTs).
In the high-energy (HE; $\SI{100}{MeV} < E < \SI{100}{GeV}$) regime, gamma rays are best detected from space.
The \fermi Large Area Telescope (\flat)~\cite{FermiLAT2009}, launched in 2008, is the leading experiment in this category, complemented by the AGILE satellite \cite{AGILE2009}.
Because the photon flux steadily decreases with increasing energy, the limited effective detection area of space-based telescopes becomes insufficient at higher energies.
Therefore, the preferred method in the very-high-energy (VHE; $E>\SI{100}{GeV}$) regime is to detect the gamma rays indirectly from ground, by measuring the extensive air shower that they launch when they hit the atmosphere.
IACTs achieve this through detection of the Cherenkov light that is emitted by secondary particles in the air shower.
The major operating IACT arrays are the High Energy Stereoscopic System (\hess) \cite{HESS_Crab_2006,Holler2015}, the Major Atmospheric Gamma Imaging Cherenkov (MAGIC) array \cite{MAGIC2016,MAGIC2016b}, and the Very Energetic Radiation Imaging Telescope Array System (VERITAS) \cite{VERITAS2002,Park2015}.

In less than two decades, these instruments have enabled the discovery of more than \num{5000} gamma-ray sources in the HE regime~(see~\cite{4FGL,4FGLDR3}) and several hundred in the VHE regime.
Very soon, the Cherenkov Telescope Array (CTA) -- the next-generation IACT observatory~\cite{CTA2019} -- is set to increase the number of VHE gamma-ray sources to beyond one thousand.
New space-based detectors in the HE range are being prepared as well (e.g.\ HERD~\cite{Gargano2021}).

Several further developments complement the success of \flat and the IACT experiments.
On the one hand, another ground-based detection technique has recently been established: the sampling of the air shower at ground level with arrays of particle detectors -- water-Cherenkov tanks, for example -- placed at a high altitude.
This technique, which excels at the highest gamma-ray energies (\SI{100}{TeV} and beyond), is employed by the High Altitude Water Cherenkov (HAWC) observatory~\cite{HAWC2017}, the Tibet air shower array~\cite{TibetASg2019}, and the Large High Altitude Air Shower Observatory (LHAASO)~\cite{LHAASO2021}.
With the Southern Wide-field Gamma-ray Observatory (SWGO)~\cite{Hinton2021}, another detector of this type is currently in planning.
On the other hand, there is a multitude of upcoming space-based missions that target the energy range below that covered by \flat (i.e.\ keV to MeV energies), for example eXTP \cite{eXTP2019}, eASTROGAM~\cite{eASTROGAM2018}, GECCO~\cite{GECCO2022}, and AMEGO~\cite{AMEGO2019}.
In combination, all of these facilities will enable unprecedented studies of the sky across an energy range that spans from several~keV to more than \SI{100}{TeV}.
This highlights the continued importance of analysing data from \flat and IACT arrays, with methods that guarantee optimal sensitivity.

It is therefore timely for us to provide a review of methods commonly employed for the analysis of gamma-ray data.
In Sect.~\ref{sec:fermi}, we describe the analysis of data from \flat, as the most relevant space-based gamma-ray detector in the HE regime.
Section~\ref{sec:vhe} then covers analysis methods for ground-based gamma-ray detectors, with a strong focus on IACT arrays.
Finally, in Sect.~\ref{sec:mwl_modeling}, we provide a very brief example of a multi-wavelength spectral modelling analysis.

\section{\flat data and spectral analysis}
\label{sec:fermi}
In this section we present a brief description of the LAT instrument onboard the \fermi satellite, and introduce the main approaches to the analysis of \flat data. Together with AGILE, \flat is the only instrument operating in the HE band as of~2023.

The \flat is a pair-conversion gamma-ray detector operating between energies of $\sim 20$~MeV and $\gtrsim 500$~GeV.\footnote{Please note, however, that the \flat data below 100~MeV is subject to large systematic uncertainties and is not used in most of the analysis types. The data above $500$~GeV can be used for the analysis, but suffers from extremely low statistics.}
It has a wide field of view (FoV) of $\sim 2.4$~sr and observes the entire sky every two orbits ($\sim 3$~h, given the average orbital altitude of $\sim 565$~km).
For a detailed description of the instrument, see~\cite{FermiLAT2009}.
In comparison to the instruments of the previous generation (e.g.\ EGRET~\cite{EGRET1989}), \flat is characterised by an effective area, FoV and energy resolution improved by an order of magnitude.

The \flat instrument consists of several subsystems, including a tracker, a calorimeter, and an anti-coincidence detector (ACD). The ACD~\cite{Moiseev2007} (comprised of plastic scintillator tiles) is designed to be the \flat outer-level defense against the charged cosmic-ray (CR) background. An efficient rejection of this background, which is by up to a factor of $10^5$ more intense than the level of astrophysical $\gamma$-ray radiation, is essential for \flat data analyses.

High-energy photons typically do not trigger the ACD and, within the detector, pass through thin layers of high-Z material called conversion foils. In these layers photons are converted into electron-positron pairs. The trajectories of the pairs are then measured by the \flat particle silicon strip tracking detectors~\cite{Baldini2013}. After the exit from the tracker the particles deposit their energy into a calorimeter~\cite{Grove2010} located below the dense layers of the tracker.

\subsection{Data structure and organisation}
\subsubsection{Raw \flat data}
The data downlinked from the \fermi spacecraft (raw data) represent information collected by the instrument's tracker, calorimeter and anti-coincidence detector. Although these data are not available to the user for direct download we briefly outline the approaches used for energy and direction reconstruction of the incoming photons.
A detailed description of the reconstruction algorithms can be found in~\cite{latp8, lat_p8_r3}.

Schematically, the arrival direction of the photon is derived from the tracks of the secondary particles measured in the tracker. The energy of the incoming photon is measured based on the amount of energy deposited in the calorimeter~\cite{latp8}. ACD data, along with a matching between tracks measured in the tracker and energy deposit in the calorimeter, are used to suppress the background from the incoming CR particles in the most recent \flat raw data processing pipelines.

\subsubsection{Access to analysis-ready data}
The high-level, ready-for-the-analysis \flat data comprise the energy and timing information of all detected events that pass certain background-rejection criteria.
These data are normally used for the data analysis and can be downloaded from the \href{https://fermi.gsfc.nasa.gov/cgi-bin/ssc/LAT/LATDataQuery.cgi}{official \flat web page}.
The user is asked to provide:
\begin{itemize}
    \item the coordinates or name (resolvable to coordinates with \href{http://simbad.u-strasbg.fr/simbad/}{SIMBAD}, \href{http://ned.ipac.caltech.edu/}{NED} or \href{http://heasarc.gsfc.nasa.gov/W3Browse/gamma-ray-bursts/grbcat.html}{HEASARC GRB catalogue}) of the centre point of the requested data selection.
    \item coordinate system (J2000, B1950 or galactic); this field is ignored if the name of the object was provided.
    \item search radius -- the radius (in degrees) of a circle around the centre point up to which events will be selected. For most types of analysis the radius should be selected large enough to cover the \flat point-spread function (PSF). This can be as broad as $15^\circ-20^\circ$ for an analysis of data at around 100~MeV energies\footnote{The \flat PSF as a function of energy is available at the \href{https://www.slac.stanford.edu/exp/glast/groups/canda/lat_Performance.htm}{\flat performance web page}}.
    \item observational dates -- the time period for which data are requested. This can be a comma-separated pair of numbers (times of start and end of observations) specified in one of the time systems described below. Any number can be replaced with a START or END keyword indicating that the data is requested from the very beginning of \flat operation or until the end of the available data at the time of the request.
    \item time system -- the time system used for observational dates (``Gregorian'', ``MJD'' or ``MET''). The format of the date specification in Gregorian format is \code{YYYY-MM-DD HH:MM:SS}. For MJD and MET the date is specified as a float number. The mission elapsed time (MET) for \flat was defined to start at 2001-01-01 00:00:00.000 UTC. The earliest available \flat data (equivalent of START keyword) start at MET 239557417 (MJD 54682.65527778).
    \item energy range -- the energy range (in MeV) in which the photons will be selected for the analysis; this could be, for example, (100, 500000) for an analysis performed in the 100~MeV -- 500~GeV energy range.
    \item LAT data type -- ``photon'', ``extended'' or ``None''. The type of the data files which are requested. Files of the photon data type contain all information and events needed for most of the analysis. Files of the extended data type contain in addition photons with looser data cuts and additional information on each photon. ``None'' indicates that the user does not request \flat photon data (can be used for a request of the file containing the spacecraft position information only, see below). For most of the analyses, the LAT data type should be set to ``photon''.
    \item Spacecraft data -- should be marked if a user would like to download a file that contain information on the location/orientation of the \flat in orbit. This file is required for most of the analyses, thus this field should be marked.
\end{itemize}
As soon as all parameters described above are specified the user may submit the request for download of the corresponding data. This will result in a web page showing the status of the queuing data. As soon as the queue request is completed the user can download the requested data. The download page contains links to the requested photon and/or spacecraft data files and summarises the parameters of the request. The displayed parameters of the request can be specifically useful if the name of the object or START/END keywords were used, as the page provides the resolved (J2000) coordinates of the centre point as well as the start and stop times of the requested data in MET seconds. We suggest the user to copy and store this information as it can be useful for the forthcoming analysis of the data. 

Once downloaded, the data are ready for the high-level scientific analysis. 
We would like to note that the approach described above is foreseen for access to \flat data of sky areas with up to $60^\circ$-radius. For the analysis of broader areas or access to all-sky \flat data we suggest the user to use the weekly-updated all-sky photon files available from the \href{https://heasarc.gsfc.nasa.gov/FTP/fermi/data/lat/weekly/photon/}{\flat FTP repository}. The spacecraft files contain only information about the \fermi satellite position and can thus be used for the analysis of sky regions of arbitrary size during the requested time period. Alternatively, weekly spacecraft files can be downloaded from the \href{https://heasarc.gsfc.nasa.gov/FTP/fermi/data/lat/weekly/spacecraft/}{\flat FTP repository}.

For completeness, we briefly describe the structure and content of the \flat data in the following section.

\subsubsection{Structure of the \flat spacecraft and event files}
The downloaded data files can be broadly divided over two classes which contain: \textit{(i)} information on the photons detected by \flat with the parameters specified in the request -- ``photon files''; \textit{(ii)} the history of the \fermi spacecraft pointing and orientation during the selected time period -- ``spacecraft file''. Typically the photon data produced in a single request is split over several photon files. These files are named \texttt{L*\_PH*.fits} if LAT data type ``photon'' was selected during the request and \texttt{L*\_EV*.fits} if the specified data type was ``extended''. One spacecraft file is typically returned for each data request, named \texttt{L*SC*.fits}.

\paragraph{\textit{Structure and content of photon files.}}

\begin{table}
    \centering
    \begin{tabularx}{\textwidth}{c|c|X}
       Column name  & Data type & Description \\ \hline
       Energy  & E,P & Reconstructed energy of the detected photon, MeV \\
       RA, DEC, L, B & E,P & Reconstructed J2000 (RA, DEC) and Galactic (L,B) coordinates of photons, degrees \\
       THETA, PHI &E,P & Reconstructed angle of incidence with respect to the \flat Z (normal to the top surface) and X (normal to the Sun-facing side of the spacecraft) axes\\
       ZENITH\_ANGLE &E,P& Angle between the event coordinates and zenith (a line from the centre of the Earth through the \fermi centre of mass) \\
       EARTH\_AZIMUTH\_ANGLE &E,P& Angle between reconstructed event coordinates and  North (line from \fermi to north celestial pole) as projected onto a plane normal to the zenith.  EARTH\_AZIMUTH\_ANGLE=$90^\circ$ indicates an event originating from west. \\
       TIME &E,P& Mission elapsed time of the event detection, seconds \\
       RUN\_ID, EVENT\_ID&E,P& Unique identifiers of \flat data acquisition period and registered event during this period.\\
       RECON\_VERSION&E,P& Version of reconstruction software used at the time of the event detection \\
       EVENT\_(CLASS,TYPE) &E,P&  Bitfields indicating event class/type, see Sec.~\ref{sec:data_quality_cuts} \\
       CONVERSION\_TYPE&E,P&  0 -- if the event induced pair production in the front (thin) layers of the tracker; 1 -- if in the back (thick) layers of the tracker. \\ \hline
       CalEnergyRaw&E&  Energy measured in a calorimeter pileup-activity removal algorithm. Always $\leq$ reconstructed energy.\\
       PtAlt&E& \fermi altitude at the time of the event detection \\
       PtDecx,PtDecz&E& Declination of \fermi X-axis/Z-axis at the time of the event detection\\
       PtLat,PtLon &E& \fermi ground point latitude/longitude at the time of the event detection \\
       PtMagLat &E& \fermi magnetic latitude at the time of the event detection\\
       PtPosx,PtPosy, PtPosz &E& X,Y,Z-components of the \fermi position vector in Earth-centred inertial coordinates \\
       PtRax,PtRaz &E&  RA of \fermi X/Z-axes at the time of the event detection \\
       PtSCzenith &E& Angle of the \fermi Z-axis from zenith \\
       Tkr1FirstLayer &E& A number of first tracker layer showing a particle hit (tracker layers are 0-17 where 0 is closest to the calorimeter).\\
       WP8Best(X,Y,Z)Dir &E& Analysis choice for the best (X,Y,Z)-direction cosine in the direction estimate \\
       WP8Best(SXX, SXY,SYY) &E& Analysis choice for the best (XX,XY,YY) slope element in the covariance matrix.\\
       WP8CTAllBkProb &E& The probability of the event to be a front-entering gamma ray (0=back-entering, 1=front-entering) \\
       WP8CTAllProb&E&  The probability of the event to be a gamma ray (0=CR-like, 1=gamma-like)\\
       WP8CTBestEnergyProb &E& Probability that the energy reconstruction of the event is correct (0=poor reconstruction quality,1=good reconstruction quality)\\
       WP8CTCalTkrProb&E& The probability of the event to be a gamma ray (different algorithms in comparison to WP8CTAllProb used). \\ 
       WP8CTCalTkrBkProb &E& The probability of the event to be a front-entering gamma ray (different algorithms in comparison to WP8CTAllBkProb used). \\ 
       WP8CTPSF(Core,Tail)&E& Probability that the direction reconstruction of the event is correct (0=poor reconstruction quality,1=good reconstruction quality). Indicates the quality of the reconstruction based on the core (68\% containment) and tails (95\% containment) of the PSF.\\
       \hline
    \end{tabularx}
    \caption{A short description of columns present in \flat photon data files. Please note that the columns CALIB\_VERSION and DIFRSP0-4 are currently not in use. The table summarises the name of the column, the data type of files in which the column can be present (P=photon, E=extended), and a short description of this column.}
    \label{tab:fermi_photon_columns}
\end{table}

Each of the downloaded photon files is a FITS\footnote{See \href{http://fits.gsfc.nasa.gov}{http://fits.gsfc.nasa.gov} for documentation about the FITS data format.} file with 3 extensions. Extension~0 is empty and contains only a header that describes the overall structure of the file and contains TSTART and TSTOP keywords, which correspond to the start and stop times of the requested data. Extension~1 contains information about the detected photons, for a brief description of the available columns see Table~\ref{tab:fermi_photon_columns}. The header of this extension contains information on the spatial, energy and time cuts used for data selection (see e.g.\ Sect.~\ref{sec:data_quality_cuts}). Extension~2 contains the list of good time intervals (GTIs) during which \flat operated in a normal mode. The two columns (START and STOP) of this extension correspond to the minimal and maximal times of the GTI in \fermi MET seconds.

\paragraph{\textit{Structure and content of spacecraft files.}}
As mentioned above, \flat spacecraft files contain the information on the pointing, location and orientation of the \fermi spacecraft. The files contain two extensions. Extension~0 is empty with a header describing the structure of the file and information on the start and stop times of the requested data. Extension~1 contains the information on the spacecraft orientation and position and comprises 31~columns, briefly described in Tab.~\ref{tab:fermi_sc_columns}.

\begin{table}
    \centering
    \begin{tabularx}{\textwidth}{c|X}
      Column name   & Description \\ \hline
      START, STOP & MET time of the beginning and end of the time interval for which further information is summarised in the following columns. \\
      SC\_POSITION & (x, y, z) position (in metres) of spacecraft in inertial coordinates at START\\
      (LAT,LON)\_GEO & Ground projection (latitude, longitude) of \fermi at START \\
      RAD\_GEO  & \fermi altitude at START, metres \\
      (RA,DEC)\_ZENITH & (RA,DEC) of zenith at START, degrees\\
      (B,L)\_MCILWAIN & McIlwain (B,L) parameters~\cite{hessbook68}, specified in Gauss and Earth radii, respectively.\\
      GEOMAG\_LAT &  Geomagnetic latitude at START; degrees \\
      LAMBDA & Effective geomagnetic latitude (sign indicates N/S hemisphere) \\
      IN\_SAA & Is \fermi in South Atlantic Anomaly (SAA) at START? True/False.\\
      RA\_SC(X,Z) &  RA of \fermi (X,Z) axes at START, degree \\
      DEC\_SC(X,Z) & DEC of \fermi (X,Z) axes at START, degree \\
      (RA,DEC)\_NPOLE & (RA,DEC) of north orbital pole at START \\
      ROCK\_ANGLE & Angle of \fermi Z-axis with respect to zenith (positive values indicate a rock toward the north) \\
      LAT\_MODE & \flat operation mode: 1: capture; 2: sunpoint; 3:inertial point; 4: maneuver; 5: zenithpoint/survey; 6,7: reentry mode\\
      LAT\_CONFIG & \flat configuration flag; 1: normal science configuration; 0: not recommended for the analysis\\
      DATA\_QUAL & \flat data quality flag: $<0$: standard IRFs do not describe the data; 0: bad data, do not use; 1: good data; 2: microsecond timing anomaly, the data cannot be used for pulsar timing analysis. \\
      LIVETIME & \flat livetime between START and STOP, seconds. Typical duration: 25-30~s.\\
      QSJ\_(1,2,3,4)& Components of spacecraft attitude quaternion\\
      (RA,DEC)\_SUN & (RA,DEC) of the Sun at START, degrees\\
      SC\_VELOCITY & 3 components of spacecraft velocity in the same coordinate frame as SC\_POSITION at START, m/s\\
      \hline    
    \end{tabularx}
    \caption{Short description of columns present in \flat spacecraft files.}
    \label{tab:fermi_sc_columns}
\end{table}

\subsection{\flat data analysis}\label{sec:flat_data_analysis}
\subsubsection{Data analysis software}
We recommend to install the \flat data analysis software, \fermitools (latest version 2.2.0 as of Feb.\ 2023), with the help of the \href{https://conda.io/docs/}{conda} package manager\footnote{Please note that the analysis of \flat data can also be performed with other software, e.g.\ \href{https://fermipy.readthedocs.io}{fermipy}, \href{https://gammapy.org}{gammapy}, \href{http://cta.irap.omp.eu/ctools}{ctools}. While the main steps of the analysis are similar in all packages, a detailed description of these software frameworks is beyond the scope of this Chapter.}. The command below creates a conda environment named \texttt{fermi}, within which the latest \fermitools and auxiliary data needed for the analysis are available:\\
\texttt{conda create -n fermi -c conda-forge -c fermi fermitools} \\
This environment can be activated with the\\
\texttt{conda activate fermi}\\
command. Additional information on the installation (or update) of \fermitools can be found at the \href{https://github.com/fermi-lat/Fermitools-conda/wiki/Quickstart-Guide}{fermitools github page}. We note that the \fermitools do not require an installation of the \code{heasoft} software. At the same time, certain stand-alone \code{heasoft} routines (e.g.\ \texttt{fv} and \texttt{ds9}) can be useful for displaying the content of the produced FITS files.

The analyses described below can be done either with the \fermitools routines, which can be executed from a terminal or shell script, or using a python interface to these routines\footnote{Please note that since version~2.0, \fermitools require Python version~3.}. We suggest the user to use the latter approach and will focus on it below. Where applicable, we will also provide the corresponding names of the stand-alone \fermitools routines. Names of arguments for the same routine agree between the two approaches. %Please note, that the names of arguments of the parameters of the same routine called from python and from a command line/script coincide.

Prior to the start of the analysis, the user has to define the time and energy interval(s) in which the analysis will be performed. This interval can be narrower than the one used for data download. For example, a user may want to produce an image of the source in several energy bands; similarly, for the spectrum extraction, the analysis can be independently launched in multiple energy bins. The user should also specify the quality cuts (event class and event type) and instrument response functions (IRFs) which will be used for the data analysis. 

The IRFs are the mapping (in a broad sense) between the physical flux of photons and the events detected by \flat. Note, that the detection of a photon by \flat is a probabilistic process, with the probability depending on the \flat hardware efficiency. The reconstruction of the type of the detected event (CR or a photon) as well as the reconstruction of energy/arrival direction can be done only with a certain accuracy, changing from one event to another. The IRFs (and different data quality cuts, see below) are thus based on a trade-off between best-possible CR background rejection; highest-possible effective area; best-possible spatial or energy resolution. While it is not possible to maximise all of these parameters, it is possible to maximise one or (up to some level) several of them. The parameters to be maximised should be chosen based on the main goals of the analysis. Technically, this is implemented in terms of the selection of data quality cuts.

\subsubsection{Data quality cuts}
\label{sec:data_quality_cuts}
We recommend the user to use for the analysis the latest available IRFs (P8R3, V3 as of Feb.\ 2023~\cite{lat_p8_r3}).
The event classes (``evclass'' parameter) are defined based on the probability of the detected event to be a photon and the quality of the reconstruction of the photon parameters. The events within the same class can be sub-divided into event types (``evtype'' parameter). This sub-division is based either on the location of the induced pair-production in the tracker layers (FRONT or BACK, see CONVERSION\_TYPE column description in Tab.~\ref{tab:fermi_photon_columns}) or on a relative quality measure of the reconstruction of the energy or direction of the photon.

The event class and event type should be selected based on the goals of the performed analysis. The \href{https://fermi.gsfc.nasa.gov/ssc/data/analysis/documentation/Cicerone/Cicerone_Data/LAT_DP.html}{\flat web page} describes the set of event classes and types with pre-calculated IRFs available for the data analysis. For most cases, the \flat team suggests to use the SOURCE event class without dedicated event type specification (evclass=128; evtype=3). This class is characterised by a good level of photon/CR separation power and provides good sensitivity for the analysis of point-like sources and moderately extended sources. The somewhat stricter CLEAN event class (evclass=256) is identical to SOURCE below 3~GeV. Above 3~GeV, it provides a 1.3--2 times lower background rate and is slightly more sensitive to hard-spectrum sources at high Galactic latitudes. The cleanest classes ULTRACLEAN and ULTRACLEANVETO (evclass=512 and evclass=1024, respectively) have up to 50\% lower background rate and are recommended for studies of diffuse emission, which require low levels of CR contamination.
Contrary to this, the TRANSIENT event classes (TRANSIENT010 and TRANSIENT020; evclass=16 and evclass=64) have a relatively high level of CR-induced background. These event classes are suggested to be used for studies of short, bright events, for example gamma-ray bursts. These event classes are available only within extended \flat photon files.

Depending on the goals of the analysis, the selected event classes can be sub-divided into event types. For example, the FRONT event type (evtype=1) corresponds to events that induced pair production in the front (thin) layers of the tracker. These events are typically characterised by a better-than-average PSF and effective area. Event types PSF0-3 (evtype=4,8,16,32) divide the events into quartiles indicating the quality of the directional reconstruction, with PSF0 corresponding to the quartile with the worst resolution and PSF3 corresponding to the quartile with the best resolution. The PSF3 event type can be of particular interest for morphological studies of bright, extended sources. The acceptance of each of the PSF0-3 types is about 1/4 of the total \flat acceptance.
Similarly, event types EDISP0-3 (evtype=64,128,256,512) correspond to the worst (EDISP0) -- best (EDISP3) quartiles in terms of energy resolution. The EDISP3 event type can be of particular interest for searches of narrow spectral features in spectra of bright objects.

All event classes and types described above need to be specified in the analyses described below in agreement with the \code{irf} parameter. The IRF name to be specified can be formed from the name of the IRFs (P8R3,V3), the event class name (EVCL), and the event type (EVTYPE) as P8R3\_EVCL\_V3:EVTYPE, or P8R3\_EVCL\_V3 if event types are not specified. For example, for the analysis of data of the SOURCE event class including all event types, the user should use irf=P8R3\_SOURCE\_V3, evclass=128, evtype=3; for the analysis with CLEAN event class and best-measured PSF (PSF3) -- irf=P8R3\_SOURCE\_V3:PSF3, evclass=256, evtype=32.

\subsubsection{Imaging analysis}
As soon as the time and energy intervals and corresponding data quality cuts are established, the user can proceed to the imaging analysis of the analysed region. This analysis allows to generate sky maps of the photons detected by \flat in the selected region. Such analyses are useful for an initial, quick look at the data, allowing one to understand the complexity of the analysed region and, if relevant, check the changes of morphology of the analysed source with energy.

The main steps of the analysis are summarised in Table~\ref{tab:imaging_analysis} (top to bottom). The first column corresponds to the name of a class of the python interface to \fermitools, while the second gives the name of the corresponding stand-alone routine in \fermitools. The third and fourth columns summarise the input and output files required and produced at the corresponding analysis step, respectively. The fifth column gives a short description of the analysis step. In case the user uses the python interface to \fermitools it is suggested that the gt\_apps are imported as\\
\texttt{import gt\_apps}\\
and each routine is run after specification of all relevant parameters, for example\\ \texttt{gt\_apps.filter.run()}.

The analysis is started with a sub-selection of the downloaded data, for which the user can provide time or energy intervals, the size of the spatial region of interest (ROI) and selected data quality cuts with the gt\_apps.filter class. The radius of the ROI (\texttt{gt\_apps.filter['roi']}) should be selected broad enough to accommodate the expected size of the produced image.
At this step it is also suggested to filter the data according to zenith angle by setting \texttt{gt\_apps.filter['zmax'] = 90}. This filter allows the rejection of events originating from the bottom half-sphere of \flat, that is, from directions close to the Earth/Earth limb. A text file with the names of the downloaded \flat photon (or extended) files and the name of the downloaded spacecraft file need to be provided as an input at this step. As an output a single FITS file with all selected events is produced.

At the next step the data are additionally filtered according to the GTIs and \flat data quality flags. The latter is based on the data quality flags described in Table~\ref{tab:fermi_photon_columns}. For most of the analyses this can be done via setting\\
\texttt{gt\_apps.maketime['filter'] = '(DATA\_QUAL>0)\&\&(LAT\_CONFIG==1)'}.\\
The FITS file with the selected events produced at the previous step, along with the \flat spacecraft file, are the inputs to this data analysis step. As an output a FITS file with the additionally filtered data is produced.

Finally, the gt\_apps.evtbin class (with \texttt{gt\_apps.evtbin['algorithm'] = 'CMAP'}) bins the GTI-filtered events into a 2D image. The size of the image in pixels, the pixel size, the coordinates of the centre of the image, and the type of projection can be provided as arguments to this class.

\begin{table}
    \centering
    \begin{tabularx}{\textwidth}{X|c|X|>{\hsize=.5\hsize}X|>{\hsize=1.5\hsize}X}
    Python name & Routine & Input & Output & Description \\ 
     & name &  &  &   \\\hline
      gt\_apps.filter   &\href{https://fermi.gsfc.nasa.gov/ssc/data/p6v11/analysis/scitools/help/gtselect.txt}{gtselect} & List of \flat photon files, spacecraft file & selected events FITS file & Initial data filtering according to user-specified parameters: minimal/maximal energy and time, zenith angle.  \\
      gt\_apps.maketime & \href{https://fermi.gsfc.nasa.gov/ssc/data/p6v11/analysis/scitools/help/gtmktime.txt}{gtmktime} &selected events FITS file, spacecraft file &GTI-filtered FITS file& Filters the events selected at the previous step according to GTI and \flat data quality selection (filter expression).\\     
      gt\_apps.evtbin  (algorithm=CMAP) & \href{https://fermi.gsfc.nasa.gov/ssc/data/p6v11/analysis/scitools/help/gtbin.txt}{gtbin}  &GTI-filtered FITS file &count map&Bins the events contained in the file produced at the previous step into a 2D sky image.\\ \hline
    \end{tabularx}
    \caption{Steps for the imaging analysis of \flat data. The columns summarise the python class name; name of the stand-alone \fermitools routine; input/output files at this step of the analysis; a short description of the analysis step. A full list of the parameters of each python class/routine can be found following the web links in the second column.}
    \label{tab:imaging_analysis}
\end{table}

The obtained image can be further explored and analysed with the help of the \href{https://docs.astropy.org/en/stable/io/fits/index.html}{astropy.io.fits} python module or stand-alone routines such as \href{https://heasarc.gsfc.nasa.gov/ftools/fv/}{fv} or \href{https://sites.google.com/cfa.harvard.edu/saoimageds9}{ds9}. The coordinates of the catalogue (e.g.\ 4FGL-DR3~\cite{4FGLDR3} as of Feb.\ 2023) can be downloaded from the \href{https://fermi.gsfc.nasa.gov/ssc/data/access/lat/12yr_catalog/}{\flat catalogue web page} in a form of all-sky ds9 region (text) files. Alternatively ds9 region files for smaller sky areas coinciding with the size of the analysed region can be generated with the \href{https://fermi.gsfc.nasa.gov/ssc/data/analysis/user/make4FGLxml.py}{make4FGLxml.py} script provided by the \flat team at the \href{https://fermi.gsfc.nasa.gov/ssc/data/analysis/user/}{user contributed software web page}.

\subsubsection{Aperture photometry analysis}
\label{sec:aperture_analysis}
The aperture photometry \flat analysis is similar to the analysis often performed in X-ray astronomy, for example in the analysis of data from the XMM-Newton or Chandra satellites. This type of analysis is based on an ``ON-OFF'' technique, for which the signal is extracted from the ``ON'' (or ``source'') region and the background (if relevant) from the ``OFF'' (or ``background'') region. The potential caveat of such an analysis is that it cannot provide reliable results in regions crowded with gamma-ray sources. In such regions several sources can be located within a PSF, which does not allow to disentangle the individual contributions of these sources to the emission in the ON region. We stress that the aperture analysis described below should be used either for isolated sources or in a case of bright sources for which the background level can be neglected. The aperture analysis can also be effective for a periodicity analysis at different time scales, as for such studies the level of the background may not be of significant importance.

The initial steps of the \flat aperture photometry data analysis are the same as in the imaging analysis. Namely, a user should produce a FITS file filtered according to user-defined energy and time ranges, GTI and \flat data quality cuts. The only difference is that the size of the region of interest should be selected small enough in order not to be affected by nearby sources. We note that this size can be smaller than the \flat PSF in the analysed energy range (e.g.\ could be $1^\circ$ at 100~MeV, while the PSF at the same energy is $\sim 10^\circ$). If the aperture analysis is intended to be used for the production of a lightcurve, the time range provided by the user should cover the whole time period (i.e.\ in this case there is no need to launch the analysis independently in several time bins, as time bins can be defined later).

As soon as an GTI-filtered file is produced the user may produce an exposure-uncorrected lightcurve file with the help of the gt\_apps.evtbin (\href{https://fermi.gsfc.nasa.gov/ssc/data/p6v11/analysis/scitools/help/gtbin.txt}{gtbin}) python class. Note that in order to produce a lightcurve, that is, to perform binning of photons in time rather than space, one should specify \texttt{gt\_apps.evtbin['algorithm'] = 'LC'}. The time bins can be selected in several ways: read from a file, fixed-width time bins, or time bins of a fixed signal-to-noise ratio (SNR; if SNR=$N$ the time bin duration will increase until it contains $N^2$ photons), see the \code{tbinalg} parameter of this routine. If the aperture analysis is used for spectrum extraction, a user may define a single time bin with a length that covers the whole analysed time interval. To produce the spectral flux points the same analysis can be repeated in several energy bins.

The exposure-uncorrected lightcurve produced at this step is stored in a FITS file containing information on how many photons (COUNTS column of the RATE extension of the file) were detected by \flat in the time interval [TIME-TIMEDEL/2; TIME+TIMEDEL/2]. Here TIME and TIMEDEL are given by the corresponding columns in the RATE extension of the lightcurve file. The ERROR column specifies the statistical uncertainty on the number of detected photons. With this file, a count-rate lightcurve (COUNTS/TIMEDEL) can be produced. Note, however, that \flat observes different sky areas with different exposure times and at different off-axis angles. This makes the exposure of \flat non-uniform and changing along the sky with time. An exposure-corrected lightcurve can be produced with the\\
\texttt{gtexposure=gt\_apps.GtApp('gtexposure')}\\
python interface to the \href{https://fermi.gsfc.nasa.gov/ssc/data/p6v11/analysis/scitools/help/gtexposure.txt}{gtexposure} \fermitools routine. This routine automatically corrects for the effects of incomplete enclosure of the ROI within the PSF if the \texttt{apcorr} parameter is set to 'yes' (default behaviour). Note also, that the \flat energy-integrated exposure depends on the spectral slope of the analysed source. While usually this dependence is quite weak we suggest the user to select the \texttt{specin} parameter according to the latest catalogue model (4FGL-DR3~\cite{4FGLDR3} as of Feb.\ 2023) for known sources. For the sources not present in the catalogue we suggest to either use values known from other (e.g.\ TeV) bands or use a relatively soft slope of $2.1-2.5$.

This routine adds an EXPOSURE column to the FITS file produced at the previous step. This column can be used to calculate the exposure-corrected lightcurve (COUNTS/EXPOSURE$\pm$ERROR/EXPOSURE), or flux level $$dN/(dEdAdt) = \mathrm{COUNTS}/\mathrm{EXPOSURE}/(E_\mathrm{max}-E_\mathrm{min})\,,$$ where $E_\mathrm{min/max}$ are the minimal/maximal analysed energies. 

If background subtraction is important for the analysis, the background level and variability can be estimated in a similar way from the analysis of a nearby source-free region. In this case the background subtraction procedure is very similar to the one typically performed in X-ray data analysis. We recommend to choose the radius of the background region equal to that of the ON region, especially if the size of the ON region was selected to be smaller than the size of the \flat PSF at the same energy. This guarantees that the correction for the incomplete enclosure in the \flat PSF applied to the background photons from the ON region is applied to the background photons from the OFF region in the same way.

\subsubsection{Likelihood analysis}\label{sec:flat_likelihood_analysis}
As mentioned in Sec.~\ref{sec:aperture_analysis} the aperture photometry analysis of the \flat data is very similar to the one used, for example, for X-ray data analysis. The caveat of this type of analysis is that it cannot provide reliable results for sources overlapping with regard to the \flat PSF.
Given that the \flat PSF is quite broad ($\sim 10^\circ$ at 100~MeV; $\sim 1^\circ$ at 1~GeV; $\sim 0.1^\circ$ at $\gtrsim 10$~GeV) and the high density of 4FGL catalogue \flat sources in the Galactic plane ($\sim 0.35$ per square degree for $|b|<5^\circ$), it is clear that in many cases the conditions for the aperture analysis may not be satisfied.

A common method to address these issues is to perform a likelihood analysis \cite{Mattox1996}. Contrary to the aperture photometry analysis this analysis relies on the fit of a spatial and spectral model of the analysed region to the available data and allows (with a certain probability) to disentangle the contributions of all model sources present in the region.
The model of the region contains spatial and spectral models of all sources of GeV emission present in the region and allows to construct a predicted flux map of the region as a function of energy. Each of the spectral and spatial source models has free parameters (e.g.\ spectral index), upon which the predicted model flux map depends. Based on information about the \flat observational exposure of the region (i.e.\ how long and under which off-axis angles the region was observed), as well as the \flat PSF and energy dispersion, the predicted model flux can be converted into a predicted count map, which can then be fitted to the observed count map (as a function of energy). 

The fitting parameters in this case are the free parameters of the spectral and spatial models of the sources included in the model of the region. The function used for the fitting is called a (log)likelihood and defined as 
\begin{align}
&\log L(\Omega) = \log\left( \prod\limits_{i} p_i(E_i,X_i | M_i(\Omega)) \right) =\sum\limits_{i}\log p_i(E_i,X_i | M_i(\Omega))\,,
\label{eq:log-like}
\end{align}
where the index $i$ iterates over all photons detected in the region (or groups thereof, see below); $p(E,X| M)$ stands for the (Poisson) probability to observe a photon (or a group of photons) with energy $E$ at spatial coordinates $X$, given that the model predicted $M(\Omega)$ photons. The $\Omega$ represents all free parameters present in the model of the region. The minimisation of this function allows to determine the best-fit spectral and spatial parameters of the model. 

Given two nested\footnote{Model $M_1$ is nested in Model $M_2$ if the parameters in Model $M_1$ are a subset of the parameters in Model $M_2$.} models (used to fit exactly the same dataset) resulting in log-likelihood values of $\log L_1$ (with $N_1$ model parameters) and $\log L_2$ ($N_2$ model parameters) one can construct a quantity $-2\Delta LL = -2(\log L_1 - \log L_2)$. This quantity follows a $\chi^2$ distribution with $N_1-N_2$ degrees of freedom~\cite{Wilks1938}\footnote{Note the caveats: the statistics of the data sample should be sufficiently Gaussian; no best-fit model parameters should be at boundaries.}. This allows to derive (statistical) uncertainties for the best-fit parameters (a $1\sigma$ range corresponds to a change in $-2\Delta LL$ of~1) and compare the quality of the fit between different nested models.
If the two models being compared differ only by one source with one free parameter (spectral normalisation), the difference $-2\Delta LL$ is called the test-statistic (TS) value of the source. In this case the significance of the source detection can be estimated as $\sqrt{TS}$.

Evidently the key ingredient of this type of analysis is a reliable model of the region. The model describes spatial shapes and spectra of all sources present in the region. \fermitools stores the models of the region in an XML format, where each source has a spectral and spatial model. The parameters that are free to vary during the analysis are marked with a \texttt{free="1"} flag.

The spectra of the sources can be described by one of the \href{https://fermi.gsfc.nasa.gov/ssc/data/analysis/scitools/source_models.html}{pre-defined spectral models}. These include commonly used 
models such as a power law (possibly with a break or cut-off) or a log-parabola. A user-defined function can be used as a \texttt{FileFunction} -- a function defined for a set of energies and interpolated between these energies. The only free parameter of such a function is its flux normalisation.

Spatial models of sources are also based on several \href{https://fermi.gsfc.nasa.gov/ssc/data/analysis/scitools/source_models.html#spatialModels}{pre-defined templates}. As of Feb.\ 2023 these include a point-like source, a radial disk or a radial Gaussian. Alternatively the model can be given in terms of a 2D (spatial) flux map of the source (\texttt{SpatialMap}; provided as a FITS file) or in terms of a 3D (spatial+energy) flux-cube of the source (\texttt{MapCubeFunction}; provided as a FITS file).

The model of the region for most analyses can be generated from the list of catalogue sources (e.g.\ 4FGL-DR3). Within \fermitools this can be done with the \href{https://fermi.gsfc.nasa.gov/ssc/data/analysis/user/make4FGLxml.py}{make4FGLxml.py} script provided by the \flat team at the \href{https://fermi.gsfc.nasa.gov/ssc/data/analysis/user/}{user contributed software web page}. The model of the region is created based on the spectral and spatial model types and parameter values of all \flat sources reported in the catalogue. In addition to the (point-like or extended) gamma-ray sources, two additional components are usually included in the model. These are: \textit{(i)} the Galactic diffuse background -- a map cube (3D spatial+energy template) of the Galactic diffuse gamma-ray background~(\cite{gll}; see also the \href{https://fermi.gsfc.nasa.gov/ssc/data/access/lat/BackgroundModels.html}{dedicated \flat web page}) and \textit{(ii)} a model of the isotropic gamma-ray background -- a sum of the contributions from unresolved extragalactic sources and CRs mis-classified as photons (see e.g.\ \cite{dimauro16}).

\begin{figure}[t]
    \centering
    \includegraphics[width=\columnwidth]{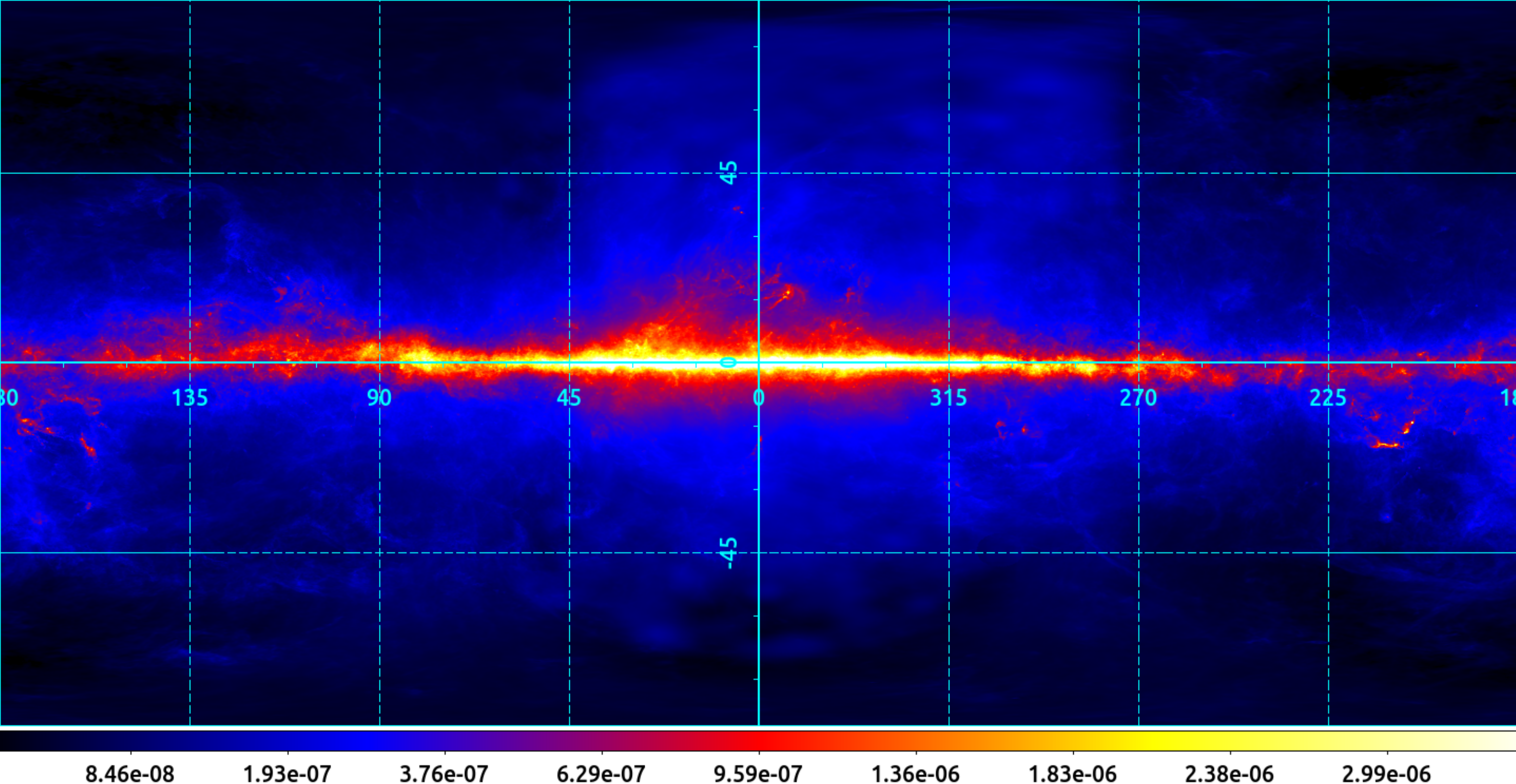}
    \caption{All-sky \flat Galactic diffuse background emission model at $\sim 100$~MeV energy, based on the \href{https://fermi.gsfc.nasa.gov/ssc/data/analysis/software/aux/4fgl/gll_iem_v07.fits}{gll\_iem\_v07.fits} template. The map is shown in Galactic coordinates, Cartesian projection}
    \label{fig:gll100MeV}
\end{figure}

The Galactic diffuse background (\href{https://fermi.gsfc.nasa.gov/ssc/data/analysis/software/aux/4fgl/gll_iem_v07.fits}{gll\_iem\_v07.fits} 3D template) is strongly spatially anisotropic, especially in the Galactic plane, see Fig.~\ref{fig:gll100MeV}. The isotropic background is assumed to be spatially isotropic and is characterised solely by its spectrum, which depends on the selected data quality cuts, see Sec.~\ref{sec:data_quality_cuts}. For example, for the SOURCE event class (with no specified event type) the spectrum of the isotropic background is given by the \href{https://fermi.gsfc.nasa.gov/ssc/data/analysis/software/aux/iso_P8R3_SOURCE_V3_v1.txt}{iso\_P8R3\_SOURCE\_V3\_v1.txt} file (as of Feb.\ 2023).

As discussed above, the \flat PSF can be as broad as $\sim 10^\circ$ at $\sim 100$~MeV energies. For the likelihood analysis we recommend to consider a region of interest of at least $15^\circ$ radius if the analysis is performed at around this energy. The user should include into the model of the region all \flat catalogue sources and templates of the Galactic and isotropic emission. Depending on the goals of the analysis the user may keep free either all spectral\footnote{Please note, that the spatial parameters of the model sources are commonly left fixed, see also ``Source detection'' paragraph below.} parameters of the sources in the model (typically done for analyses encompassing long time periods and/or broad energy ranges) or allow to vary only the flux normalisations of the spectral models of the sources (typically done for analyses in narrow energy or time ranges). In the latter case the rest of the spectral parameters should be fixed to either catalogue values or to best-fit values derived from a dedicated analysis (e.g.\ over a broader energy range). For most of the analyses we suggest to keep free the normalisations of the Galactic diffuse and isotropic backgrounds. To avoid potential effects connected to the presence of bright \flat sources located just beyond the boundaries of the selected ROI, we recommend to include in the model of the region all catalogue sources within $10^\circ$ around the ROI with their spectral parameters (including flux normalisations) fixed to their catalogue values. 

\paragraph{\textit{Unbinnned and binned likelihood analysis}}
The approach described above is implemented in terms of the standard binned and unbinned analyses of \flat data. The unbinned analysis operates on a photon-by-photon basis, that is, the sum in Eq.~\ref{eq:log-like} goes over all photons detected by \flat in the region. This type of analysis is computationally expensive, as its complexity increases proportionally to the number of detected photons. On the other hand, the method is accurate and provides reliable results even if the number of detected photons is small. The \href{https://fermi.gsfc.nasa.gov/ssc/data/analysis/scitools/likelihood_tutorial.html}{\flat collaboration suggests} to mainly use this analysis for datasets with small number of detected photons.

In a binned analysis, the detected photons are binned into 3D (2D spatial+energy) cubes. The spatial size of each ``cell'' in the cube should be small enough compared to the \flat PSF. Typical choices for the cell size are $0.05^\circ - 0.1^\circ$. The size of the cell in the energy dimension should be selected such that the number of cells is sufficient to perform a meaningful fit of the selected spectral models of all sources present in the model of the region. For example, if the model comprises only power-law spectra (2 free parameters), the number of energy bins could be selected to be $\geq 3$. If a source with a broken power-law model with a cut-off (4 free parameters) is present in the model, the number of energy bins should be selected $\geq 5$. 
Furthermore, if the analysis is performed over a broad energy range, the number of energy bins should be selected large enough to handle the possible rapid changes of fluxes of the sources present in the model, for example due to an exponential cut-off. For most of the analyses a good choice of the number of energy bins is three bins per decade of energy (but at least 5).

In the binned likelihood analysis, the product/sum in Eq.~\ref{eq:log-like} goes over all cells of the 3D cube described above. A detailed description of \href{https://fermi.gsfc.nasa.gov/ssc/data/analysis/scitools/binnededisp_tutorial.html}{binned} and \href{https://fermi.gsfc.nasa.gov/ssc/data/analysis/scitools/likelihood_tutorial.html}{unbinned} analyses of \flat data are given on the respective \flat web pages. Table~\ref{tab:spectral_analysis} briefly summarises the steps/routines used in these analyses. Note that all the steps described in Table~\ref{tab:imaging_analysis} are prerequisites for both the binned and unbinned data analyses described below.

\begin{table}
    \centering
    \begin{tabularx}{\textwidth}{X|c|X|>{\hsize=.5\hsize}X|>{\hsize=1.5\hsize}X}
    Python name & Routine name& Input files & Output files & Description \\  \hline
     gt\_apps.evtbin (algorithm= CCUBE)$^\dagger$ & \href{https://fermi.gsfc.nasa.gov/ssc/data/p6v11/analysis/scitools/help/gtbin.txt}{gtbin} & GTI-filtered event file & 3D count cube & Bins photons in 3D (2D spatial+energy cube) \\
     gt\_apps.expCube$^{\dagger\star}$ & \href{https://raw.githubusercontent.com/fermi-lat/fermitools-fhelp/master/fhelp_files/gtltcube.txt}{gtltcube} &  GTI-filtered event file, spacecraft file & livetime cube &  Livetime cube (all-sky HealPix table; integrated livetime vs. inclination with respect to  the \flat z-axis) \\
     gt\_apps.GtApp( 'gtexpcube2', 'Likelihood')$^\dagger$ & \href{https://raw.githubusercontent.com/fermi-lat/fermitools-fhelp/master/fhelp_files/gtexpcube2.txt}{gtexpcube2} & livetime cube & exposure map & Binned map of \flat exposure of the  analysed region\\
    gt\_apps.srcMaps$^\dagger$ &\href{https://raw.githubusercontent.com/fermi-lat/fermitools-fhelp/master/fhelp_files/gtsrcmaps.txt}{gtsrcmaps}& spacecraft file, livetime cube, exposure map, region model& source map & Predicted count map of all sources present in the model of the region\\
    gt\_apps.diffResps$^\star$ & \href{https://raw.githubusercontent.com/fermi-lat/fermitools-fhelp/master/fhelp_files/gtdiffrsp.txt}{gtdiffrsp} & GTI-filtered event file & gti-filtered  & Adds a column to GTI-filtered event file, containing\\
    &&spacecraft file, region model &event file*& integral over solid angle of diffuse sources in the model, convolved with the IRFs\\ \hline
    binnedAnalysis$^\dagger$, UnbinnedAnalysis$^\star$ & \href{https://raw.githubusercontent.com/fermi-lat/fermitools-fhelp/master/fhelp_files/gtlike.txt}{gtlike} & GTI-filtered event file(*), region model, spacecraft file, livetime cube, exposure map & analysis object A& An object used for for the (un)binned likelihood analysis; gtlike routine performs fitting directly \\
    A.fit()$^{\dagger\star}$ &&&& performs actual log-likelihood minimisation given analysis object A\\
    A.writeXml()$^{\dagger\star}$ &&&& allows to save best-fit model to xml file \\
    A.flux( $src, E_1, E_2$)$^{\dagger\star}$&&&& returns best-fit flux (ph/cm$^2$/s) for the model source $src$ in energy range $E_1$, $E_2$ MeV \\
    A.fluxError($src$, $E_1$, $E_2$)$^{\dagger\star}$&&&& returns best-fit flux uncertainty\\
    A.Ts($src$)$^{\dagger\star}$&&&& returns test-statistic value of the source $src$\\
    A.deleteSource( $src$)$^{\dagger\star}$&&&& deletes source $src$ from analysis object\\ \hline
    \end{tabularx}
    \caption{Steps for the binned/unbinned likelihood analysis of \flat data. Steps marked with $^\dagger$ are specific for the binned analysis, those with $^\star$ for the unbinned one. Please note, binnedAnalysis class is included into BinnedAnalysis python module provided within \fermitools; UnbinnedAnalysis is provided within UnbinnedAnalysis module.}
    \label{tab:spectral_analysis}
\end{table}

The steps summarised in the table allow the determination of the best-fit flux and its uncertainty for any source in the model, thus enabling a spectral analysis (in narrow energy bins or over a broader energy range) and a light-curve analysis (in narrow time bins). We recommend to perform the fitting in an iterative way: \textit{(i)} perform initial fitting; \textit{(ii)} remove all sources with low TS (e.g.\ $TS<1$) and re-do the fit until no insignificant sources remain. This approach allows a more accurate determination of the profile of the log-likelihood function around its minimum and results in a more accurate determination of the uncertainties of the best-fit values.

The uncertainties returned via \texttt{analysis.fluxError} (see Table~\ref{tab:spectral_analysis}) may not be reliable for sources detected with a low significance. In this case the user may want to calculate an upper limit on the flux of the source. We recommend to perform upper limit computations with the help of the \texttt{calc\_int} function of the \texttt{IntegralUpperLimit} python module (a part of the standard \fermitools). Specifically, with the defined \texttt{analysis} object (see Table~\ref{tab:spectral_analysis}) one can calculate an upper limit on the flux of the source \texttt{src} in the energy range $(E_1;E_2)$~MeV as\\
\texttt{upperlimit,results = calc\_int(analysis,src,emin=$E_1$,emax=$E_2$)},\\
where the units of the \texttt{upperlimit} flux are ph/cm$^2$/s.

We note, that the uncertainties derived within the binned/unbinned analyses are statistical only. These uncertainties can be small in comparison to the level of the \flat systematic uncertainties, which can be as large as $\sim 5-10$\% of the flux level for typically analysed energy ranges\footnote{See description of the \href{https://fermi.gsfc.nasa.gov/ssc/data/analysis/scitools/Aeff_Systematics.html}{\flat effective area systematic uncertainties}.}.

\paragraph{\textit{Source detection}}
If the source of interest is present in the \flat catalogue the likelihood analysis can be performed as described above. If, however, the source of interest is a weak or transient source not present in the catalogue, a user may want to perform the data analysis to investigate the presence of the source in the data with a morphology consistent with the assumed one. We would like to note that the inclusion of an additional source into the model of the region and a measurement of its flux by means of a binned or unbinned analysis cannot serve as a proof of a firm detection of the source. For example, under the presence of nearby bright sources or in complicated source regions in the Galactic plane, the detection could be spurious and connected to imperfect modelling of the nearby sources or Galactic diffuse background.

In these cases we suggest the user to generate a TS map of the region. This map shows the TS value of 
a hypothetical additional point-like source (with a power-law spectrum with index fixed to $-2$), which is moved along a grid of locations on the sky, minimising the log-likelihood at each location.
Such a map allows to assess the significance of residual emission not yet included in the model and to test whether that residual emission is consistent with the hypothesis of an additional source with the expected spatial morphology (e.g. point-like source).

The procedure described above can be performed with the \href{https://raw.githubusercontent.com/fermi-lat/fermitools-fhelp/master/fhelp_files/gttsmap.txt}{gttsmap} \fermitools routine. 
For a perfectly modelled region (e.g. if a source is already included in the catalogue), the TS map should be zero everywhere.

\begin{figure}[th]
    \centering
    \includegraphics[width=\textwidth]{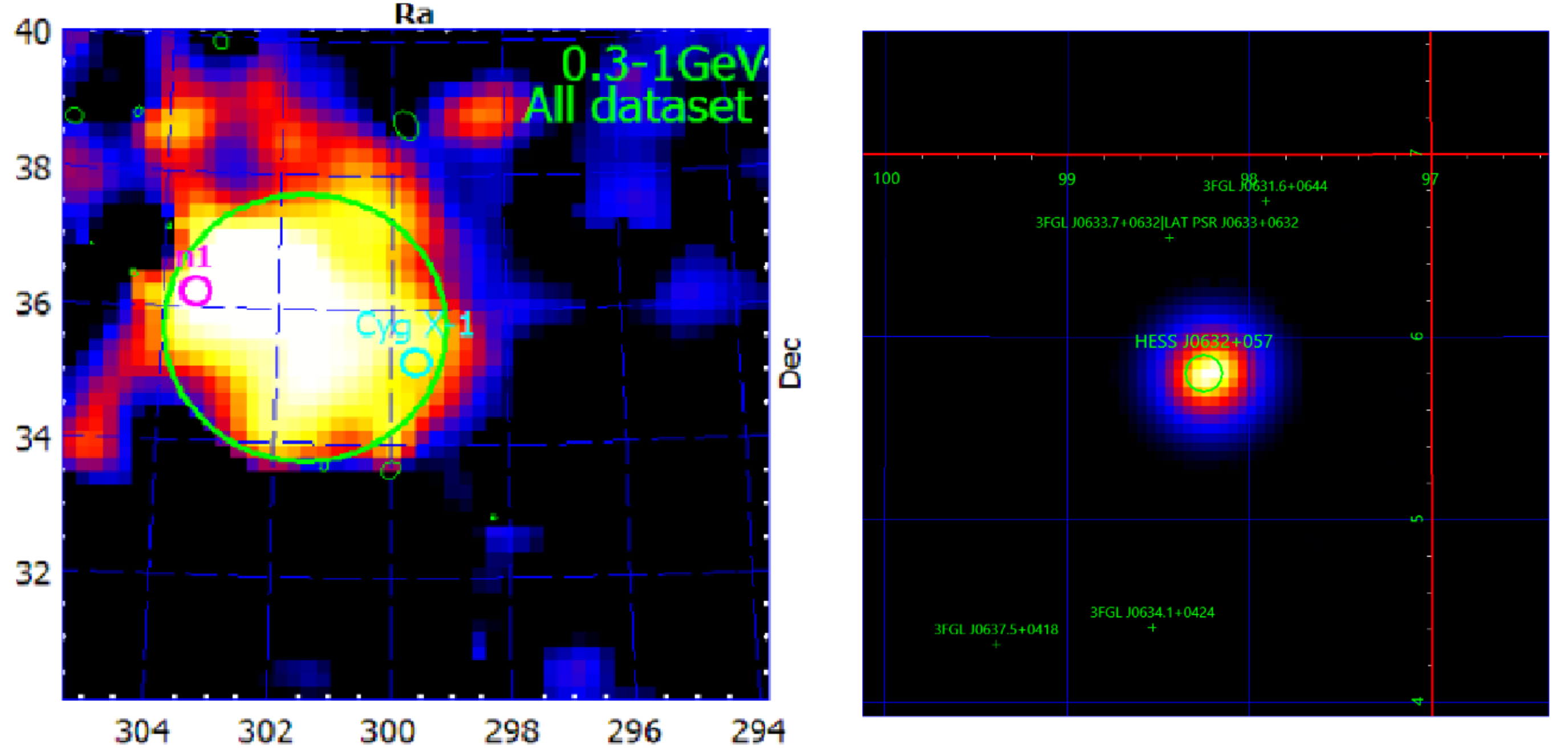}
    \caption{TS maps of the Cygnus X-1 region (left panel,~\cite{cygx1}) and of HESS~J0632+057 (right panel, \cite{j0632}). Residual emission can be seen around the Cygnus X-1 position, suggesting an imperfect modelling of the Galactic diffuse emission. The residuals at the HESS~J0632+057 position illustrate the presence of a point-like source not present in the catalogue (please note that no source at this position was included in the model). Reproduced with permission from the authors \copyright.}
    \label{fig:ts_maps}
\end{figure}

Visual inspection of the TS map can allow the user to estimate the spatial morphology of sources not present in the model, see for example Fig.~\ref{fig:ts_maps}. This figure illustrates two cases: (i) strong diffuse residuals that are not consistent with a point-like source (Cygnus X-1 region, left panel~\cite{cygx1}); (ii) residuals with a morphology that is consistent with a point-like source (HESS~J0632+057 region, right panel, \cite{j0632}). The coordinates of a (point-like) source visible on a TS map can be estimated based on the location of the maximum of the TS value. Alternatively, the coordinates of such a source and uncertainties on its position can be assessed with the help of the \href{https://raw.githubusercontent.com/fermi-lat/fermitools-fhelp/master/fhelp_files/gtfindsrc.txt}{gtfindsrc} routine.

\paragraph{\textit{Concluding remarks}}
The analysis scheme described above allows to perform the basic analysis of the \flat data, including spectral, lightcurve and spatial analyses. While this scheme covers the basic aspects of the analysis we suggest the reader to follow any updates on the analysis/software that may be announced at the \flat \href{https://fermi.gsfc.nasa.gov/ssc/data/analysis/}{website}. We furthermore recommend to familiarise oneself with the \href{https://fermi.gsfc.nasa.gov/ssc/data/analysis/LAT_caveats.html}{web page} summarising known caveats of the \flat data analysis.

%%%%%%%%%%%%%%%%%%%%%%%%%%%%%%%%%%%%%%%%%%%%%%%%%%%%%%%%%%%%%%%%%%%%%%%%%%%%%%%%%%%%%%%%%%%%%%%%%%%%%%%
%%%%%%%%%%%%%%%%%%%%%%%%%%%%%%%%%%%%%%%%%%%%%%%%%%%%%%%%%%%%%%%%%%%%%%%%%%%%%%%%%%%%%%%%%%%%%%%%%%%%%%%
%%%%%%%%%%%%%%%%%%%%%%%%%%%%%%%%%%%%%%%%%%%%%%%%%%%%%%%%%%%%%%%%%%%%%%%%%%%%%%%%%%%%%%%%%%%%%%%%%%%%%%%
%%%%%%%%%%%%%%%%%%%%%%%%%%%%%%%%%%%%%%%%%%%%%%%%%%%%%%%%%%%%%%%%%%%%%%%%%%%%%%%%%%%%%%%%%%%%%%%%%%%%%%%
%%%%%%%%%%%%%%%%%%%%%%%%%%%%%%%%%%%%%%%%%%%%%%%%%%%%%%%%%%%%%%%%%%%%%%%%%%%%%%%%%%%%%%%%%%%%%%%%%%%%%%%
%%%%%%%%%%%%%%%%%%%%%%%%%%%%%%%%%%%%%%%%%%%%%%%%%%%%%%%%%%%%%%%%%%%%%%%%%%%%%%%%%%%%%%%%%%%%%%%%%%%%%%%

\newpage
\section{Analysis methods for ground-based gamma-ray instruments}
\label{sec:vhe}

This section provides a summary of standard methods employed in the analysis of data from ground-based gamma-ray detectors.
A strong focus lies on analysis methods for IACTs; methods for ground-level particle detector arrays will only briefly be covered at the end.

In contrast to \flat and other space-based gamma-ray telescopes, IACTs detect gamma rays indirectly through imaging the air showers that are initiated when they hit the atmosphere of the Earth.
They do so by measuring the short (few~ns) and faint flash of Cherenkov light that is emitted from the secondary particles in the air shower.
This requires telescopes with large mirrors (to collect enough Cherenkov photons) as well as sensitive cameras with fast sampling (to be able to detect the short signal on top of the steady background of night-sky photons).
As a consequence, IACTs can only operate during sufficiently dark nights, and have a limited field of view of a few degrees in radius.
For more information about the IACT detection technique, see for example the chapter ``Detecting gamma-rays with high resolution and moderate field of view: the air Cherenkov technique'' in this handbook, or one of the existing reviews on this subject (e.g.\ \cite{deNaurois2015,Holder2021,Sitarek2022}).

There are three major operating arrays of IACTs: \hess~\cite{HESS_Crab_2006,Holler2015}, consisting of five telescopes; MAGIC~\cite{MAGIC2016,MAGIC2016b}, consisting of two telescopes; and VERITAS~\cite{VERITAS2002,Park2015}, consisting of four telescopes.
Currently under construction as the major successor experiment in this field is the CTA \cite{CTA2019}, which will consist of 13 telescopes at its Northern site on La Palma, Spain, and of $\sim$50 telescopes at its Southern site in the Atacama desert in Chile.
This handbook contains chapters about each of these observatories, to which the reader is referred for more details.
In the following, we provide an overview of analysis methods for IACT data.
Because the methods employed by the different collaborations operating the above-mentioned instruments are in principle very similar, those used within the \hess Collaboration are taken as example cases.
We try to point out relevant differences to other methods where we are aware of them.
Note also that many of the software tools used by the collaborations are proprietary and not openly available.
With the advent of CTA, which will be operated as an open observatory, this situation is however currently changing.

The section is structured as follows:
We introduce data levels and formats relevant for IACT data in Sect.~\ref{sec:iact_data_levels_formats}.
The low-level data processing is briefly summarised in Sect.~\ref{sec:iact_low_level_data}, while Sect.~\ref{sec:iact_event_reco} covers event reconstruction methods.
Sections~\ref{sec:iact_gamma_hadron_separation} and~\ref{sec:iact_background_modelling} deal with the suppression and modelling of cosmic ray-induced background events, respectively.
The generation of instrument response functions (IRFs) is described in Sect.~\ref{sec:iact_irf_generation}, before we discuss methods for high-level data analysis (creation of maps, extraction of energy spectra, source modelling) in Sect.~\ref{sec:iact_high_level_analysis}.
Finally, in Sect.~\ref{sec:particle_detectors} we briefly summarise similarities and differences to analysis methods used for ground-level particle-detector gamma-ray observatories.

\subsection{Data levels and formats}
\label{sec:iact_data_levels_formats}

\begin{figure}
    \centering
    \includegraphics[width=0.98\textwidth]{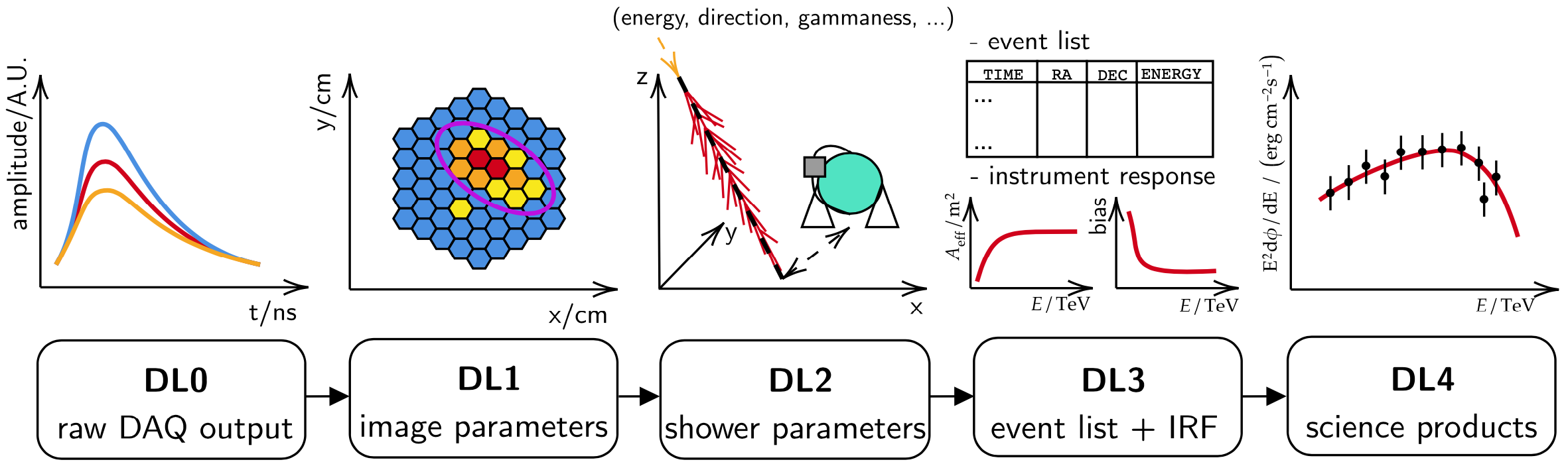}
    \caption{
        Sketch illustrating different data levels (DL0 -- DL4) of data recorded with IACTs.
        Figure taken from \cite{Nigro2021}.
        Reproduced with permission from the authors \copyright.
    }
    \label{fig:iact_data_formats}
\end{figure}

Following the data processing chain, IACT data can be categorised into different levels.
We follow here the definition of data levels (DLs) foreseen for CTA \cite{Contreras2015}, which are illustrated in Fig.~\ref{fig:iact_data_formats}.
Specifically, the levels are:
\begin{itemize}
    \item \textbf{DL0} ---
          Raw output of the data acquisition system.
          At this level, the data consist of the still uncalibrated signals measured in the photo-sensors in each camera pixel.
          Note that some cameras already perform a time integration of the signal in the hardware, whereas most modern cameras provide a full sampled waveform.
    \item \textbf{DL1} ---
          Calibrated signals and derived per-telescope parameters.
          The calibration converts the pixel signals into physical quantities (charge in photo-electrons (p.e.), typically).
          Parameters derived from the full camera image (``Hillas parameters'', see Sect.~\ref{sec:iact_event_reco}) also belong to this level.
    \item \textbf{DL2} ---
          Reconstructed shower parameters.
          In the event reconstruction, the data recorded by the different telescopes are combined to infer properties (direction, energy, etc.) of the air shower, typically under the assumption of a gamma-ray primary particle.
          This level also includes parameters that discriminate between different primary particles.
    \item \textbf{DL3} ---
          List of gamma-ray candidate events and associated IRFs.
          This level provides the reconstructed properties of events that pass certain event selection criteria, as well as the IRFs (effective area, point spread function, etc.\ -- typically obtained from simulation data) that detail the performance of the system with respect to these criteria.
    \item \textbf{DL4} ---
          High-level science data products.
          From the list of selected events and the IRFs, ``astronomical'' science products such as sky maps, energy spectra, or light curves may be derived.
\end{itemize}
Physical models of gamma-ray sources can be fit to either DL3 or DL4 level data (see Sect.~\ref{sec:iact_high_level_analysis}).
The definition of DLs for CTA includes an additional level DL5, which is used to refer to legacy products of the observatory, such as survey maps or catalogues.
This level is however not relevant for the review presented here.

The data at DL0--DL2 are typically not shared publicly, and thus stored in closed, custom data formats (often based on the \code{ROOT} framework widespread in particle physics \cite{Brun1997}) developed by the collaborations that operate the different instruments.
At DL3, data from the currently operating IACT experiments are generally also not publicly available.
The CTA Observatory, however, will make all CTA data at the DL3 level publicly available in an open data format after a one-year proprietary period, and provide software tools for their analysis \cite{Zanin2021}.
The obvious advantage of an open data format is that it can be defined such that data from different instruments can be represented, and thus enable analyses that combine the data at this level.
With this goal in mind, a prototype for the CTA DL3 data format -- labelled ``GADF'', for ``gamma astro data formats'' \cite{Deil_GADF_2017,Deil_GADFv3_2022} -- has been developed by the community, and small test data sets of the current-generation instruments have been released in this format (e.g.\ \cite{HESS_FITS_2018,Nigro2019}).
For a more detailed overview about the evolution of data formats in gamma-ray astronomy, see the detailed review in \cite{Nigro2021}.

%%%%%%  DL0 --> DL1  %%%%%%
\subsection{Low-level data processing}
\label{sec:iact_low_level_data}

The low-level data processing concerns the transition from DL0 to DL1.
Besides a calibration of the recorded signals in each camera pixel, this step also comprises a cleaning of the camera image as well as the derivation of image parameters.

\subsubsection{Calibration}
As the calibration procedure is rather specific to the respective instruments, it will not be covered in detail here.
At the camera level, the procedure typically includes the determination of the amplification gain of the photo-sensors for a single photo-electron, the measurement of the typical noise (night-sky background and electronic noise), and the derivation of correction factors to homogenise the response of the sensors across the camera (``flat-fielding'').
It also comprises a step that identifies pixels that should be removed from the analysis because of a malfunction or too bright illumination from stars.
Finally, the signal is reduced (e.g.\ integrated over time) to estimate the amount of light received from the air shower.
For more details, see for example \cite{HESS_Calib_2004}.
Further steps related to the calibration aim at measuring the absolute throughput of the entire telescopes, including in addition to the camera the mirrors and possibly light guides in front of the photo-sensors -- this is necessary to be able to reliably reconstruct the energy of the primary gamma ray.
The throughput calibration can be achieved using light emitted by muons generated in air showers (primarily used for \hess \cite{Mitchell2016} and proposed for CTA \cite{Gaug2019}), but also with dedicated calibration devices (primarily used for VERITAS \cite{VERITAS_Throughput_2022}).

\subsubsection{Image cleaning}
\begin{figure}
    \centering
    \includegraphics[width=0.9\textwidth]{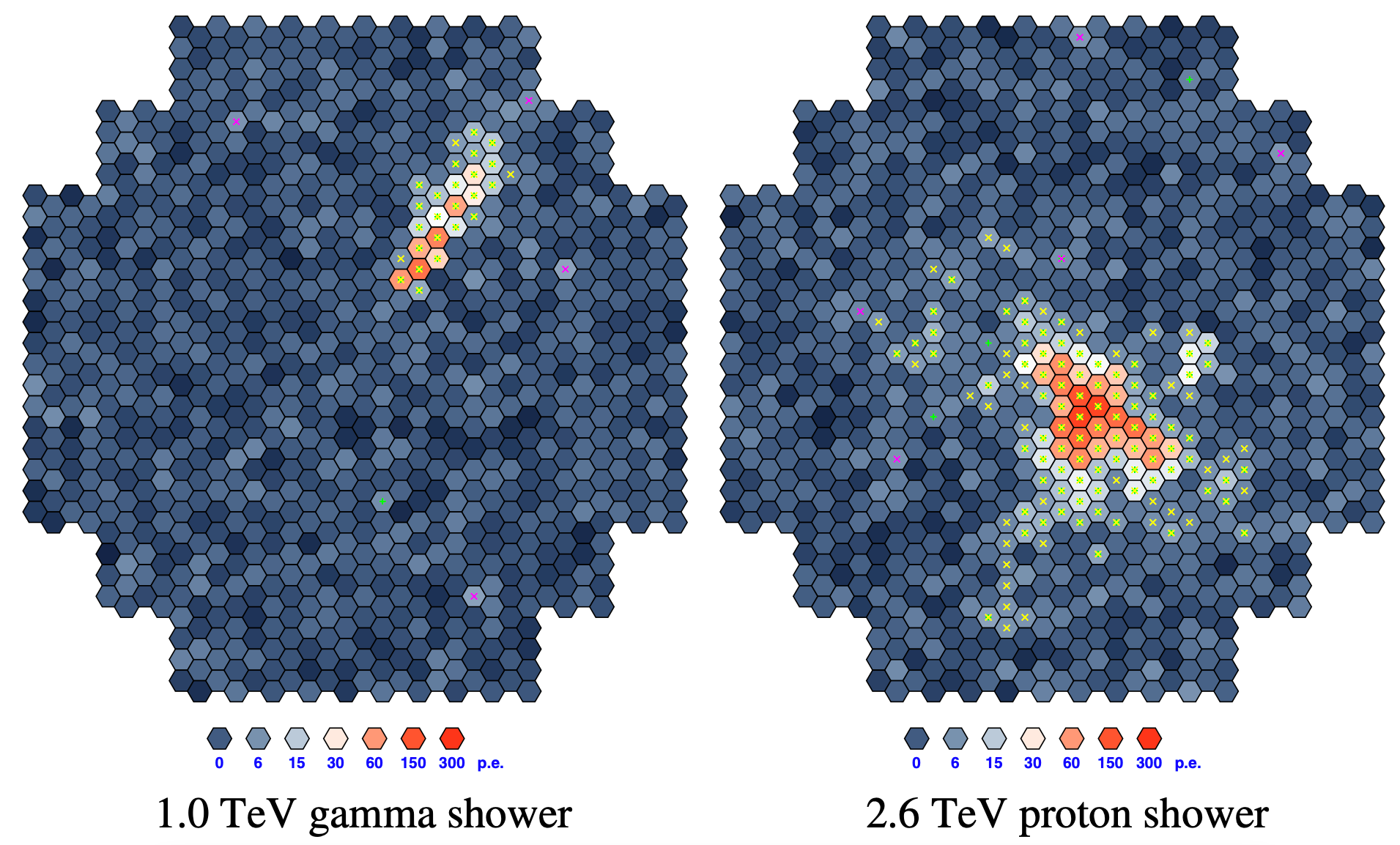}
    \caption{
        Camera images showing the expected signal from a simulated \SI{1}{TeV} gamma-ray shower (left) and from a simulated \SI{2.6}{TeV} proton shower (right).
        The colour code expresses the measured signal in each pixel in units of photo-electrons (p.e.).
        Figure taken from \cite{Voelk2009}.
        Reproduced with permission from the authors \copyright.
    }
    \label{fig:showers_gamma_hadron}
\end{figure}

After the calibration of the signals recorded in the individual camera pixels, a calibrated camera image is obtained.
Fig.~\ref{fig:showers_gamma_hadron} shows camera images obtained from simulated data, for a gamma ray-initiated shower and a proton-initiated shower.
Besides the air shower signal, the images contain noise, that is, signals that are due to night-sky background (NSB) photons or electronic noise.
Before further processing, the images need to be ``cleaned'' from these noise signals.
In contrast to the shower, which appears concentrated in a group of adjacent pixels, the noise signals occur randomly across the camera, typically in isolated pixels.
A commonly used cleaning method that makes use of this is the ``tail-cut cleaning'', which is a filtering algorithm with two signal thresholds; 5\,p.e. and 10\,p.e.\ are commonly used values \cite{HESS_Crab_2006}.
For a pixel to be retained, it requires a signal above 5\,p.e.\ if an adjacent pixel has a signal of at least 10\,p.e., or vice versa.
Naturally, this procedure can also inadvertently remove pixels with only a marginal signal from the air shower.
Lower thresholds (e.g.\ 4\,p.e./7\,p.e.) can be used to rectify this, at the expense of a larger susceptibility to NSB variations.
Similar to the spatial clustering of pixels containing a shower signal, the shower signals are also correlated in time, whereas noise signals are not.
For cameras that provide a time-resolved signal, this can be exploited to further optimise the image cleaning (e.g.\ \cite{MAGIC_Timing_2009,Lombardi2011,Shayduk2013}).
After cleaning, the camera images should consist only of pixels that contain a signal from the air shower.

\subsubsection{Hillas parameters}
\begin{figure}
    \centering
    \includegraphics[width=0.5\textwidth]{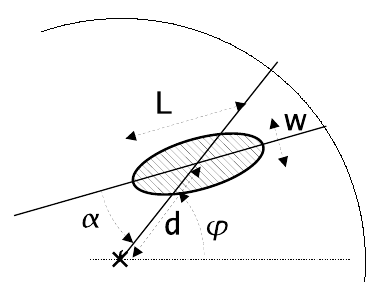}
    \caption{
        Sketch illustrating the definition of the Hillas parameters.
        Figure taken from \cite{deNaurois2012}.
        Reproduced with permission from the author \copyright.
    }
    \label{fig:hillas_parameters}
\end{figure}

The final step of the low-level data processing is the computation of camera image parameters.
First introduced by Hillas \cite{Hillas1985}, these are commonly referred to as ``Hillas parameters''.
The parametrisation makes use of the fact that gamma ray-initiated showers appear in the camera as elongated ellipses (cf.\ Fig.~\ref{fig:showers_gamma_hadron}, left).
The main Hillas parameters are defined as illustrated in the sketch in Fig.~\ref{fig:hillas_parameters}.
They include:
\begin{itemize}
    \item the width ($W$) and length ($L$) of the ellipse;
    \item the total image amplitude (sometimes referred to as the image ``size'');
    \item the nominal distance $d$ between the centre of the camera and the (amplitude-weighted) centroid of the image\footnote{Note that in the original paper by Hillas and sometimes also elsewhere, the ``distance'' parameter is defined as the angular distance between the \emph{source position} and the image centroid. That definition, however, requires a priori knowledge about the source.};
    \item the azimuthal angle $\varphi$ of the image centroid with respect to the camera centre;
    \item the orientation angle $\alpha$ of the ellipse in the camera.
\end{itemize}

These parameters can be computed analytically from the measured signal amplitude $q_i$ and position $(x_i,y_i)$ of each pixel as follows \cite{deNaurois2012}:\\
Starting from the first- and second-order moments
\begin{align*}
    \langle x\rangle = \frac{\sum_i x_iq_i}{\sum_i q_i}\,,&\quad \langle y\rangle = \frac{\sum_i y_iq_i}{\sum_i q_i}\,,\\
    \langle x^2 \rangle = \frac{\sum_i x_i^2q_i}{\sum_i q_i}\,,\quad \langle y^2 \rangle &= \frac{\sum_i y_i^2q_i}{\sum_i q_i}\,,\quad \langle xy \rangle = \frac{\sum_i x_iy_iq_i}{\sum_i q_i}
\end{align*}
and the variances and covariances
\begin{equation*}
    \sigma_{x^2} = \langle x^2 \rangle - \langle x\rangle^2\,,\quad \sigma_{y^2} = \langle y^2 \rangle - \langle y\rangle^2\,,\quad \sigma_{xy}=\langle xy\rangle - \langle x\rangle\langle y\rangle\,,
\end{equation*}
one can define the intermediate variables
\begin{align*}
    \chi &= \sigma_{x^2} - \sigma_{y^2}\\
    z &= \sqrt{\chi^2+4\sigma_{xy}}\\
    b &= \sqrt{\frac{(1+\chi/z)\langle x\rangle^2 + (1-\chi/z)\langle y \rangle^2 - 2\sigma_{xy}\langle x\rangle\langle y\rangle}{2}}\,,
\end{align*}
which then yield the Hillas parameters:
\begin{align*}
    L &= \sigma_{x^2} + \sigma_{y^2} + z\,,& W &= \sigma_{x^2} + \sigma_{y^2} - z\,,\\
    d &= \sqrt{\langle x\rangle^2 + \langle y\rangle^2}\,,& \alpha &= \arcsin\left(\frac{b}{d}\right)\,.
\end{align*}
These analytic expressions allow the computation of the Hillas parameters independently of the shape of the image, that is, they are well defined even if the image is not ellipse-shaped.
It is possible to define further parameters, such as the image ``skewness'' and ``kurtosis'', which are based on the third-order moments of the signal distribution.
These are particularly helpful when only a single telescope image is available, as the parametrisation as a symmetric ellipse leads to two degenerate solutions for the direction of the incoming shower (see following section).

%%%%%%  DL1 --> DL2  %%%%%%
\subsection{Event reconstruction}
\label{sec:iact_event_reco}

In the event reconstruction, the calibrated and cleaned camera images are used to infer the properties of the detected air shower -- it thus represents the transition from DL1 to DL2.
Of primary interest are the incoming direction of the shower and the total deposited energy -- both of which immediately transfer to the primary particle initiating the shower, which is usually assumed to be a gamma ray.
Further shower properties that are typically derived include the impact position on the ground and the depth of shower maximum.

\subsubsection{Event reconstruction with Hillas parameters}
\begin{figure}
    \centering
    \includegraphics[width=0.98\textwidth]{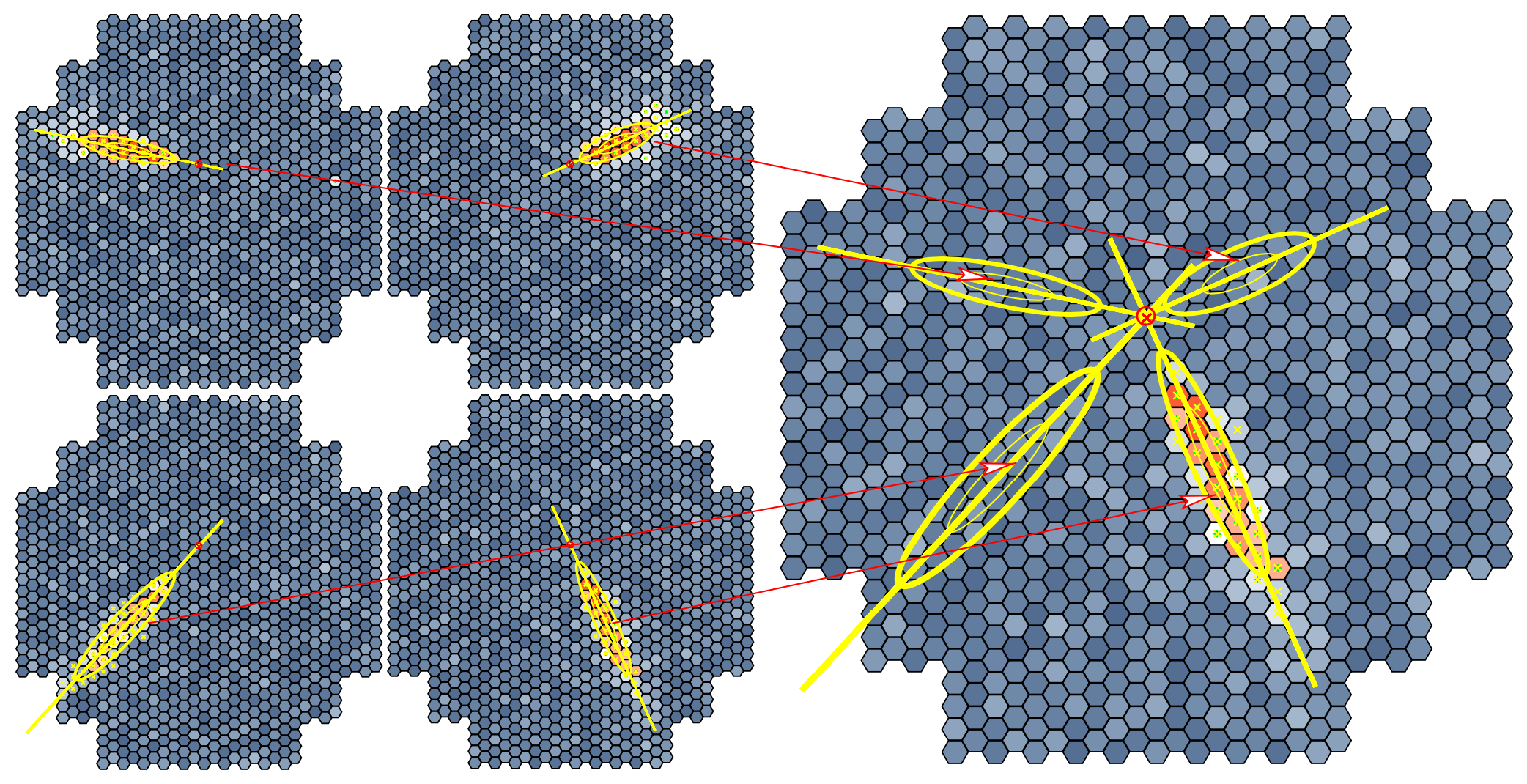}
    \caption{
        Illustration of the stereoscopic event reconstruction method.
        The shower direction is reconstructed by parameterising the shower images as ellipses and intersecting their major axes.
        Figure taken from \cite{Voelk2009}.
        Reproduced with permission from the authors \copyright.
    }
    \label{fig:stereo_reco}
\end{figure}

Traditionally, the event reconstruction is based on the Hillas parameters introduced in the previous section (see e.g.\ \cite{Hofmann1999}).
In the camera coordinate system, the direction of the shower is constrained to lie along the major axis of the ellipse.
When only a single image is available, this can lead to degenerate solutions.
Nevertheless, it is still possible to estimate the direction of the incoming shower by exploiting the small expected asymmetry of the image (e.g.\ \cite{Lessard2001,DomingoSantamaria2005,Murach2015}).
A more reliable way to break the degeneracy, however, is to observe the shower with multiple telescopes (i.e.\ to perform ``stereoscopic'' observations) -- a technique now employed by all current-generation instruments.
In this case, the direction of the shower can be reconstructed by intersecting the major axes of the ellipses obtained in the different camera images, as illustrated in Fig.~\ref{fig:stereo_reco}.
Once the direction of the shower is known, also its impact position on the ground can be determined.
In combination with the image amplitudes, this can in turn be used to infer the energy of the primary gamma ray, typically with the help of look-up tables generated from simulation data.

An important drawback of reconstructing the shower properties with the Hillas parameters is that the major axis of the ellipse can only be reliably determined if the shower image is fully contained in the camera.
For partly contained shower images, the direction reconstruction can be heavily biased.
Therefore, a selection cut on the nominal distance parameter is usually applied, retaining only those images for which the centroid of the signal distribution is not too close to the camera edges.
This reduces the effective area (i.e.\ the gamma-ray detection efficiency) in particular at high energies, where the shower images can fill a substantial fraction of the full camera and are thus more likely not to pass the nominal distance cut.

\subsubsection{Event reconstruction with image templates}
\begin{figure}
    \centering
    \subfigure[]{
        \includegraphics[width=0.55\textwidth]{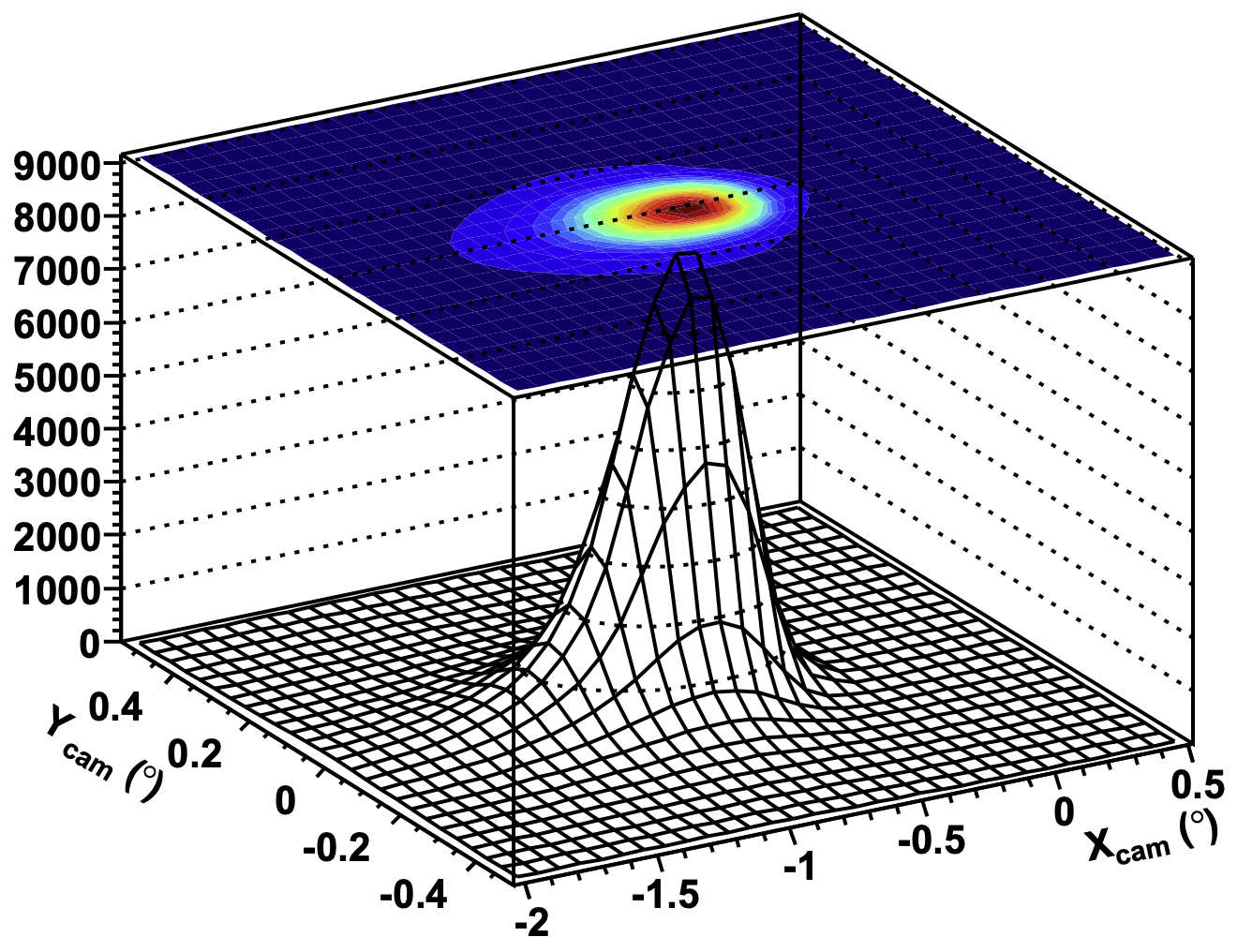}
        \label{fig:impact_template}
    }
    \subfigure[]{
        \includegraphics[width=0.41\textwidth]{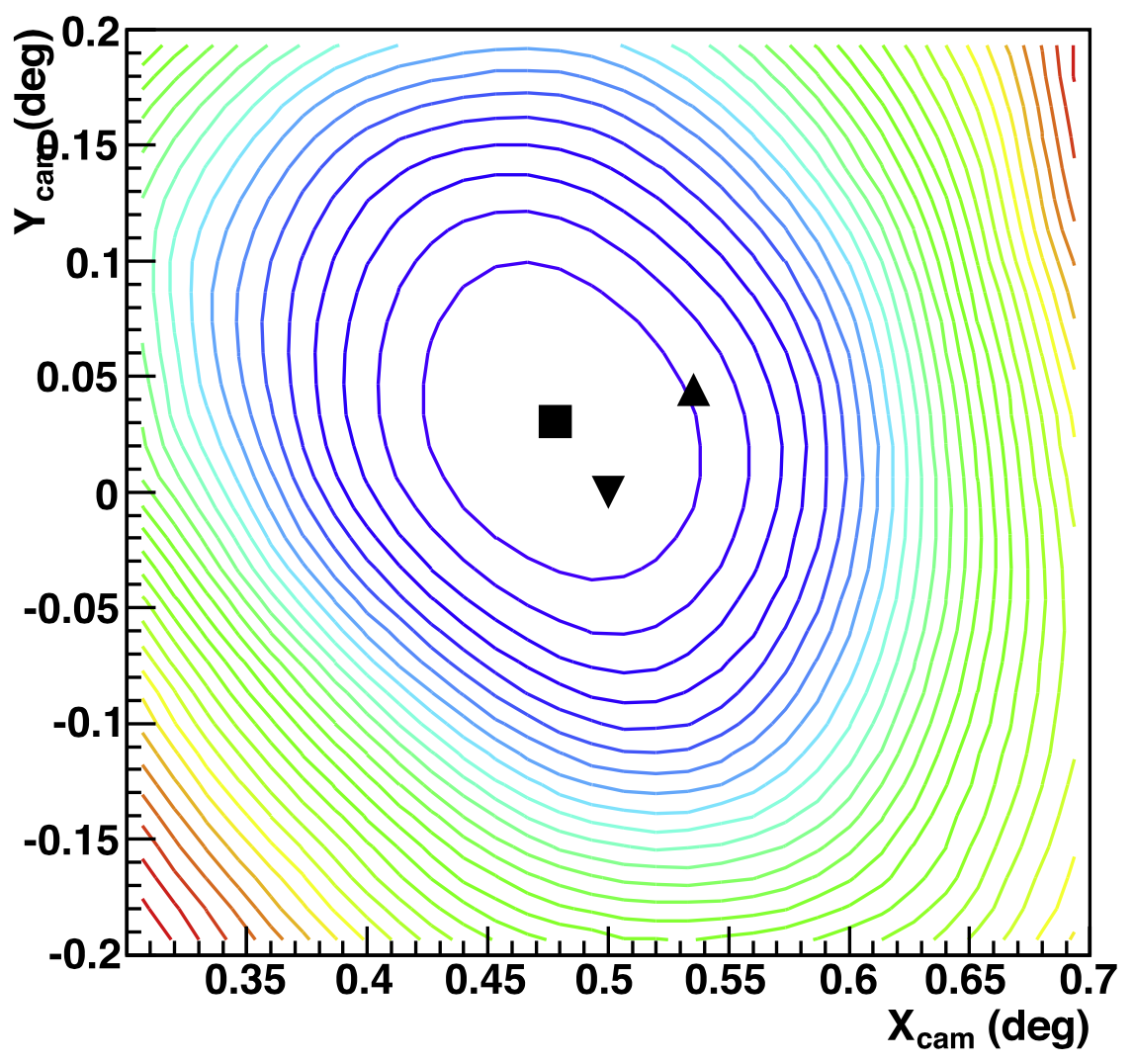}
        \label{fig:impact_likeliood}
    }
    \caption{
        Illustration of the \impct event reconstruction method.
        (a)~Example of a simulated image template, for a \SI{1}{TeV} primary gamma ray with a depth of shower maximum of \SI{300}{\gram\per\square\centi\meter} and an impact point at a distance of \SI{100}{\meter} from the array centre.
        The x- and y-axes display camera coordinates in degrees, the vertical axis is shown in units of photoelectrons per square degree.
        (b)~Two-dimensional projection of a likelihood surface for a simulated example event in the plane of the camera.
        The upward-pointing triangle denotes the direction obtained with a traditional ``Hillas'' reconstruction, whereas the square shows the result of the \impct reconstruction.
        The true simulated shower direction is indicated by the downward-pointing triangle.
        Figures taken from \cite{Parsons2014}.
        Reproduced with permission \copyright\ Elsevier.
    }
    \label{fig:impact_reco}
\end{figure}

A more advanced event reconstruction method is to fit the observed camera images with a model, or template, of a gamma-ray air shower.
This technique has been pioneered for the CAT telescope \cite{LeBohec1998}, where a semi-analytical shower model (based on one originally proposed by Hillas \cite{Hillas1982}) has been obtained from Monte-Carlo simulations and fitted to the observed images by means of a $\chi^2$-minimisation.
The approach has been further developed for application to \hess data \cite{deNaurois2009}, using a more refined shower model as well as employing a likelihood fit in order to account for the Poisson nature of the distribution of the number of recorded Cherenkov photons in each pixel.
A similar method, based on a purely analytical, three-dimensional Gaussian parametrisation of the air shower, has also been applied to \hess data \cite{LemoineGoumard2006}.
Finally, the \impct reconstruction method \cite{Parsons2014}, again developed for \hess, is also based on a likelihood fit but employs a library of smoothed, simulated shower images instead of modelling the air shower.
Inspired by this, a similar algorithm has been developed for VERITAS as well \cite{Vincent2015,Christiansen2017}.
Fig.~\ref{fig:impact_template} displays an example \impct image template, while Fig.~\ref{fig:impact_likeliood} shows a projection of the likelihood for a simulated example event.
Note that, for a stable performance, these algorithms typically require the presence of pixels without a shower signal in the images; this can for example be achieved by restoring a few rings of pixels around the cleaned shower image.

The most evident advantage of these advanced reconstruction algorithms is that they provide an enhanced capability of reconstructing the direction and energy of the primary gamma rays.
For example, at a true gamma-ray energy of \SI{1}{TeV} and the telescopes pointing at $20^\circ$ from zenith, the \impct reconstruction improves the angular resolution (defined as the 68\% containment radius around the true direction) from $\sim$0.09$^\circ$ to $\sim$0.055$^\circ$ compared to a traditional ``Hillas'' reconstruction, and the energy resolution (defined as the RMS of the fractional deviation from the true energy) from $\sim$15\% to $\sim$9\% \cite{Parsons2014}.
Moreover, because it is able to take into account the statistical fluctuations in every pixel, the method is less prone to an overestimation of the energy near the detection threshold of the instrument (which typically occurs with the traditional method because only showers for which the recorded signal over-fluctuates pass the threshold).
Additional advantages of the model-/template-based reconstructions are that they can better handle images with missing information (i.e.\ due to malfunctioning pixels or partly contained showers), and that a goodness-of-fit value can be derived and used to reject cosmic-ray induced air showers (see the following section on gamma/hadron separation).
Their successful application, however, relies on a proper convergence of the fit, which in all mentioned cases proceeds via numerical minimisation algorithms.
This requires a good starting point (``seed position'') for the fit, which is typically obtained from the standard, Hillas parameters-based reconstruction method.
For events with a heavily biased seed position, it is possible that the fit algorithm cannot find the global optimum in the multi-dimensional parameter space.
This can be ameliorated by using multiple seed positions, and taking the result with the largest likelihood value.

\subsubsection{Event reconstruction with deep-learning techniques}
Lastly, inspired by the tremendous success of convolutional neural networks (also referred to as ``deep learning'') in image recognition, there have been multiple recent attempts to apply this technique to IACT event reconstruction.
The networks are typically trained using full (simulated) camera images as input, which has the advantage that one does not need to rely on a particular image parametrisation or shower model.
To give a few examples, the performance of convolutional neural networks at event reconstruction has been assessed using simulated \hess\ \cite{Holch2017,Shilon2019}, CTA \cite{Mangano2018,Miener2021,Aschersleben2021}, and TAIGA \cite{Polyakov2021} data.
First successful applications to real observational data from the CTA LST-1 telescope \cite{Vuillaume2021} and from the MAGIC telescopes \cite{Miener2022} have also recently been achieved.
However, deep learning-based reconstruction algorithms have not yet been demonstrated to be able to outperform the established techniques.
Generally, the application of convolutional neural networks to the task of event classification (i.e.\ background rejection) seems more advanced (see the following section).

%%%%%%  DL2 --> DL3  %%%%%%
\subsection{Gamma/hadron separation}
\label{sec:iact_gamma_hadron_separation}

One of the biggest challenges in the analysis of IACT data is the suppression of the background of cosmic ray-initiated events.
The largest contribution to this background is made by cosmic-ray nuclei (``hadronic'' cosmic rays), which usually outnumber gamma-ray events by several orders of magnitude.
Gamma-ray showers are essentially purely electromagnetic showers, which normally develop very regularly, leading to regular, ellipse-shaped images of the shower in the camera (see Fig.~\ref{fig:showers_gamma_hadron}, left).
In contrast to this, the first interactions in cosmic-ray showers are dominated by hadronic interactions, which are subject to much larger shower-to-shower fluctuations.
Hadronic showers can therefore consist of multiple sub-showers, contain muons that can reach the ground, and generally develop more irregularly compared to gamma-ray showers.
As a consequence, the recorded shower images also appear irregular, exhibiting fringes or multiple clusters of illuminated pixels (see Fig.~\ref{fig:showers_gamma_hadron}, right).
Background rejection methods hence mainly rely on separating the regular, ellipse-shaped images of gamma-ray showers from the more irregular cosmic-ray shower images.

\subsubsection{Event selection with Hillas parameters}
\begin{figure}
    \centering
    \subfigure[]{
        \includegraphics[width=0.63\textwidth]{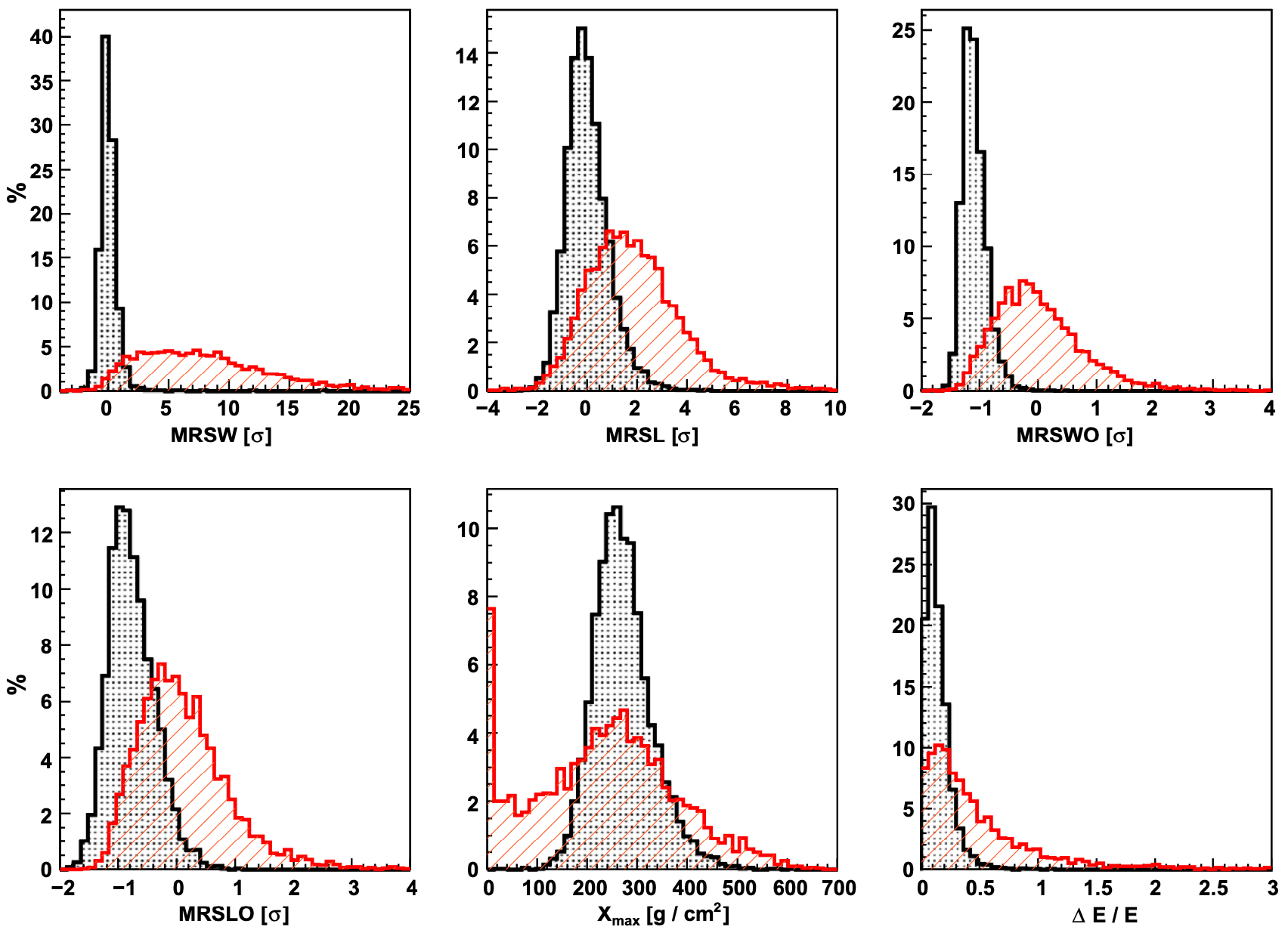}
        \label{fig:tmva_inputs}
    }
    \subfigure[]{
        \includegraphics[width=0.33\textwidth]{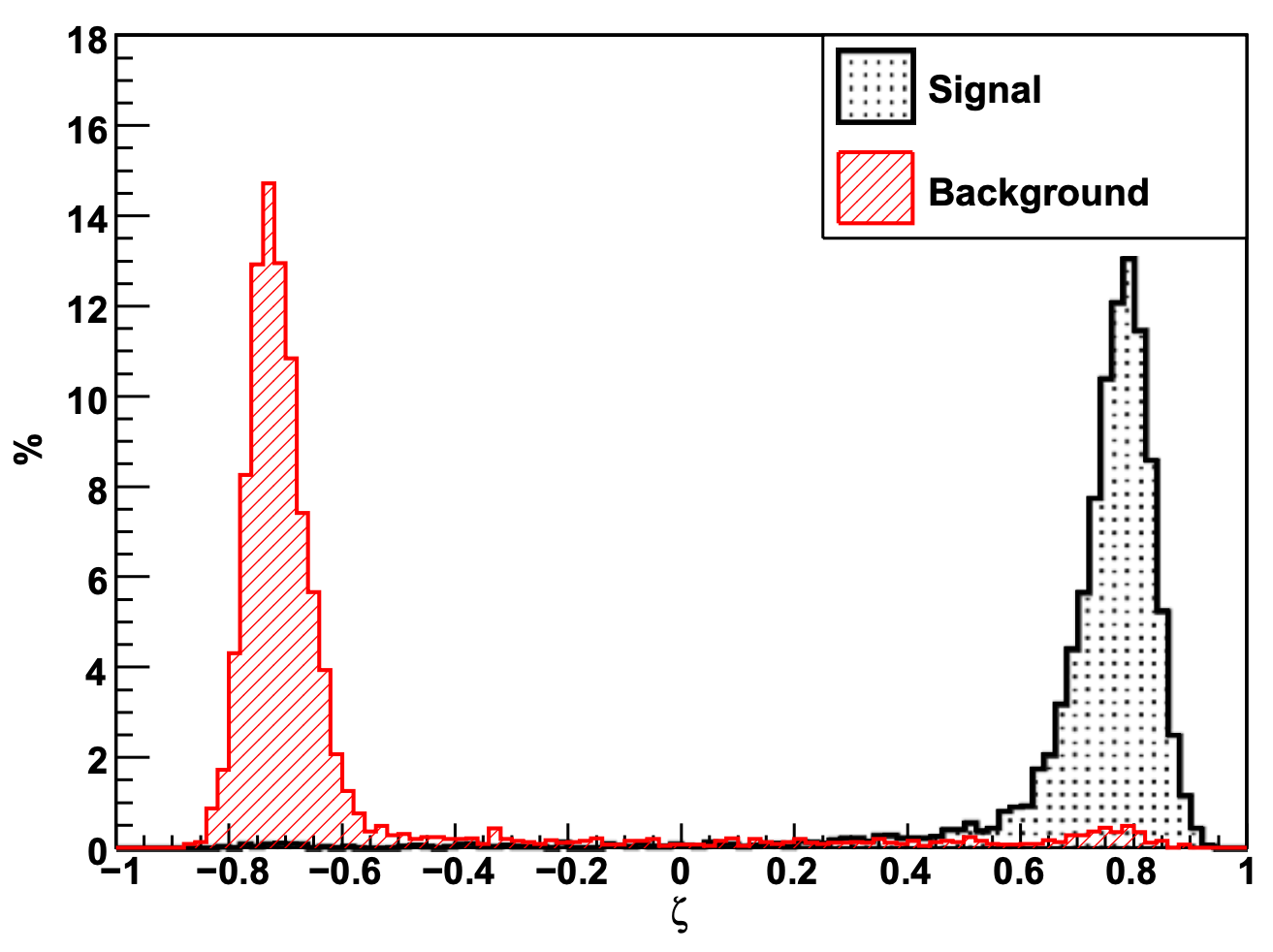}
        \label{fig:tmva_output}
    }
    \caption{
        Distributions of input (a) and output (b) parameters of the machine learning-based gamma/hadron separation method introduced in \cite{Ohm2009}.
        The black histograms show the distributions obtained for simulated gamma rays whereas the red histograms are for ``off'' data, i.e.\ real background events.
        Note that the distributions have been normalised, meaning that they do no reflect the true expected ratio between gamma-ray events and background events in a typical observation.
        Figures taken from \cite{Ohm2009}.
        Reproduced with permission \copyright\ Elsevier.
    }
    \label{fig:gamma_hadron_tmva}
\end{figure}

The simplest method to reject hadronic background events is to apply selection cuts directly on the Hillas parameters, in particular the Hillas width and length.
Because these parameters also depend on the total image amplitude, the observation zenith angle, and the shower impact position, however, so-called ``scaled parameters'' are introduced \cite{HEGRA1997,HESS_Crab_2006}.
For a parameter value $p$, they are defined as
\begin{equation*}
    p_\mathrm{scaled}=(p-\bar{p})/\sigma_p\,,
\end{equation*}
where $\bar{p}$ denotes the mean and $\sigma_p$ the standard deviation of a distribution of parameter values obtained from Monte Carlo simulations of gamma-ray showers.
These quantities are derived for multiple image amplitudes, zenith angles, and impact positions and stored in look-up tables.
Finally, the scaled parameter values are averaged over all telescope images to obtain a \emph{mean reduced scaled width} (MRSW) and a \emph{mean reduced scaled length} (MRSL) for every recorded event.
The first two panels of Fig.~\ref{fig:tmva_inputs} show distributions of MRSW and MRSL for simulated gamma-ray events (black histograms) and background events from real data (red histograms).
It is evident that in particular the MRSW parameter provides relatively good separation power between signal and background events.

The background rejection can be further improved by considering more discriminating parameters, and by exploiting the correlations between these.
Current state-of-the-art methods employ machine-learning techniques to achieve this, namely boosted decision trees (BDTs) in the case of \hess and VERITAS \cite{Ohm2009,Krause2017} and random forests in the case of MAGIC \cite{MAGIC_RF_2008}.
As an example, Fig.~\ref{fig:tmva_inputs} shows distributions of the parameters used as input for the BDT-based gamma/hadron separation method described in \cite{Ohm2009}.
In addition to the MRSW/MRSL parameters, they include corresponding versions scaled to the expectations from ``off'' events (i.e.\ real data from sky regions without gamma-ray sources; MRSWO/MRSLO), the reconstructed depth of shower maximum ($X_\mathrm{max}$), and the relative spread of the shower energy estimated with the different telescopes ($\Delta E/E$).
The BDT classifier is trained using simulated gamma-ray showers as signal and off events as background.
It produces an output variable, labelled $\zeta$ in Fig.~\ref{fig:tmva_output}, that provides an improved discrimination power between signal and background compared to any of the input parameters.
An enhanced suppression of the background can then simply be achieved by cutting on this output parameter.

\subsubsection{Event selection with other approaches}
Other parameters that are not immediately based on the Hillas parameters exist and can also be used to reject hadronic background events (either solely or in combination with the approaches outlined above).
For example, similarly to the image cleaning, additional information can be gained and used for background rejection by exploiting the time evolution of the recorded signals (if provided by the camera).
This has been proposed long ago (e.g.\ \cite{Chitnis2001}) but is currently used routinely only for the analysis of MAGIC data \cite{MAGIC_Timing_2009}, although the approach has been investigated for the upcoming CTA as well \cite{Spencer2021}.
Further options exist if a model- or template-based event reconstruction method is used.
For these methods, it is usually possible to derive a goodness-of-fit parameter that indicates how well the fitted model matches the observed data.
Because the model (or template) assumes a gamma-ray primary, the goodness-of-fit will typically be worse for cosmic-ray showers, and can thus be used to reject such events.
This approach is used as the primary background rejection method in one of the analysis pipelines maintained by the \hess Collaboration \cite{deNaurois2009}.
A slightly different ansatz, dubbed ``ABRIR'' (\emph{Algorithm for Background Rejection using Image Residuals}), has recently been proposed for IACT arrays that include at least one telescope with a very large mirror area (such as \hess, with its central telescope featuring a \SI{28}{m}-diameter mirror) \cite{OliveraNieto2021,OliveraNieto2022}.
In that work, the image recorded by the large telescope is compared to the best-fit predicted shower image resulting from the \impct reconstruction (which uses only the images from the smaller telescopes of the array as input).
Events for which the large-telescope image still exhibits residual signals after subtraction of the predicted image -- for example due to muons or sub-showers -- can then be rejected, which enhances the background suppression in particular at the highest energies ($>\SI{10}{TeV}$).

Finally, as for the case of event reconstruction, multiple groups are engaging in applying deep-learning techniques to the task of gamma/hadron separation for IACTs.
This is a potentially very powerful approach as the full recorded camera images can be fed into the neural networks, thus avoiding the inevitable loss of information pertaining to any parametrisation of the images.
A non-exhaustive list of recent works includes:
\begin{itemize}
    \item \cite{Holch2017,Shilon2019,Parsons2020,Glombitza2023}: application to simulated and real \hess data;
    \item \cite{De2023}: application to simulated \hess data, also investigating the separation of different primary cosmic-ray nuclei as well as anomaly detection;
    \item \cite{Mangano2018,Miener2021,Aschersleben2021,Spencer2021,Nieto2017,Lyard2020}: application to simulated CTA data;
    \item \cite{Polyakov2021}: application to simulated data from the TAIGA experiment;
    \item \cite{Vuillaume2021}: application to simulated and real data from the CTA LST-1 telescope;
    \item \cite{Miener2022}: application to simulated and real MAGIC data.
\end{itemize}
In all of these works, different network architectures are used -- sometimes using solely the camera images as input, sometimes supplying additional parameters.
Explaining the differences between the approaches is beyond the scope of this review; the reader is referred to the listed references.
In summary, the studies on simulated data show that deep learning-based methods have the potential to match or even exceed the performance of established methods for gamma/hadron separation.
However, they have only sparsely been validated with real observational data so far.
Some of the challenges arising from that are described and discussed in \cite{Parsons2022}, where it is found that in particular the noise level in the camera needs to be reproduced extremely well in the Monte Carlo simulations to be able to produce accurate IRFs from them.
Further studies in this direction are needed to be able to fully exploit the potential of convolutional neural networks for background rejection in IACT data analyses.

\subsection{Background modelling}
\label{sec:iact_background_modelling}

It is impossible for any background rejection algorithm to completely eliminate cosmic ray-initiated shower events.
For instance, hadronic air showers in which -- by chance -- a large fraction of the primary-particle energy is channelled into an electromagnetic sub-shower can mimic very closely a shower initiated by a gamma ray.
Moreover, there is an irreducible background of events that are due to cosmic-ray electrons and positrons, which initiate showers that develop identically to gamma-ray showers.
Therefore, the residual background that remains after the gamma/hadron separation procedure needs to be modelled in the analysis.
While it would theoretically be possible to obtain a background model from simulations of hadronic air showers, this approach is not followed in practice for two main reasons: first, because only a very small fraction of showers will pass the event selection, it would require an extraordinary amount of showers to be simulated, which is generally not feasible; second, even small uncertainties in the hadronic interaction models employed in the simulations -- which are not well constrained at the relevant energies \cite{Parsons2019} -- can lead to grossly different background estimates.
Hence, the residual background is essentially always estimated from real observations.

\subsubsection{First success: the On/Off method}
Historically, the first background modelling method that led to the detection of a gamma-ray source (the Crab nebula) in the TeV energy domain by an IACT was the ``On/Off'' method \cite{Weekes1989}.
With this method, the source is tracked by the telescope(s) during an ``On'' observation (typically around \SI{30}{\minute}), followed by an ``Off'' observation of the same duration and taken under approximately the same conditions (same zenith angle, in particular), but targeted at an empty sky region devoid of gamma-ray sources.
The rate of events detected during the Off observation then simply serves as an estimate of the cosmic-ray background for the On observation.
While this method has enabled the first detection of a gamma-ray source with an IACT, it has the obvious disadvantage that it requires spending half of the observation time for the observation of empty sky regions.
Nowadays, it is nonetheless occasionally used for the analysis of extremely extended gamma-ray sources that fill (or exceed) the field of view of the telescopes (e.g.\ \cite{Hona2021,Mitchell2021}).
In these cases, however, the Off observations are typically not taken immediately in succession to the On observations, but are selected from (appropriately matched) archival observations of regions that do not contain a significantly detected gamma-ray source -- thus avoiding the surplus observation time.

\subsubsection{Estimating the background from the observation itself}
\begin{figure}
    \centering
    \includegraphics[width=0.98\textwidth]{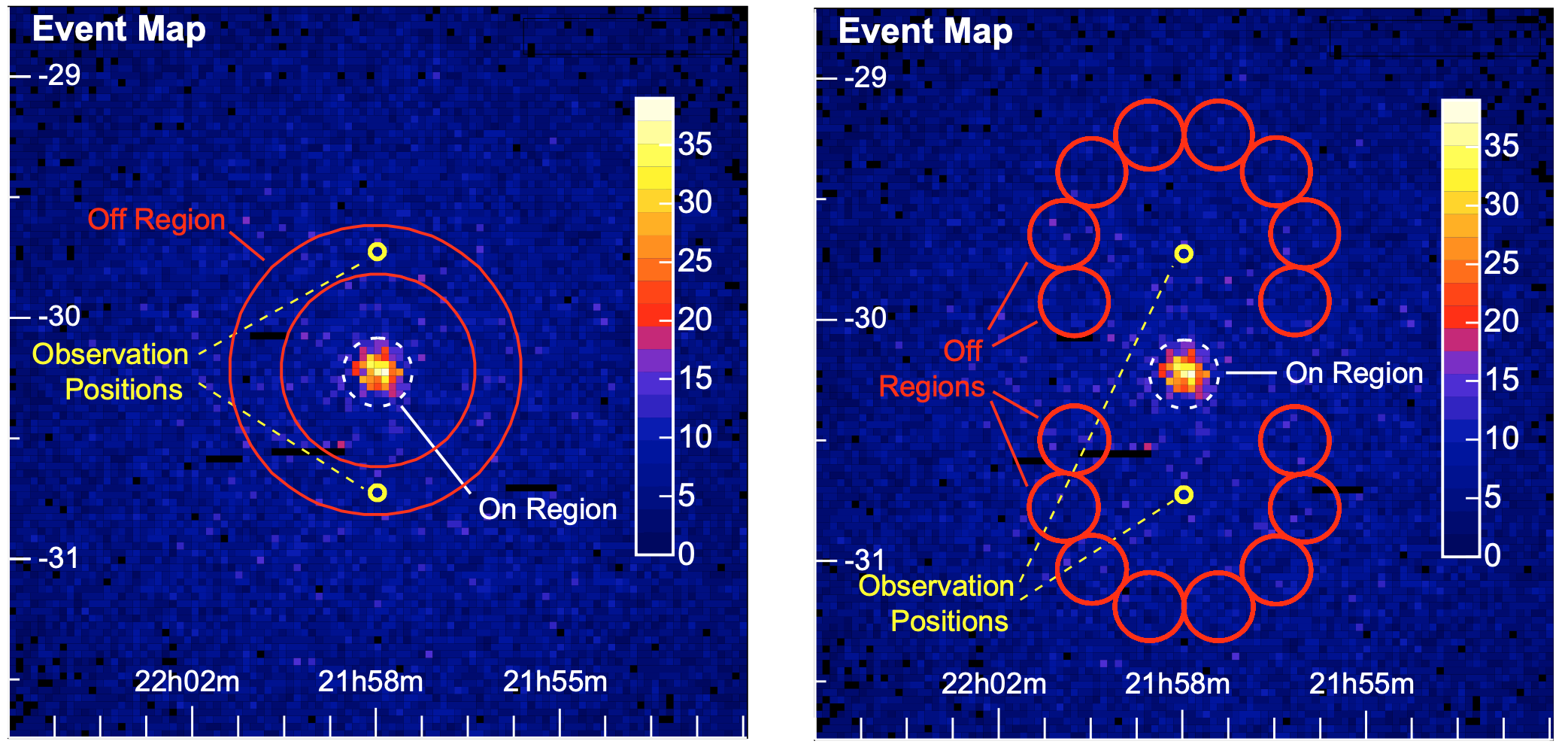}
    \caption{
        Illustrations of the ring background method (left) and the reflected background method (right).
        In the ring method, the background level is estimated from an annulus around the test position.
        In the reflected method, it is taken from multiple Off regions that are placed at the same angular offset from the pointing position as that of the On region, which contains the gamma-ray source.
        The images show observations of the blazar PKS~2155$-$304, which lies in the On region at the centre.
        Figure taken from \cite{Berge2007}.
        Reproduced with permission \copyright\ ESO.
    }
    \label{fig:ring_reflected_background}
\end{figure}

Many more background modelling methods have been developed in the meantime; a comprehensive overview is given in \cite{Berge2007}.
In most of the approaches, the residual hadronic background is estimated from the observation itself.
For instance, in the two most widely applied methods -- the ``ring background'' method and the ``reflected background'' method -- it is derived from ``Off'' regions in the observed field of view that do not contain significant gamma-ray emission.
Both methods are illustrated in Fig.~\ref{fig:ring_reflected_background}.
The ring background method can be used to estimate the background rate anywhere in the field of view, but requires a model of the background acceptance to correct for the difference in exposure between the ring region and the source region.
The reflected background method, on the other hand, provides a background estimate for a specific region only, but does not need a background acceptance model.
Note, however, that it can only be applied if the targeted source is not in the centre of the field of view, that is, the telescopes need to point at a position that is offset (by typically $0.5^\circ$ or $0.7^\circ$) from the source position (this was first proposed in \cite{Fomin1994}).
Furthermore, it is important that for both the ring background method and the reflected background method, regions that are known to contain (previously detected) gamma-ray sources are excluded from the background determination.
Another option of estimating the background from the observation itself is to define the Off region not spatially, but using another parameter that provides different acceptance for signal and background events.
For instance, one can obtain a template for the hadronic background by selecting events from an interval of the MRSW parameter that is dominated by cosmic-ray events rather than gamma-ray events \cite{Rowell2003,Berge2007}.
To account for the difference in acceptance to background events, this template then needs to be normalised within some source-free region before applying it as a background model for the gamma-ray candidate events.

\subsubsection{Background model from archival observations}
\begin{figure}
    \centering
    \includegraphics{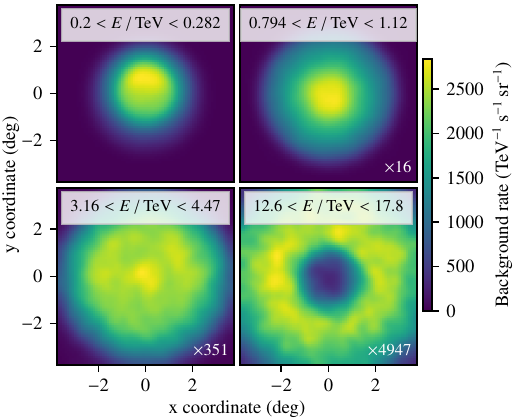}
    \caption{
        Two-dimensional projections in different energy bands of the hadronic background model introduced in \cite{Mohrmann2019}.
        The model is shown in field-of-view coordinates, in which the telescope pointing direction is in the centre and the y-axis is aligned with the altitude angle.
        Figure taken from \cite{Mohrmann2019}.
        Reproduced with permission \copyright\ ESO.
    }
    \label{fig:3d_background_model}
\end{figure}

Finally, it is also possible to generate a model of the residual hadronic background not from the observation of interest itself, but from archival observations (in the terminology of \cite{Berge2007}, this is known as the ``field-of-view background'').
In contrast to the above-mentioned On/Off method, which may also employ archival observations to estimate the background, here the model is built from \emph{multiple} observations, which reduces statistical fluctuations.
As is the case for the On/Off method, however, the archival observations used to produce the model need to be chosen carefully, to match as closely as possible the observation conditions of the observation to be ana\-lysed.
Even then, the model usually still needs to be normalised to the data, using sky regions without gamma-ray contamination.
Traditionally, field-of-view background models have been constructed assuming a rotational symmetry -- that is, they were two-dimensional, depending on the reconstructed energy and the radial offset from the pointing axis.
Recently, also three-dimensional models, featuring two spatial coordinates, have been developed.
Asymmetries in the background acceptance can for example be present if the array of telescopes exhibits a preferred axis, as is the case for the MAGIC telescopes \cite{DaVela2018}.
They may, however, also occur for symmetric arrays, as is shown in Fig.~\ref{fig:3d_background_model} for a model created for the \hess telescopes \cite{Mohrmann2019}, where an asymmetry is clearly visible in the lowest displayed energy range (caused most likely by a dependency of the background acceptance on the altitude angle).
Field-of-view background models have experienced an increased interest in the last years, as they are a necessary ingredient for likelihood-based high-level analyses of IACT data, which are becoming more and more popular (see \cite{Mohrmann2019} and Sect.~\ref{sec:iact_high_level_analysis}).

\subsection{Generation of instrument response functions}
\label{sec:iact_irf_generation}

The result of the event reconstruction (Sect.~\ref{sec:iact_event_reco}) and gamma/hadron separation (Sect.~\ref{sec:iact_gamma_hadron_separation}) steps -- a list of selected gamma-ray candidate events with reconstructed properties -- constitutes the first part of the DL3 level data.
To be able to derive physical results from these measured events, however, it is also necessary to ascertain the performance of the instrument -- that is, to derive its IRFs -- with respect to the applied event selection criteria.
The IRFs depend on the (true or reconstructed) gamma-ray energy ($E_\mathrm{true}/E_\mathrm{reco}$) and direction ($d_\mathrm{true}/d_\mathrm{reco}$).
Normally considered are:
\begin{itemize}
    \item the effective area $A_\mathrm{eff}(E_\mathrm{true}, d_\mathrm{true})$, which is the product of the detector collection area with the gamma-ray detection efficiency;
    \item the point-spread function (PSF) $f_\mathrm{PSF}(d_\mathrm{reco}|E_\mathrm{true},d_\mathrm{true})$, which gives the probability of measuring a direction $d_\mathrm{reco}$ for an event with energy $E_\mathrm{true}$ and true direction $d_\mathrm{true}$;
    \item the energy dispersion (EDISP) $f_\mathrm{EDISP}(E_\mathrm{reco}|E_\mathrm{true}, d_\mathrm{true})$, which gives the probability of reconstructing an energy $E_\mathrm{reco}$ for an event with true energy $E_\mathrm{true}$ and direction $d_\mathrm{true}$;
    \item the background model $R_\mathrm{bkg}(E_\mathrm{reco}, d_\mathrm{reco})$, which specifies the predicted rate of background events at a given reconstructed energy $E_\mathrm{reco}$ and direction $d_\mathrm{reco}$.
\end{itemize}
The first three of this list can be derived from end-to-end Monte-Carlo simulations of air showers and their detection by the instrument.
The most wide-spread program for this is the \texttt{sim$\_$telarray} package \cite{Bernloehr2008}, but simulations based on the \texttt{KASKADE} package \cite{Kertzman1994} are also still being used.
Typically, look-up tables for the IRFs are generated for a grid of parameter values (e.g.\ zenith angle, azimuth angle, etc.), which are then interpolated to obtain the IRFs for a specific observation (see e.g.\ \cite{HESS_Crab_2006}).
However, also an approach in which dedicated simulations are performed for every observation carried out (``run-wise simulations'') has been proposed and is employed regularly for one of the analysis pipelines in use in the \hess Collaboration \cite{Holler2020}.
This enables a fine-tuning of parameters characterising the observation conditions in the simulations -- something that may increasingly be called for with CTA, which, owing to its many planned telescopes, will likely carry out observations in many different array configurations.
Finally, the background model is normally derived from real data as described in Sect.~\ref{sec:iact_background_modelling}.

A major systematic uncertainty in the derivation of the IRFs lies in the usually uncertain conditions of the atmosphere -- in particular the level of aerosols, which affects the transparency to Cherenkov photons and thus directly impacts the reconstruction of the gamma-ray energy.
An estimate of the atmospheric conditions, the Cherenkov transparency coefficient (CTC), can be obtained from the trigger rates of the telescopes, however \cite{Hahn2014}.
This quantity can then be used to take the atmospheric conditions into account when deriving the IRFs -- either as a simulation parameter in the case that run-wise simulations are used, or as a correction to the IRFs generated from simulations that assume an atmosphere with some nominal transparency.

%%%%%%  DL3 --> DL4  %%%%%%
\subsection{High-level data analysis}
\label{sec:iact_high_level_analysis}

The high-level data analysis produces the scientific output of the analysis, for example sky maps or energy spectra (i.e.\ DL4 data).
Similarly to the \flat data analysis (cf.\ Sect.~\ref{sec:flat_data_analysis}), the high-level analysis methods for IACTs can be distinguished into two broad classes: \emph{aperture photometry}, that is, the generation of sky maps or energy spectra on the basis of event counting, and \emph{three-dimensional (3D) likelihood analyses}, which attempt to simultaneously model the observed data spectrally and morphologically by means of a likelihood fitting procedure.
Both will be discussed in turn in the following.

\subsubsection{Aperture photometry}
Aperture photometry techniques have predominantly been used in the field since its initiation until a few years ago, and are still widely being applied.
Given a number of measured events in an ``On'' region around the (putative) gamma-ray source, $N_\mathrm{ON}$, and a number of events in one or multiple ``Off'' regions that are supposed to be devoid of gamma-ray emission, $N_\mathrm{OFF}$, they compute the gamma-ray excess as
\begin{equation*}
    N_\gamma = N_\mathrm{ON} - \alpha\cdot N_\mathrm{OFF}\,,
\end{equation*}
where $\alpha$ is a normalisation factor that corrects for the difference in exposure between the On and Off region(s).
The significance of the observed excess can be determined following the Li\&Ma formula \cite{Li1983}:
\begin{align*}
    S \,\,=\,\,& \sqrt{2}\left\{N_\mathrm{ON}\cdot \ln\left[\frac{1+\alpha}{\alpha}\left(\frac{N_\mathrm{ON}}{N_\mathrm{ON}+N_\mathrm{OFF}}\right)\right]\right. + \\
    &\left. \qquad N_\mathrm{OFF}\cdot \ln\left[(1+\alpha)\left(\frac{N_\mathrm{OFF}}{N_\mathrm{ON}+N_\mathrm{OFF}}\right)\right]\right\}^{1/2}\,.
\end{align*}

A typical application of this is the computation of sky maps that denote the significance of an observed gamma-ray excess in a certain sky region.
Most often, such maps are generated with the ring background estimation method, which allows an estimation of the residual hadronic background in the entire field of view of the telescopes.
In this case, the normalisation factor $\alpha$ needs to be computed by integrating the background acceptance in the On and Off region and taking the ratio of the two.

The extraction of energy spectra, on the other hand, typically relies on the reflected background method.
Because the background acceptance can be assumed to be equal in all regions in this case, $\alpha$ is simply given by the inverse of the number of Off regions used.
To convert from a number of measured gamma-ray excess events -- which still depends, for example, on the efficiency of the instrument -- to a physical quantity (i.e.\ a gamma-ray flux), a procedure called ``forward folding'' is normally employed \cite{Piron2001}.
In it, a model for the gamma-ray source flux, $\Phi(\hat{\theta})$, depending on some parameters $\hat{\theta}$, is folded with the IRFs to obtain a predicted number of gamma-ray events,
\begin{equation}\label{eq:npred}
    N_\mathrm{pred}(\hat{\theta}) = f_\mathrm{EDISP} \cdot (f_\mathrm{PSF} \cdot (A_\mathrm{eff}\cdot t_\mathrm{obs}\cdot \Phi(\hat{\theta})))\,,
\end{equation}
where $t_\mathrm{obs}$ is the observation time.
The optimal parameters of the model can then be determined by comparing the predicted number of gamma-ray events with the observed excess, usually with a Poisson likelihood fit.
This method is also commonly employed in X-ray spectroscopy, see for example the chapter ``Statistical Aspects of X-ray Spectral Analysis'' in this book.

In addition to a spectral model, it is common to compute flux points that denote the measured gamma-ray flux in a sequence of small energy ranges.
A standard method to achieve this is to repeat the forward-folding fit in separate ranges of reconstructed energy, re-fit the flux normalisation in each range (keeping other model parameters fixed), and take the best-fit normalisation as a measurement of the flux in the respective range.
While this method is simple and straightforward to apply, it implicitly assumes that the energy reconstruction of the events exhibits no (or only a negligible) bias, as the flux normalisation obtained for a range in reconstructed energy is commonly quoted for a range in true gamma-ray energy.
A way of avoiding this is to apply an unfolding algorithm for the computation of flux points (see e.g.\ \cite{HEGRA1999,MAGIC_Unfolding_2007,Milke2013}).
For these methods, however, it is necessary to specify regularisation terms that contain an assumption about the correlation between the different flux points.

Aperture photometry works very well for isolated, only marginally extended gamma-ray sources.
Because it relies on the counting of events in a fixed sky region, however, it is difficult to apply when multiple gamma-ray sources contribute to the emission in that region, as their individual contributions cannot easily be separated.
Furthermore, because it requires the availability of regions without gamma-ray emission in the observed field of view for background estimation (at least when the ring or reflected background methods are applied), it can often not be employed for the analysis of largely extended gamma-ray sources, or large-scale diffuse gamma-ray emission.
In such cases, a 3D likelihood analysis -- as introduced in the following section -- is the better choice of analysis method.

\subsubsection{3D likelihood analysis}\label{sec:iact_likelihood_analysis}
In a 3D likelihood analysis, the observed data are fitted with source models that simultaneously specify the energy spectra and spatial morphologies of all sources in the region of interest.
The analysis is again based on the forward-folding method, that is, proceeds via a calculation of the number of events predicted by the model(s) following Eq.~\ref{eq:npred}, but now taking into account two spatial dimensions in addition to the energy dimension.
A Poisson likelihood fit is then used to determine the optimal source parameters.
This approach is essentially identical to that usually used for the analysis of \flat data, see Sect.~\ref{sec:flat_likelihood_analysis} for more details.
An important caveat to note is that the 3D likelihood analysis method requires an accurate model for the residual hadronic background (see also Sect.~\ref{sec:iact_background_modelling}).

\begin{figure}
    \centering
    \includegraphics[width=0.98\textwidth]{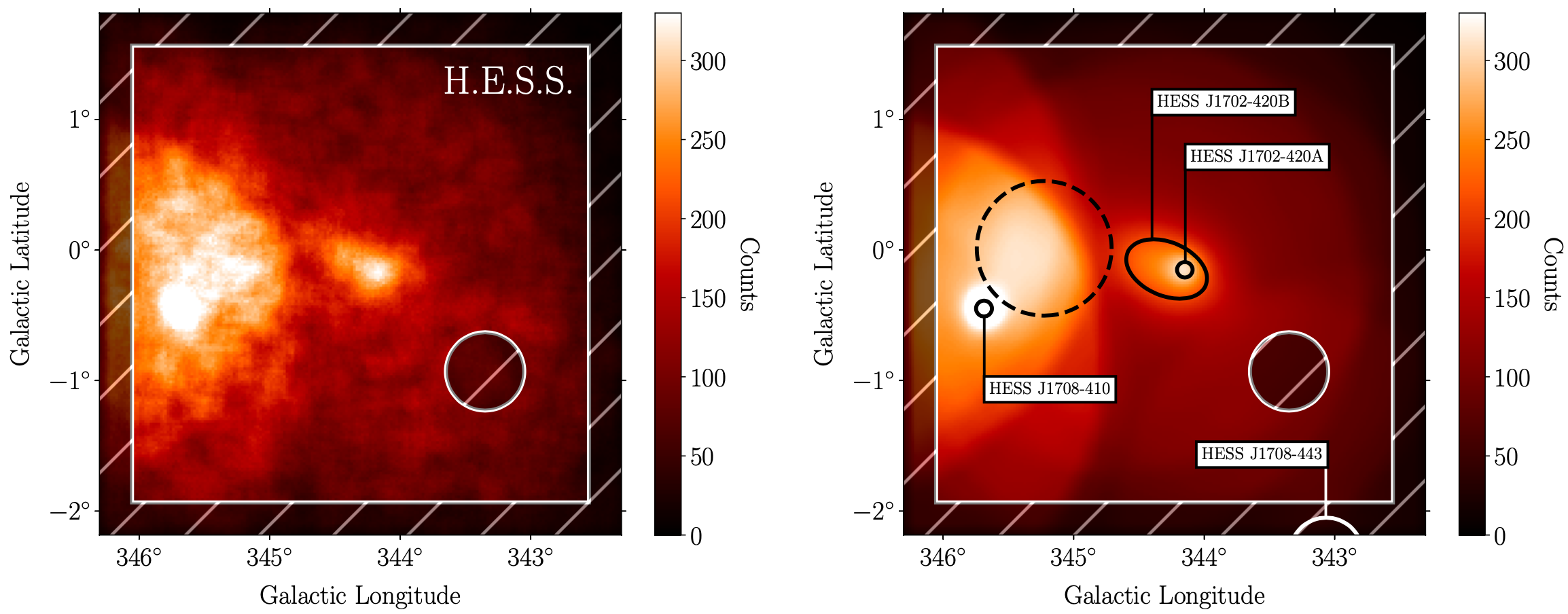}
    \caption{
        Illustration of the 3D~likelihood analysis method.
        The image on the left shows events recorded with \hess in a region around HESS~J1702$-$420, where events with reconstructed energies above \SI{2}{TeV} have been selected and the image has been correlated with a top-hat kernel of $0.1^\circ$ radius.
        The image on the right shows the number of events predicted by the best-fit model obtained in the likelihood analysis, which includes various source components (for HESS~J1702$-$420 and other nearby gamma-ray sources) as well as a model for the residual hadronic background.
        Hatched regions have been excluded from the fit.
        Figure taken from \cite{HESS_J1702_2021}.
        Reproduced with permission from the authors \copyright.
    }
    \label{fig:3d_analysis_j1702}
\end{figure}

In the past, the 3D likelihood method has only rarely been applied to IACT data.
To a large part, this is due to the difficulty in constructing a reliable background model.
This is because, contrary to \flat, which normally operates under relatively stable conditions, the observation conditions of IACTs can vary drastically from observation to observation.
A 3D background model has recently been developed for \hess, however, and its use with \hess data validated \cite{Mohrmann2019}.
As a result, 3D likelihood analyses are now routinely being applied to \hess data (e.g.\ \cite{HESS_J1702_2021,HESS_Wd1_2022,HESS_J1809_2023}).
An example of this is shown in Fig.~\ref{fig:3d_analysis_j1702}, which illustrates the resolution of the gamma-ray source HESS~J1702$-$420 into two distinct, overlapping components \cite{HESS_J1702_2021} -- a result that would not have been possible to obtain with established aperture photometry techniques.
Efforts to establish the 3D likelihood analysis for MAGIC \cite{Vovk2018} and VERITAS \cite{Cardenzana2017} have also been made.

\subsubsection{Open software tools for IACT data analysis}
As already mentioned in Sect.~\ref{sec:iact_data_levels_formats}, CTA, the major upcoming IACT facility, will be operated as an observatory, and make its data publicly available.
Besides an open data format, this has sparked the development of open-source software tools for the analysis of CTA data.
Specifically, the packages \code{ctools} \cite{Knoedlseder2016} and \code{Gammapy} \cite{Deil_Gammapy_2017} have been proposed as science analysis tools for CTA, and in 2021 the CTA observatory has selected \code{Gammapy} as its official science analysis tool\footnote{see \href{https://www.cta-observatory.org/ctao-adopts-the-gammapy-software-package-for-science-analysis}{this https URL}.}.
Both packages can also be applied to the analysis of data from the current generation of IACT instruments, and have been validated using an open \hess test data set \cite{Mohrmann2019,Knoedlseder2019}.
Today, \code{Gammapy} is used as a standard analysis tool in the \hess Collaboration.
A detailed explanation of how to use the open software tools is beyond the scope of this review chapter.
Instead, we refer the reader to the comprehensive online tutorials provided for each of the tools\footnote{See \href{http://cta.irap.omp.eu/ctools/users/tutorials/index.html}{http://cta.irap.omp.eu/ctools/users/tutorials/index.html} for the \code{ctools} package and \href{https://docs.gammapy.org/1.1/tutorials/index.html}{https://docs.gammapy.org/1.1/tutorials/index.html} for \code{Gammapy}.}.

The use of open-source software tools like \code{ctools} and \code{Gammapy} has several advantages.
First, from the beginning of their development, both packages supported 3D likelihood analyses (which is not true for all of the closed-source software so far used for \hess, MAGIC, or VERITAS data analysis).
Second, in combination with a shared and open data format (cf.\ Sect.~\ref{sec:iact_data_levels_formats}), they enable multi-instrument analyses, in which DL3 data from several instruments can be combined and fitted jointly -- as demonstrated, for example, in \cite{Nigro2019}.
Recently, this work has been extended to include, for the first time, data from a ground-level particle-detector gamma-ray experiment, namely the HAWC observatory \cite{HAWC2022}.
Finally, the tools can easily be used together with other open-source software.
For example, \code{Gammapy} includes a wrapper class for the \code{naima} package \cite{Zabalza2015}, which allows one to fit physical gamma-ray models (e.g.\ a model for inverse Compton emission) directly to the DL3 level data.

\subsection{Similarities and differences for ground-level particle detector arrays}
\label{sec:particle_detectors}

Instead of imaging the air shower, ground-level particle detector arrays (GPDAs) effectively take a snapshot of it when it hits the ground.
They do so by detecting the secondary particles of the shower with an array of detectors placed at high altitude.
HAWC employs an array of water-Cherenkov tanks for this purpose, whereas scintillation detectors are used in the Tibet array.
The LHAASO observatory uses a combination of the two approaches.
Because the detector stations can be operated with a duty cycle of close to 100\%, and because GPDAs can simultaneously observe a large part of the overhead sky (i.e.\ compared to IACTs they have a very large field of view), they typically offer a larger exposure.
This is particularly relevant at energies of around \SI{100}{TeV} and above, where the currently operating IACT arrays often run out of statistics.
The increase in exposure comes, however, at the expense of a worse resolution both in terms of the incoming direction and the energy of the primary gamma ray.
For a summary of the particle-detector array technique, we refer the reader to the chapter ``Detecting gamma-rays with moderate resolution and large field of view: Particle detector arrays and water Cherenkov technique'' in this book.

In the following, iterating through the topics discussed for IACT arrays in the preceding sections, we highlight which aspects also apply to GPDAs, and for which ones there are differences.
\begin{itemize}
    \item \emph{Data levels and formats} ---
    With only a few generalisations, the data levels defined in Sect.~\ref{sec:iact_data_levels_formats} can also be used to describe GPDA data.
    In particular, levels DL2--DL4 apply identically.
    This implies that the open GADF DL3 data format can -- once appropriately extended -- be used to store GPDA data as well (while lower-level data formats are obviously not the same).
    \item \emph{Low-level data processing} ---
    The calibration of the photo-sensors in GPDA detector stations proceeds in a similar manner as for those in IACT cameras.
    Instead of the Hillas parameters, events are characterised by other parameters, such as the fraction of detector stations that have triggered on an air shower event, $f_\mathrm{hit}$.
    \item \emph{Event reconstruction} ---
    The event reconstruction is arguably where the largest differences between GPDAs and IACTs lie.
    For instance, while with IACTs the shapes of the recorded images of the air shower can be used to infer its incoming direction, GPDAs need to rely on the time differences between the signals recorded in the individual detector stations for this task.
    Similarly, the primary gamma-ray energy is reconstructed in a different way.
    Using the HAWC detector as an example, approaches to energy reconstruction include simple estimates using the $f_\mathrm{hit}$ parameter \cite{HAWC2017}, but also more advanced methods that are based on the lateral distribution of the recorded air shower signals or that employ neural networks \cite{HAWC2019}.
    A shower template-based algorithm -- similar to the \code{ImPACT} reconstruction for IACTs discussed above -- that can simultaneously infer the core position and energy of the shower has also been proposed~\cite{Joshi2019}.
    \item \emph{Gamma/hadron separation} ---
    Like for IACTs, cosmic ray-induced air showers strongly dominate over gamma-ray showers for GPDAs and need to be suppressed, exploiting the differences in the shower development.
    As is the case for IACT camera images, hadronic air showers usually leave an irregular footprint on the ground.
    One method to reject such events is therefore to define parameters that capture the sub-structures and use these as discriminating variables.
    Another technique that is very powerful for GPDAs is to reject cosmic-ray events through the identification of muons, which occur much more frequently compared to gamma-ray showers.
    Because muons penetrate deeper than other secondary particles, this can for example be achieved by using double-layered detector stations, where the signals in the lower layer are more likely to be due to muons.
    This technique works especially well at high energies, for instance, the LHAASO array reaches rejection factors exceeding $10^4$ above \SI{100}{TeV} \cite{LHAASO2021a}.
    \item \emph{Background modelling} ---
    As for IACTs, the residual hadronic background remaining after the gamma/hadron separation needs to be estimated from data.
    The standard algorithm for doing this is called ``direct integration'', in which the all-sky event rate is convolved with the distribution of arrival directions in detector coordinates within a \SI{2}{\hour} time window \cite{MILAGRO2012,HAWC2017}.
    \item \emph{Generation of instrument response functions} ---
    The analysis of GPDA data requires the same set of IRFs that is also necessary for IACT data.
    Furthermore, as is usually the case for IACTs, the effective area, PSF, and energy dispersion are normally derived from simulations, while the residual hadronic background is modelled from real data.
    One advantage of GPDAs is that they are less susceptible to variations in the atmospheric conditions compared to IACTs.
    \item \emph{High-level data analysis} ---
    The standard high-level data analysis method applied for GPDA data is the 3D likelihood analysis, outlined in Sects.~\ref{sec:flat_likelihood_analysis} and~\ref{sec:iact_likelihood_analysis}.
    For the most part, these are being carried out with custom, proprietary software tools.
    Nevertheless, HAWC data have been combined with that of other instruments using the 3ML framework \cite{Vianello2015}.
    As the DL3 level data can be stored in the same format as that of the IACT arrays, however, it is also possible to analyse them with the above-mentioned open-source tools, as recently demonstrated for the case of HAWC and \code{Gammapy} \cite{HAWC2022}.
    However, as is the case for current IACT arrays, data from HAWC, Tibet, and LHAASO are generally not publicly available.
\end{itemize}

% \appendix
\newpage
\section{Multi-wavelength spectral modelling}
\label{sec:mwl_modeling}
This section briefly describes the approach typically used for modelling of the HE/VHE spectra obtained as a result of the analysis of data from corresponding instruments, as described above.
Contrary to the typical X-ray spectral modelling (see the chapter ``Modelling and simulating spectra'' by L.\ Ducci and C.\ Malacaria), the analysis usually already provides spectra that represent a measurement of the gamma-ray flux (in physical units) as a direct function of gamma-ray energy.
These spectra can thus in principle be analysed by any third-party software able to read the spectral data points and fit them with a certain model. 

Such a fitting procedure can be used to find the best-fit parameters of an empirical model that describes the gamma-ray spectrum (e.g., determine a spectral index in case of a power-law model).
However, in many cases a modelling of the spectral data aims at constraining the properties of the primary particle population that is responsible for the production of the gamma-ray emission, and recovering the physical conditions in the emitting region.
Such an analysis requires the development of a physically motivated model that can predict the gamma-ray spectrum as a function of the parameters of the primary particle population and the properties of the source region.
It is not uncommon, however, that when taking into account data in the HE and VHE regime only, parameter degeneracies occur and no unambiguous modelling solution exists.
For instance, if the HE and VHE emission originates from inverse Compton (IC) scattering of a population of relativistic electrons, its intensity is proportional to the product of the electron density and the density of the IC seed photon field in the region.
Hence, these quantities cannot independently be determined.
Such a degeneracy can be broken by including into the modelling observations of the source at other wavelengths, where the emission is produced by another physical mechanisms.
Staying with our example above, the addition of measurements of the synchrotron emission of the relativistic electron population, which is proportional to the electron density and the magnetic field strength in the region, may allow to constrain, for example, the IC photon field density.

Thus, the development of a physically motivated model for the emission from the observed gamma-ray source often requires a self-consistent modelling of its multi-wavelength spectrum.
As an example, we present in what follows a multi-wavelength modelling analysis of the spectrum of the Crab nebula with a one-zone synchrotron self-Compton (SSC) model.
We do so by employing the widely-used \code{naima} Python package \cite{Zabalza2015}, and follow a corresponding tutorial provided as part of its online documentation.\footnote{See \href{https://naima.readthedocs.io/en/latest/examples.html\#crab-nebula-ssc-model}{https://naima.readthedocs.io/en/latest/examples.html\#crab-nebula-ssc-model}.}
Given a primary particle spectrum, \code{naima} allows the computation of photon spectra produced via non-thermal bremsstrahlung, synchrotron emission, IC scattering, and hadronic interactions (i.e.\ pion decay).
Higher-level processes such as the production of secondary electrons are currently not implemented in \code{naima}, but other packages such as \code{aafragpy}\footnote{\href{https://github.com/aafragpy/aafragpy}{https://github.com/aafragpy/aafragpy}.}~\cite{koldobskiy21} can be used for this purpose.

\begin{figure}[th]
    \centering
    \includegraphics[width=\textwidth]{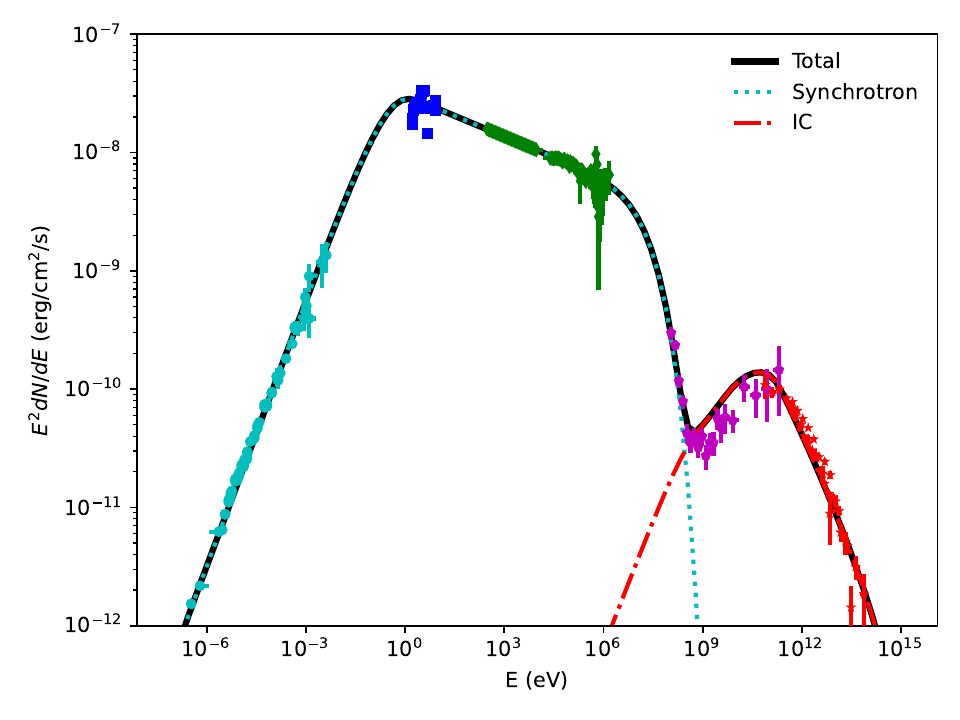}
    \caption{
        The multi-wavelength spectrum of the Crab nebula, fitted with a synchrotron self-Compton model.
        Cyan, blue, green, purple, and red points represent radio, optical, X-ray, HE, and VHE data, respectively.
        The dotted cyan and dot-dashed red curves present the individual contributions from synchrotron and IC emission of the model, while the black curve displays the sum of these two components.
        The parameters of the model are adopted from the \code{naima} online tutorial, while the data points are taken from~\cite{HESS_Crab_2006,meyer10,buhler12,kuiper01,sollerman00,tziamtis09,aleksic11,aliu11,abdo13}.
    }
    \label{fig:crab_spectrum}
\end{figure}

The multi-wavelength (radio-to-VHE) spectrum of the Crab nebula is shown in Fig~\ref{fig:crab_spectrum}.
The data points display measurements in the radio, optical, X-ray, HE, and VHE domains\footnote{They are available online at \href{https://github.com/zblz/naima/blob/main/examples/CrabNebula_spectrum.ecsv}{https://github.com/zblz/naima/blob/main/examples/CrabNebula\_spectrum.ecsv}.}.
The black curve shows an SSC model fitted to the data points.
In this model, the emission from radio frequencies to X-ray energies corresponds to synchrotron radiation from a population of relativistic electrons in the nebula.
The HE and VHE emission, on the other hand, is produced by IC scattering of the same population of electrons, where the synchrotron photons serve as one of the IC seed photon fields.

Typically, a spectral modelling with \code{naima} proceeds along the following steps:
\begin{itemize}
    \item Definition of a spectral model describing the population of primary particles (either electrons or protons).
    This model can be selected either from a set of predefined functions (e.g.\ power law, broken power law, log-parabola), or be provided by the user as a tabulated function.
    The parameters of the spectral model can be fitted to the data in the subsequent fitting procedure.
    \item Selection of a physical emission model.
    This can, for example, be a synchrotron radiation model or an IC emission model for the case of electrons as primary particles, or a pion-decay model for primary protons.
    Properties of the emitting region, such as the ambient magnetic field strength, photon field densities, or the ambient gas density, also need to be specified at this stage.
    These properties can, however, also be free parameters in the modelling.
    \item Fit of the model parameters to a data set.
    This is achieved by computing, for a given set of parameter values, the gamma-ray spectrum predicted by the model and comparing it to the observational data.
    The best-fitting parameter values can be determined by various means, for example with a simple $\chi^2$ fit.
    The \code{naima} package itself offers an optimisation with a Markov chain Monte Carlo method, using the \code{emcee} package \cite{ForemanMackey2013}.
    Furthermore, it provides an \code{InteractiveModelFitter}, which can be used to interactively vary the model parameters and display the resulting photon spectra.
\end{itemize}

For the model shown in Fig.~\ref{fig:crab_spectrum}, the spectrum of the primary electrons follows a broken power law with exponential cut-off (with spectral indices $\Gamma_1=1.5$ and $\Gamma_2=3.233$, break energy \SI{0.265}{TeV}, and cut-off energy \SI{1863}{TeV}).
The synchrotron radiation is calculated for a region with magnetic field of $B=\SI{125}{\micro G}$.
The IC component is computed with the cosmic microwave background, optical photons, and synchrotron photons as seed photon fields.

\newpage
\section*{Acknowledgements}
LM acknowledges helpful discussion with Vincent Marandon and Jim Hinton, and thanks Werner Hofmann for reading parts of the manuscript.
The authors acknowledge support by the state of Baden-W\"urttemberg through bwHPC.
The work of DM was supported by DLR through grant 50OR2104 and by DFG through grant MA 7807/2-1.

\section*{Cross-References}
\begin{itemize}
    \item ``Pair Production Detectors for Gamma-Ray Astrophysics''\\ by Thompson, D., Moiseev, A.
    \item ``The Fermi Large Area Telescope''\\by Rando, R.
    \item ``Detecting gamma-rays with high resolution and moderate field of view: the air Cherenkov technique''\\by Cortina, J., Delgado, C.
    \item ``Detecting gamma-rays with moderate resolution and large field of view: Particle detector arrays and water Cherenkov technique''\\by DuVernois, M., Di Sciascio, G.
    \item ``The High Energy Stereoscopic System (H.E.S.S.)''\\by Leuschner, F., Pühlhofer, G., Salzmann, H.
    \item ``The Major Gamma-ray Imaging Cherenkov Telescopes (MAGIC)''\\by Blanch, O., Sitarek, J.
    \item ``The Very Energetic Radiation Imaging Telescope Array System (VERITAS)''\\by Hanna, D., Mukherjee, R.
    \item ``The Cherenkov Telescope array (CTA): a worldwide endeavor for the next level of ground-based gamma-ray astronomy''\\by Hofmann, W., Zanin, R.
    \item ``Modeling and Simulating Spectra''\\by Ducci, L., Malacaria, C.
\end{itemize}

% Bibliography and bibfile
\def\aj{AJ\xspace}%
          % Astronomical Journal
\def\actaa{Acta Astron.\xspace}%
          % Acta Astronomica
\def\araa{ARA\&A\xspace}%
          % Annual Review of Astron and Astrophys
\def\aph{Astropart.\ Phys.\xspace}%
          % Astroparticle Physics
\def\apj{ApJ\xspace}%
          % Astrophysical Journal
\def\apjl{ApJ\xspace}%
          % Astrophysical Journal, Letters
\def\apjs{ApJS\xspace}%
          % Astrophysical Journal, Supplement
\def\ao{Appl.~Opt.\xspace}%
          % Applied Optics
\def\apss{Ap\&SS\xspace}%
          % Astrophysics and Space Science
\def\aap{A\&A\xspace}%
          % Astronomy and Astrophysics
\def\aapr{A\&A~Rev.\xspace}%
          % Astronomy and Astrophysics Reviews
\def\aaps{A\&AS\xspace}%
          % Astronomy and Astrophysics, Supplement
\def\azh{AZh\xspace}%
          % Astronomicheskii Zhurnal
\def\baas{BAAS\xspace}%
          % Bulletin of the AAS
\def\bac{Bull. astr. Inst. Czechosl.\xspace}%
          % Bulletin of the Astronomical Institutes of Czechoslovakia
\def\caa{Chinese Astron. Astrophys.\xspace}%
          % Chinese Astronomy and Astrophysics
\def\cjaa{Chinese J. Astron. Astrophys.\xspace}%
          % Chinese Journal of Astronomy and Astrophysics
\def\icarus{Icarus\xspace}%
          % Icarus
\def\jcap{JCAP\xspace}%
          % Journal of Cosmology and Astroparticle Physics
\def\jrasc{JRASC\xspace}%
          % Journal of the RAS of Canada
\def\mnras{MNRAS\xspace}%
          % Monthly Notices of the RAS
\def\memras{MmRAS\xspace}%
          % Memoirs of the RAS
\def\na{New A\xspace}%
          % New Astronomy
\def\nar{New A Rev.\xspace}%
          % New Astronomy Review
\def\nima{Nucl.\ Instr.\ Methods Phys.\ Res.\ A\xspace}
          % Nuclear Instruments and Methods in Physics Research A
\def\pasa{PASA\xspace}%
          % Publications of the Astron. Soc. of Australia
\def\pra{PRA\xspace}%
          % Physical Review A: General Physics
\def\prb{PRB\xspace}%
          % Physical Review B: Solid State
\def\prc{PRC\xspace}%
          % Physical Review C
\def\prd{PRD\xspace}%
          % Physical Review D
\def\pre{PRE\xspace}%
          % Physical Review E
\def\prl{PRL\xspace}%
          % Physical Review Letters
\def\pasp{PASP\xspace}%
          % Publications of the ASP
\def\pasj{PASJ\xspace}%
          % Publications of the ASJ
\def\qjras{QJRAS\xspace}%
          % Quarterly Journal of the RAS
\def\rmxaa{Rev. Mexicana Astron. Astrofis.\xspace}%
          % Revista Mexicana de Astronomia y Astrofisica
\def\skytel{S\&T\xspace}%
          % Sky and Telescope
\def\solphys{Sol.~Phys.\xspace}%
          % Solar Physics
\def\sovast{Soviet~Ast.\xspace}%
          % Soviet Astronomy
\def\ssr{Space~Sci.~Rev.\xspace}%
          % Space Science Reviews
\def\zap{ZAp\xspace}%
          % Zeitschrift fuer Astrophysik
\def\nat{Nature\xspace}%
          % Nature
\def\iaucirc{IAU~Circ.\xspace}%
          % IAU Cirulars
\def\aplett{Astrophys.~Lett.\xspace}%
          % Astrophysics Letters
\def\apspr{Astrophys.~Space~Phys.~Res.\xspace}%
          % Astrophysics Space Physics Research
\def\bain{Bull.~Astron.~Inst.~Netherlands\xspace}%
          % Bulletin Astronomical Institute of the Netherlands
\def\fcp{Fund.~Cosmic~Phys.\xspace}%
          % Fundamental Cosmic Physics
\def\gca{Geochim.~Cosmochim.~Acta\xspace}%
          % Geochimica Cosmochimica Acta
\def\grl{Geophys.~Res.~Lett.\xspace}%
          % Geophysics Research Letters
\def\jcp{J.~Chem.~Phys.\xspace}%
          % Journal of Chemical Physics
\def\jgr{J.~Geophys.~Res.\xspace}%
          % Journal of Geophysics Research
\def\jqsrt{J.~Quant.~Spec.~Radiat.~Transf.\xspace}%
          % Journal of Quantitiative Spectroscopy and Radiative Trasfer
\def\memsai{Mem.~Soc.~Astron.~Italiana\xspace}%
          % Mem. Societa Astronomica Italiana
\def\nphysa{Nucl.~Phys.~A\xspace}%
          % Nuclear Physics A
\def\physrep{Phys.~Rep.\xspace}%
          % Physics Reports
\def\physscr{Phys.~Scr\xspace}%
          % Physica Scripta
\def\planss{Planet.~Space~Sci.\xspace}%
          % Planetary Space Science
\def\procspie{Proc.~SPIE\xspace}%
          % Proceedings of the SPIE
\let\astap=\aap
\let\apjlett=\apjl
\let\apjsupp=\apjs
\let\applopt=\ao
\bibliographystyle{spbasic}
\bibliography{references}

\begin{thebibliography}{130}
\providecommand{\natexlab}[1]{#1}
\providecommand{\url}[1]{{#1}}
\providecommand{\urlprefix}{URL }
\providecommand{\doi}[1]{DOI~\discretionary{}{}{}\href{https://doi.org/#1}{#1}}
\providecommand{\eprint}[2][]{[\href{https://arxiv.org/abs/#2}{#2}]}

\bibitem[{{\flat Collaboration}(2009)}]{FermiLAT2009}
{\flat Collaboration} (2009) {The Large Area Telescope on the Fermi Gamma-ray
  Space Telescope Mission}. \apj 697:1071, \doi{10.1088/0004-637X/697/2/1071},
  \eprint{0902.1089}

\bibitem[{{Tavani} et~al(2009){Tavani}, {Barbiellini}, {Argan}
  et~al}]{AGILE2009}
{Tavani} M, {Barbiellini} G, {Argan} A, et~al (2009) {The AGILE Mission}. \aap
  502:995--1013, \doi{10.1051/0004-6361/200810527}, \eprint{0807.4254}

\bibitem[{{H.E.S.S.\ Collaboration}(2006)}]{HESS_Crab_2006}
{HESS\ Collaboration} (2006) {Observations of the Crab nebula with HESS}. \aap
  457:899--915, \doi{10.1051/0004-6361:20065351}, \eprint{astro-ph/0607333}

\bibitem[{Holler et~al(2015)Holler, Berge, van Eldik, Lenain, Marandon, Murach,
  de~Naurois, Parsons, Prokoph, and Zaborov}]{Holler2015}
Holler M, Berge D, van Eldik C, Lenain JP, Marandon V, Murach T, de~Naurois M,
  Parsons RD, Prokoph H, Zaborov D (2015) {Observations of the Crab Nebula with
  H.E.S.S.\ phase II}. In: Proc.\ 34th Int.\ Cosmic Ray Conf.\ (ICRC2015), vol
  236, p 847, \doi{10.22323/1.236.0847}, \eprint{1509.02902}

\bibitem[{{MAGIC Collaboration}(2016{\natexlab{a}})}]{MAGIC2016}
{MAGIC Collaboration} (2016{\natexlab{a}}) {The major upgrade of the MAGIC
  telescopes, Part I: The hardware improvements and the commissioning of the
  system}. \aph 72:61--75, \doi{10.1016/j.astropartphys.2015.04.004},
  \eprint{1409.6073}

\bibitem[{{MAGIC Collaboration}(2016{\natexlab{b}})}]{MAGIC2016b}
{MAGIC Collaboration} (2016{\natexlab{b}}) {The major upgrade of the MAGIC
  telescopes, Part II: A performance study using observations of the Crab
  Nebula}. \aph 72:76--94, \doi{10.1016/j.astropartphys.2015.02.005},
  \eprint{1409.5594}

\bibitem[{{VERITAS Collaboration}(2002)}]{VERITAS2002}
{VERITAS Collaboration} (2002) {VERITAS: the Very Energetic Radiation Imaging
  Telescope Array System}. \aph 17:221--243,
  \doi{10.1016/S0927-6505(01)00152-9}, \eprint{astro-ph/0108478}

\bibitem[{Park(2015)}]{Park2015}
Park N (2015) {Performance of the VERITAS experiment}. In: Proc.\ 34th Int.\
  Cosmic Ray Conf.\ (ICRC2015), vol 236, p 771, \doi{10.22323/1.236.0771},
  \eprint{1508.07070}

\bibitem[{{\flat Collaboration}(2020)}]{4FGL}
{\flat Collaboration} (2020) {Fermi Large Area Telescope Fourth Source
  Catalog}. \apjs 247:33, \doi{10.3847/1538-4365/ab6bcb}, \eprint{1902.10045}

\bibitem[{{\flat Collaboration}(2022)}]{4FGLDR3}
{\flat Collaboration} (2022) {Incremental Fermi Large Area Telescope Fourth
  Source Catalog}. \apjs 260:53, \doi{10.3847/1538-4365/ac6751},
  \eprint{2201.11184}

\bibitem[{{CTA Consortium}(2019)}]{CTA2019}
{CTA Consortium} (2019) {Science with the Cherenkov Telescope Array}. World
  Scientific Publishing, \doi{10.1142/10986}, \eprint{1709.07997}

\bibitem[{Gargano(2021)}]{Gargano2021}
Gargano F (2021) {The High Energy cosmic-Radiation Detection (HERD) facility on
  board the Chinese Space Station: hunting for high-energy cosmic rays}. In:
  Proc.\ 37th Int.\ Cosmic Ray Conf.\ (ICRC2021), vol 395, p 026,
  \doi{10.22323/1.395.0026}

\bibitem[{{HAWC Collaboration}(2017)}]{HAWC2017}
{HAWC Collaboration} (2017) {Observation of the Crab Nebula with the HAWC
  Gamma-Ray Observatory}. \apj 843:39, \doi{10.3847/1538-4357/aa7555},
  \eprint{1701.01778}

\bibitem[{{Tibet AS$\gamma$ Collaboration}(2019)}]{TibetASg2019}
{Tibet AS$\gamma$ Collaboration} (2019) {First Detection of Photons with Energy
  beyond 100 TeV from an Astrophysical Source}. \prl 123:051101,
  \doi{10.1103/PhysRevLett.123.051101}, \eprint{1906.05521}

\bibitem[{{LHAASO Collaboration}(2021)}]{LHAASO2021}
{LHAASO Collaboration} (2021) {Observation of the Crab Nebula with LHAASO-KM2A
  -- a performance study}. Chinese Phys\ C 45:025002,
  \doi{10.1088/1674-1137/abd01b}, \eprint{2010.06205}

\bibitem[{Hinton(2021)}]{Hinton2021}
Hinton J (2021) {The Southern Wide-field Gamma-ray Observatory: Status and
  Prospects}. In: Proc.\ 37th Int.\ Cosmic Ray Conf.\ (ICRC2021), vol 395, p
  023, \doi{10.22323/1.395.0023}, \eprint{2111.13158}

\bibitem[{{in 't Zand} et~al(2019){in 't Zand}, {Bozzo}, {Qu} et~al}]{eXTP2019}
{in 't Zand} JJM, {Bozzo} E, {Qu} J, et~al (2019) {Observatory science with
  eXTP}. Science China Physics, Mechanics, and Astronomy 62:29506,
  \doi{10.1007/s11433-017-9186-1}, \eprint{1812.04023}

\bibitem[{{De Angelis} et~al(2018){De Angelis}, {Tatischeff}, {Grenier}
  et~al}]{eASTROGAM2018}
{De Angelis} A, {Tatischeff} V, {Grenier} IA, et~al (2018) {Science with
  e-ASTROGAM. A space mission for MeV-GeV gamma-ray astrophysics}. JHEAp
  19:1--106, \doi{10.1016/j.jheap.2018.07.001}, \eprint{1711.01265}

\bibitem[{{Orlando} et~al(2022){Orlando}, {Bottacini}, {Moiseev}
  et~al}]{GECCO2022}
{Orlando} E, {Bottacini} E, {Moiseev} AA, et~al (2022) {Exploring the MeV sky
  with a combined coded mask and Compton telescope: the Galactic Explorer with
  a Coded aperture mask Compton telescope (GECCO)}. \jcap 2022:036,
  \doi{10.1088/1475-7516/2022/07/036}, \eprint{2112.07190}

\bibitem[{{McEnery} et~al(2019){McEnery}, {van der Horst}, {Dominguez}
  et~al}]{AMEGO2019}
{McEnery} J, {van der Horst} A, {Dominguez} A, et~al (2019) {All-sky Medium
  Energy Gamma-ray Observatory: Exploring the Extreme Multimessenger Universe}.
  In: Bulletin of the American Astronomical Society, vol~51, p 245,
  \eprint{1907.07558}

\bibitem[{{Kanbach} et~al(1989){Kanbach}, {Bertsch}, {Fichtel}, {Hartman},
  {Hunter}, {Kniffen}, {Hughlock}, {Favale}, {Hofstadter}, and
  {Hughes}}]{EGRET1989}
{Kanbach} G, {Bertsch} DL, {Fichtel} CE, {Hartman} RC, {Hunter} SD, {Kniffen}
  DA, {Hughlock} BW, {Favale} A, {Hofstadter} R, {Hughes} EB (1989) {The
  project EGRET (energetic gamma-ray experiment telescope) on NASA's Gamma-Ray
  Observatory GRO}. \ssr 49:69--84, \doi{10.1007/BF00173744}

\bibitem[{{Moiseev} et~al(2007){Moiseev}, {Hartman}, {Ormes}, {Thompson},
  {Amato}, {Johnson}, {Segal}, and {Sheppard}}]{Moiseev2007}
{Moiseev} AA, {Hartman} RC, {Ormes} JF, {Thompson} DJ, {Amato} MJ, {Johnson}
  TE, {Segal} KN, {Sheppard} DA (2007) {The anti-coincidence detector for the
  GLAST large area telescope}. \aph 27:339--358,
  \doi{10.1016/j.astropartphys.2006.12.003}, \eprint{astro-ph/0702581}

\bibitem[{Baldini(2013)}]{Baldini2013}
Baldini L (2013) {The Silicon Strip Tracker of the Fermi Large Area Telescope:
  the first five years in orbit}. PoS Vertex2013:039, \doi{10.22323/1.198.0039}

\bibitem[{{Grove} and {Johnson}(2010)}]{Grove2010}
{Grove} JE, {Johnson} WN (2010) {The calorimeter of the Fermi Large Area
  Telescope}. In: {Arnaud} M, {Murray} SS, {Takahashi} T (eds) Space Telescopes
  and Instrumentation 2010: Ultraviolet to Gamma Ray, Society of Photo-Optical
  Instrumentation Engineers (SPIE) Conference Series, vol 7732, p 77320J,
  \doi{10.1117/12.857839}

\bibitem[{{Atwood} et~al(2013){Atwood}, {Albert}, {Baldini}, {Tinivella},
  {Bregeon}, {Pesce-Rollins}, {Sgr{\`o}}, {Bruel}, {Charles}, {Drlica-Wagner},
  {Franckowiak}, {Jogler}, {Rochester}, {Usher}, {Wood}, {Cohen-Tanugi}, and
  {Zimmer}}]{latp8}
{Atwood} W, {Albert} A, {Baldini} L, {Tinivella} M, {Bregeon} J,
  {Pesce-Rollins} M, {Sgr{\`o}} C, {Bruel} P, {Charles} E, {Drlica-Wagner} A,
  {Franckowiak} A, {Jogler} T, {Rochester} L, {Usher} T, {Wood} M,
  {Cohen-Tanugi} J, {Zimmer} S (2013) {Pass 8: Toward the Full Realization of
  the Fermi-LAT Scientific Potential}. arXiv e-prints \eprint{1303.3514}

\bibitem[{{Bruel} et~al(2018){Bruel}, {Burnett}, {Digel}, {Johannesson},
  {Omodei}, and {Wood}}]{lat_p8_r3}
{Bruel} P, {Burnett} TH, {Digel} SW, {Johannesson} G, {Omodei} N, {Wood} M
  (2018) {Fermi-LAT improved Pass~8 event selection}. arXiv e-prints
  \eprint{1810.11394}

\bibitem[{{Hess}(1968)}]{hessbook68}
{Hess} WN (1968) {The radiation belt and magnetosphere}. Blaisdell

\bibitem[{Mattox et~al(1996)Mattox, Bertsch, Chiang, Dingus, Digel, Esposito,
  Fierro, Hartman, Hunter, Kanbach, Kniffen, Lin, Macomb, Mayer-Hasselwander,
  Michelson, von Montigny, Mukherjee, Nolan, Ramanamurthy, Schneid, Sreekumar,
  Thompson, and Willis}]{Mattox1996}
Mattox JR, Bertsch DL, Chiang J, Dingus BL, Digel SW, Esposito JA, Fierro JM,
  Hartman RC, Hunter SD, Kanbach G, Kniffen DA, Lin YC, Macomb DJ,
  Mayer-Hasselwander HA, Michelson PF, von Montigny C, Mukherjee R, Nolan PL,
  Ramanamurthy PV, Schneid E, Sreekumar P, Thompson DJ, Willis TD (1996) {The
  Likelihood Analysis of EGRET Data}. \apj 461:396--407, \doi{10.1086/177068}

\bibitem[{Wilks(1938)}]{Wilks1938}
Wilks SS (1938) {The Large-Sample Distribution of the Likelihood Ratio for
  Testing Composite Hypotheses}. Annals of Mathematical Statistics 9:60--62,
  \doi{10.1214/aoms/1177732360}

\bibitem[{{\flat Collaboration}(2016)}]{gll}
{\flat Collaboration} (2016) {Development of the Model of Galactic Interstellar
  Emission for Standard Point-source Analysis of Fermi Large Area Telescope
  Data}. \apjs 223:26, \doi{10.3847/0067-0049/223/2/26}, \eprint{1602.07246}

\bibitem[{{Di Mauro}(2016)}]{dimauro16}
{Di Mauro} M (2016) {The origin of the Fermi-LAT $\gamma$-ray background}.
  arXiv e-prints \eprint{1601.04323}

\bibitem[{{Malyshev} et~al(2013){Malyshev}, {Zdziarski}, and
  {Chernyakova}}]{cygx1}
{Malyshev} D, {Zdziarski} AA, {Chernyakova} M (2013) {High-energy gamma-ray
  emission from Cyg X-1 measured by Fermi and its theoretical implications}.
  \mnras 434:2380--2389, \doi{10.1093/mnras/stt1184}, \eprint{1305.5920}

\bibitem[{{Malyshev} and {Chernyakova}(2016)}]{j0632}
{Malyshev} D, {Chernyakova} M (2016) {Constraints on the spectrum of HESS
  J0632+057 from Fermi-LAT data}. \mnras 463:3074--3077,
  \doi{10.1093/mnras/stw2173}, \eprint{1601.08216}

\bibitem[{de~Naurois and Mazin(2015)}]{deNaurois2015}
de~Naurois M, Mazin D (2015) {Ground-based detectors in very-high-energy
  gamma-ray astronomy}. C R Phys 16:610--627, \doi{10.1016/j.crhy.2015.08.011},
  \eprint{1511.00463}

\bibitem[{Holder(2021)}]{Holder2021}
Holder J (2021) {Atmospheric Cherenkov Gamma-Ray Telescopes}. In: Burrows DN
  (ed) {The WSPC Handbook of Astronomical Instrumentation Volume 5: Gamma-Ray
  and Multimessenger Astronomical Instrumentation}, World Scientific,
  \doi{10.1142/9446-vol5}, \eprint{1510.05675}

\bibitem[{Sitarek(2022)}]{Sitarek2022}
Sitarek J (2022) {TeV Instrumentation: Current and Future}. Galaxies 10:21,
  \doi{10.3390/galaxies10010021}, \eprint{2201.08611}

\bibitem[{Contreras et~al(2015)Contreras, Satalecka, Bernl\"ohr, Boisson,
  Bregeon, Bulgarelli, De~Cesare, de~los Reyes, Fioretti, Kosack, Lavalley,
  Lyard, Marx, Rico, Sanguillot, Servillat, Walter, Ward, and
  Zoli}]{Contreras2015}
Contreras JL, Satalecka K, Bernl\"ohr K, Boisson C, Bregeon J, Bulgarelli A,
  De~Cesare G, de~los Reyes R, Fioretti V, Kosack K, Lavalley C, Lyard E, Marx
  R, Rico J, Sanguillot M, Servillat M, Walter R, Ward JE, Zoli A (2015) {Data
  model issues in the Cherenkov Telescope Array project}. In: Proc. 34th Int.
  Cosmic Ray Conf. (ICRC2015), vol 236, p 960, \doi{10.22323/1.236.0960},
  \eprint{1508.07584}

\bibitem[{Nigro et~al(2021)Nigro, Hassan, and Olivera-Nieto}]{Nigro2021}
Nigro C, Hassan T, Olivera-Nieto L (2021) {Evolution of Data Formats in
  Very-High-Energy Gamma-Ray Astronomy}. Universe 7:374,
  \doi{10.3390/universe7100374}, \eprint{2109.14661}

\bibitem[{{Brun} and {Rademakers}(1997)}]{Brun1997}
{Brun} R, {Rademakers} F (1997) {ROOT {\textemdash} An object oriented data
  analysis framework}. {\nima} 389:81--86, \doi{10.1016/S0168-9002(97)00048-X}

\bibitem[{{Zanin, R. for the CTA Consortium}(2021)}]{Zanin2021}
{Zanin, R for the CTA Consortium} (2021) {CTA – the World’s largest
  ground-based gamma-ray observatory}. In: Proc.\ 37th Int.\ Cosmic Ray Conf.\
  (ICRC2021), vol 395, p 005, \doi{10.22323/1.395.0005}

\bibitem[{Deil et~al(2017)Deil, Boisson, Kosack, Perkins, King, Eger, Mayer,
  Wood, Zabalza, Kn\"odlseder, Hassan, Mohrmann, Ziegler, Kh\'elifi, Dorner,
  Maier, Pedaletti, Rosado, Contreras, Lefaucheur, Br\"ugge, Servillat,
  Terrier, Walter, and Lombardi}]{Deil_GADF_2017}
Deil C, Boisson C, Kosack K, Perkins J, King J, Eger P, Mayer M, Wood M,
  Zabalza V, Kn\"odlseder J, Hassan T, Mohrmann L, Ziegler A, Kh\'elifi B,
  Dorner D, Maier G, Pedaletti G, Rosado J, Contreras JL, Lefaucheur J,
  Br\"ugge K, Servillat M, Terrier R, Walter R, Lombardi S (2017) {Open
  high-level data formats and software for gamma-ray astronomy}. AIP Conf Proc
  1792:070006, \doi{10.1063/1.4969003}, \eprint{1610.01884}

\bibitem[{Deil et~al(2022)}]{Deil_GADFv3_2022}
Deil C, et~al (2022) {Data formats for gamma-ray astronomy - version 0.3}.
  \doi{10.5281/zenodo.7304668},
  \urlprefix\url{https://gamma-astro-data-formats.readthedocs.io}

\bibitem[{{H.E.S.S.\ Collaboration}(2018)}]{HESS_FITS_2018}
{HESS\ Collaboration} (2018) {H.E.S.S.\ first public test data release}.
  \doi{10.5281/zenodo.1421098},
  \urlprefix\url{https://www.mpi-hd.mpg.de/hfm/HESS/pages/dl3-dr1/},
  \eprint{1810.04516}

\bibitem[{{Nigro} et~al(2019){Nigro}, {Deil}, {Zanin}, {Hassan}, {King},
  {Ruiz}, {Saha}, {Terrier}, {Br{\"u}gge}, {N{\"o}the}, {Bird}, {Lin},
  {Aleksi{\'c}}, {Boisson}, {Contreras}, {Donath}, {Jouvin}, {Kelley-Hoskins},
  {Khelifi}, {Kosack}, {Rico}, and {Sinha}}]{Nigro2019}
{Nigro} C, {Deil} C, {Zanin} R, {Hassan} T, {King} J, {Ruiz} JE, {Saha} L,
  {Terrier} R, {Br{\"u}gge} K, {N{\"o}the} M, {Bird} R, {Lin} TTY,
  {Aleksi{\'c}} J, {Boisson} C, {Contreras} JL, {Donath} A, {Jouvin} L,
  {Kelley-Hoskins} N, {Khelifi} B, {Kosack} K, {Rico} J, {Sinha} A (2019)
  {Towards open and reproducible multi-instrument analysis in gamma-ray
  astronomy}. \aap 625:A10, \doi{10.1051/0004-6361/201834938},
  \eprint{1903.06621}

\bibitem[{{H.E.S.S.\ Collaboration}(2004)}]{HESS_Calib_2004}
{HESS\ Collaboration} (2004) {Calibration of cameras of the H.E.S.S.\
  detector}. \aph 22:109--125, \doi{10.1016/j.astropartphys.2004.06.006},
  \eprint{astro-ph/0406658}

\bibitem[{Mitchell(2016)}]{Mitchell2016}
Mitchell AMW (2016) {Optical Efficiency Calibration for Inhomogeneous IACT
  Arrays and a Detailed Study of the Highly Extended Pulsar Wind Nebula HESS
  J1825$-$137}. {PhD thesis}, Ruprecht-Karls-Universit\"at Heidelberg,
  \urlprefix\url{https://hdl.handle.net/11858/00-001M-0000-002B-1DD2-F}

\bibitem[{Gaug et~al(2019)Gaug, Fegan, Mitchell, Maccarone, Mineo, and
  Okumura}]{Gaug2019}
Gaug M, Fegan S, Mitchell AMW, Maccarone MC, Mineo T, Okumura A (2019) {Using
  Muon Rings for the Calibration of the Cherenkov Telescope Array: A Systematic
  Review of the Method and Its Potential Accuracy}. \apjs 243:11,
  \doi{10.3847/1538-4365/ab2123}, \eprint{1907.04375}

\bibitem[{{VERITAS Collaboration}(2022)}]{VERITAS_Throughput_2022}
{VERITAS Collaboration} (2022) {The throughput calibration of the VERITAS
  telescopes}. \aap 658:A83, \doi{10.1051/0004-6361/202142275},
  \eprint{2111.04676}

\bibitem[{{MAGIC Collaboration}(2009)}]{MAGIC_Timing_2009}
{MAGIC Collaboration} (2009) {Improving the performance of the single-dish
  Cherenkov telescope MAGIC through the use of signal timing}. \aph
  30:293--305, \doi{10.1016/j.astropartphys.2008.10.003}, \eprint{0810.3568}

\bibitem[{{Lombardi, S. for the MAGIC Collaboration}(2011)}]{Lombardi2011}
{Lombardi, S for the MAGIC Collaboration} (2011) {Advanced stereoscopic
  gamma-ray shower analysis with the MAGIC telescopes}. In: Proc.\ 32nd Int.\
  Cosmic Ray Conf.\ (ICRC2011), \eprint{1109.6195}

\bibitem[{{Shayduk, M. for the CTA Consortium}(2013)}]{Shayduk2013}
{Shayduk, M for the CTA Consortium} (2013) {Optimized Next-neighbour Image
  Cleaning Method for Cherenkov Telescopes}. In: Proc.\ 33rd Int.\ Cosmic Ray
  Conf.\ (ICRC2013), \eprint{1307.4939}

\bibitem[{V\"{o}lk and Bernl\"ohr(2009)}]{Voelk2009}
V\"{o}lk HJ, Bernl\"ohr K (2009) {Imaging very high energy gamma-ray
  telescopes}. Exp Astron 25:173--191, \doi{10.1007/s10686-009-9151-z},
  \eprint{0812.4198}

\bibitem[{Hillas(1985)}]{Hillas1985}
Hillas AM (1985) {Cerenkov light images of EAS produced by primary gamma rays
  and nuclei}. In: Proc.\ 19th Int.\ Cosmic Ray Conf., vol~3, pp 445--448,
  \urlprefix\url{https://ui.adsabs.harvard.edu/abs/1985ICRC....3..445H}

\bibitem[{de~Naurois(2012)}]{deNaurois2012}
de~Naurois M (2012) {L'astronomie $\gamma$ de tr\`es haute \'energie. Ouverture
  d'une nouvelle fen\^{e}tre astronomique sur l'Univers non thermique.}
  {Habilitation thesis}, Universit\'e de Paris,
  \urlprefix\url{https://inspirehep.net/record/1122589/files/these_short.pdf}

\bibitem[{Hofmann et~al(1999)Hofmann, Jung, Konopelko, Krawczynski, Lampeitl,
  and P\"{u}hlhofer}]{Hofmann1999}
Hofmann W, Jung I, Konopelko A, Krawczynski H, Lampeitl H, P\"{u}hlhofer G
  (1999) {Comparison of techniques to reconstruct VHE gamma-ray showers from
  multiple stereoscopic Cherenkov images}. \aph 12:135--143,
  \doi{10.1016/S0927-6505(99)00084-5}, \eprint{astro-ph/9904234}

\bibitem[{Lessard et~al(2001)Lessard, Buckley, Connaughton, and
  LeBohec}]{Lessard2001}
Lessard RW, Buckley JH, Connaughton V, LeBohec S (2001) {A new analysis method
  for reconstructing the arrival direction of TeV gamma rays using a single
  imaging atmospheric Cherenkov telescope}. \aph 15:1--18,
  \doi{10.1016/S0927-6505(00)00133-X}, \eprint{astro-ph/0005468}

\bibitem[{Domingo-Santamar\'ia et~al(2005)Domingo-Santamar\'ia, Flix,
  Scalzotto, Wittek, and Rico}]{DomingoSantamaria2005}
Domingo-Santamar\'ia E, Flix J, Scalzotto V, Wittek W, Rico J (2005) {The DISP
  analysis method for point-like or extended $\gamma$ source searches/studies
  with the MAGIC Telescope}. In: Proc.\ 29th Int.\ Cosmic Ray Conf.\
  (ICRC2005), \eprint{astro-ph/0508274}

\bibitem[{Murach et~al(2015)Murach, Gajdus, and Parsons}]{Murach2015}
Murach T, Gajdus M, Parsons RD (2015) {A Neural Network-based Reconstruction
  Algorithm for monoscopically detected Air Showers observed with the H.E.S.S.\
  Experiment}. In: Proc. 34th Int. Cosmic Ray Conf. (ICRC2015), vol 236, p
  1022, \doi{10.22323/1.236.1022}, \eprint{1509.00794}

\bibitem[{{Le Bohec} et~al(1998){Le Bohec}, {Degrange}, {Punch}, {Barrau},
  {Bazer-Bachi}, {Cabot}, {Chounet}, {Debiais}, {Dezalay}, {Djannati-Atai},
  {Dumora}, {Espigat}, {Fabre}, {Fleury}, {Fontaine}, {George}, {Ghesquiere},
  {Goret}, {Gouiffes}, {Grenier}, {Iacoucci}, {Malet}, {Meynadier}, {Munz},
  {Palfrey}, {Pare}, {Pons}, {Quebert}, {Ragan}, {Renault}, {Rivoal}, {Rob},
  {Schovanek}, {Smith}, {Tavernet}, and {Vrana}}]{LeBohec1998}
{Le Bohec} S, {Degrange} B, {Punch} M, {Barrau} A, {Bazer-Bachi} R, {Cabot} H,
  {Chounet} LM, {Debiais} G, {Dezalay} JP, {Djannati-Atai} A, {Dumora} D,
  {Espigat} P, {Fabre} B, {Fleury} P, {Fontaine} G, {George} R, {Ghesquiere} C,
  {Goret} P, {Gouiffes} C, {Grenier} IA, {Iacoucci} L, {Malet} I, {Meynadier}
  C, {Munz} F, {Palfrey} TA, {Pare} E, {Pons} Y, {Quebert} J, {Ragan} K,
  {Renault} C, {Rivoal} M, {Rob} L, {Schovanek} P, {Smith} D, {Tavernet} JP,
  {Vrana} J (1998) {A new analysis method for very high definition imaging
  atmospheric Cherenkov telescopes as applied to the CAT telescope}. \nima
  416:425--437, \doi{10.1016/S0168-9002(98)00750-5}, \eprint{astro-ph/9804133}

\bibitem[{Hillas(1982)}]{Hillas1982}
Hillas AM (1982) {Angular and energy distributions of charged particles in
  electron-photon cascades in air}. J\ Phys\ G Nucl\ Phys 8:1461,
  \doi{10.1088/0305-4616/8/10/016}

\bibitem[{de~Naurois and Rolland(2009)}]{deNaurois2009}
de~Naurois M, Rolland L (2009) {A high performance likelihood reconstruction of
  $\gamma$-rays for imaging atmospheric Cherenkov telescopes}. \aph
  32:231--252, \doi{10.1016/j.astropartphys.2009.09.001}, \eprint{0907.2610}

\bibitem[{Lemoine-Goumard et~al(2006)Lemoine-Goumard, Degrange, and
  Tluczykont}]{LemoineGoumard2006}
Lemoine-Goumard M, Degrange B, Tluczykont M (2006) {Selection and
  3D-reconstruction of gamma-ray-induced air showers with a stereoscopic system
  of atmospheric Cherenkov telescopes}. \aph 25:195--211,
  \doi{10.1016/j.astropartphys.2006.01.005}, \eprint{astro-ph/0601373}

\bibitem[{Parsons and Hinton(2014)}]{Parsons2014}
Parsons RD, Hinton JA (2014) {A Monte Carlo template based analysis for
  air-Cherenkov arrays}. \aph 56:26--34,
  \doi{10.1016/j.astropartphys.2014.03.002}, \eprint{1403.2993}

\bibitem[{Vincent(2015)}]{Vincent2015}
Vincent S (2015) {A Monte Carlo template-based analysis for very high
  definition imaging atmospheric Cherenkov telescopes as applied to the VERITAS
  telescope array}. In: Proc. 34th Int. Cosmic Ray Conf. (ICRC2015), vol 236, p
  844, \doi{10.22323/1.236.0844}, \eprint{1509.01980}

\bibitem[{Christiansen(2017)}]{Christiansen2017}
Christiansen J (2017) {Characterization of a Maximum Likelihood Gamma-Ray
  Reconstruction Algorithm for VERITAS}. In: Proc. 35th Int. Cosmic Ray Conf.
  (ICRC2017), vol 301, p 789, \doi{10.22323/1.301.0789}, \eprint{1708.05684}

\bibitem[{Holch et~al(2017)Holch, Shilon, B\"uchele, Fischer, Funk, Groeger,
  Jankowsky, Lohse, Schwanke, and Wagner}]{Holch2017}
Holch TL, Shilon I, B\"uchele M, Fischer T, Funk S, Groeger N, Jankowsky D,
  Lohse T, Schwanke U, Wagner P (2017) {Probing Convolutional Neural Networks
  for Event Reconstruction in Gamma-Ray Astronomy with Cherenkov Telescopes}.
  In: Proc.\ 35th Int.\ Cosmic Ray Conf.\ (ICRC2017), vol 301, p 795,
  \doi{10.22323/1.301.0795}, \eprint{1711.06298}

\bibitem[{Shilon et~al(2019)Shilon, Kraus, B\"uchele, Egberts, Fischer, Holch,
  Lohse, Schwanke, Steppa, and Funk}]{Shilon2019}
Shilon I, Kraus M, B\"uchele M, Egberts K, Fischer T, Holch TL, Lohse T,
  Schwanke U, Steppa C, Funk S (2019) {Application of deep learning methods to
  analysis of imaging atmospheric Cherenkov telescopes data}. \aph 105:44--53,
  \doi{10.1016/j.astropartphys.2018.10.003}, \eprint{1803.10698}

\bibitem[{Mangano et~al(2018)Mangano, Delgado, Bernardos, Lallena, and
  Rodr{\'i}guez~V{\'a}zquez}]{Mangano2018}
Mangano S, Delgado C, Bernardos MI, Lallena M, Rodr{\'i}guez~V{\'a}zquez JJ
  (2018) {Extracting Gamma-Ray Information from Images with Convolutional
  Neural Network Methods on Simulated Cherenkov Telescope Array Data}. In:
  Pancioni L, Schwenker F, Trentin E (eds) {Artificial Neural Networks in
  Pattern Recognition}, Springer International Publishing, pp 243--254,
  \eprint{1810.00592}

\bibitem[{Miener et~al(2021)Miener, Nieto, Brill, Spencer, and
  Contreras}]{Miener2021}
Miener T, Nieto D, Brill A, Spencer ST, Contreras JL (2021) {Reconstruction of
  stereoscopic CTA events using deep learning with CTLearn}. In: Proc.\ 37th
  Int.\ Cosmic Ray Conf.\ (ICRC2021), vol 395, p 730,
  \doi{10.22323/1.395.0730}, \eprint{2109.05809}

\bibitem[{Aschersleben et~al(2021)Aschersleben, Peletier, Vecchi, and
  Wilkinson}]{Aschersleben2021}
Aschersleben J, Peletier R, Vecchi M, Wilkinson M (2021) {Application of
  Pattern Spectra and Convolutional Neural Networks to the Analysis of
  Simulated Cherenkov Telescope Array Data}. In: Proc.\ 37th Int.\ Cosmic Ray
  Conf.\ (ICRC2021), vol 395, p 697, \doi{10.22323/1.395.0697},
  \eprint{2108.00834}

\bibitem[{Polyakov et~al(2021)Polyakov, Demichev, Kryukov, and
  Postnikov}]{Polyakov2021}
Polyakov S, Demichev A, Kryukov A, Postnikov E (2021) {The use of convolutional
  neural networks for processing images from multiple IACTs in the TAIGA
  experiment}. In: Proc.\ 37th Int.\ Cosmic Ray Conf.\ (ICRC2021), vol 395, p
  753, \doi{10.22323/1.395.0753}

\bibitem[{Vuillaume et~al(2021)Vuillaume, Jacquemont, de~Bony~de Lavergne,
  Sanchez, Poireau, Maurin, Benoit, Lambert, and Lamanna}]{Vuillaume2021}
Vuillaume T, Jacquemont M, de~Bony~de Lavergne M, Sanchez DA, Poireau V, Maurin
  G, Benoit A, Lambert P, Lamanna G (2021) {Analysis of the Cherenkov Telescope
  Array first Large Size Telescope real data using convolutional neural
  networks}. In: Proc.\ 37th Int.\ Cosmic Ray Conf.\ (ICRC2021), vol 395, p
  703, \doi{10.22323/1.395.0703}, \eprint{2108.04130}

\bibitem[{{Miener} et~al(2022){Miener}, {Nieto}, {L{\'o}pez-Coto}, {Contreras},
  {Green}, {Green}, and {Mariotti}}]{Miener2022}
{Miener} T, {Nieto} D, {L{\'o}pez-Coto} R, {Contreras} JL, {Green} JG, {Green}
  D, {Mariotti} E (2022) {The performance of the MAGIC telescopes using deep
  convolutional neural networks with CTLearn}. arXiv e-prints
  \eprint{2211.16009}

\bibitem[{{HEGRA Collaboration}(1997)}]{HEGRA1997}
{HEGRA Collaboration} (1997) {First results on the performance of the HEGRA
  IACT array}. \aph 8:1--11, \doi{10.1016/S0927-6505(97)00031-5},
  \eprint{astro-ph/9704098}

\bibitem[{Ohm et~al(2009)Ohm, van Eldik, and Egberts}]{Ohm2009}
Ohm S, van Eldik C, Egberts K (2009) {$\gamma$/hadron separation in
  very-high-energy $\gamma$-ray astronomy using a multivariate analysis
  method}. \aph 31:383--391, \doi{10.1016/j.astropartphys.2009.04.001},
  \eprint{0904.1136}

\bibitem[{Krause et~al(2017)Krause, Pueschel, and Maier}]{Krause2017}
Krause M, Pueschel E, Maier G (2017) {Improved $\gamma$/hadron separation for
  the detection of faint $\gamma$-ray sources using boosted decision trees}.
  \aph 89:1--9, \doi{10.1016/j.astropartphys.2017.01.004}, \eprint{1701.06928}

\bibitem[{{MAGIC Collaboration}(2008)}]{MAGIC_RF_2008}
{MAGIC Collaboration} (2008) {Implementation of the Random Forest method for
  the Imaging Atmospheric Cherenkov Telescope MAGIC}. \nima 588:424--432,
  \doi{10.1016/j.nima.2007.11.068}, \eprint{0709.3719}

\bibitem[{{Chitnis} and {Bhat}(2001)}]{Chitnis2001}
{Chitnis} VR, {Bhat} PN (2001) {Possible discrimination between gamma rays and
  hadrons using {\v{C}}erenkov photon timing measurements}. \aph 15:29--47,
  \doi{10.1016/S0927-6505(00)00137-7}, \eprint{astro-ph/0006133}

\bibitem[{Spencer et~al(2021)Spencer, Armstrong, Watson, Mangano, Renier, and
  Cotter}]{Spencer2021}
Spencer S, Armstrong T, Watson J, Mangano S, Renier Y, Cotter G (2021) {Deep
  learning with photosensor timing information as a background rejection method
  for the Cherenkov Telescope Array}. \aph 129:102579,
  \doi{10.1016/j.astropartphys.2021.102579}, \eprint{2103.06054}

\bibitem[{Olivera-Nieto et~al(2021)Olivera-Nieto, Mitchell, Bernl\"ohr, and
  Hinton}]{OliveraNieto2021}
Olivera-Nieto L, Mitchell AMW, Bernl\"ohr K, Hinton JA (2021) {Muons as a tool
  for background rejection in imaging atmospheric Cherenkov telescope arrays}.
  Eur\ Phys\ J\ C 81:1101, \doi{10.1140/epjc/s10052-021-09869-0},
  \eprint{2111.12041}

\bibitem[{Olivera-Nieto et~al(2022)Olivera-Nieto, Ren, Mitchell, Marandon, and
  Hinton}]{OliveraNieto2022}
Olivera-Nieto L, Ren HX, Mitchell AMW, Marandon V, Hinton JA (2022) {Background
  rejection using image residuals from large telescopes in imaging atmospheric
  Cherenkov telescope arrays}. Eur\ Phys\ J\ C 82:1118,
  \doi{10.1140/epjc/s10052-022-11067-5}, \eprint{2211.13167}

\bibitem[{Parsons and Ohm(2020)}]{Parsons2020}
Parsons RD, Ohm S (2020) {Background rejection in atmospheric Cherenkov
  telescopes using recurrent convolutional neural networks}. Eur Phys J C
  80:363, \doi{10.1140/epjc/s10052-020-7953-3}, \eprint{1910.09435}

\bibitem[{{Glombitza} et~al(2023){Glombitza}, {Joshi}, {Bruno}, and
  {Funk}}]{Glombitza2023}
{Glombitza} J, {Joshi} V, {Bruno} B, {Funk} S (2023) {Application of Graph
  Networks to background rejection in Imaging Air Cherenkov Telescopes}. arXiv
  e-prints \eprint{2305.08674}

\bibitem[{De et~al(2023)De, Maitra, Rentala, and Thalapillil}]{De2023}
De S, Maitra W, Rentala V, Thalapillil AM (2023) {Deep learning techniques for
  imaging air Cherenkov telescopes}. \prd 107:083026,
  \doi{10.1103/PhysRevD.107.083026}, \eprint{2206.05296}

\bibitem[{Nieto Casta\~no et~al(2017)Nieto Casta\~no, Brill, Kim, and
  Humensky}]{Nieto2017}
Nieto Casta\~no D, Brill A, Kim B, Humensky TB (2017) {Exploring deep learning
  as an event classification method for the Cherenkov Telescope Array}. In:
  Proc.\ 35th Int.\ Cosmic Ray Conf.\ (ICRC2017), vol 301, p 809,
  \doi{10.22323/1.301.0809}, \eprint{1709.05889}

\bibitem[{{Lyard} et~al(2020){Lyard}, {Walter}, {Sliusar}, and
  {Produit}}]{Lyard2020}
{Lyard} E, {Walter} R, {Sliusar} V, {Produit} N (2020) {Probing Neural Networks
  for the Gamma/Hadron Separation of the Cherenkov Telescope Array}. J\ Phys\
  Conf\ Ser 1525:012084, \doi{10.1088/1742-6596/1525/1/012084},
  \eprint{1907.02428}

\bibitem[{{Parsons} et~al(2022){Parsons}, {Mitchell}, and {Ohm}}]{Parsons2022}
{Parsons} RD, {Mitchell} AMW, {Ohm} S (2022) {Investigations of the Systematic
  Uncertainties in Convolutional Neural Network Based Analysis of Atmospheric
  Cherenkov Telescope Data}. arXiv e-prints \eprint{2203.05315}

\bibitem[{Parsons and Schoorlemmer(2019)}]{Parsons2019}
Parsons RD, Schoorlemmer H (2019) {Systematic differences due to high energy
  hadronic interaction models in air shower simulations in the 100 GeV-100 TeV
  range}. PRD 100:023010, \doi{10.1103/PhysRevD.100.023010},
  \eprint{1904.05135}

\bibitem[{Weekes et~al(1989)Weekes, Cawley, Fegan, Gibbs, Hillas, Kowk, Lamb,
  Lewis, Macomb, Porter, Reynolds, and Vacanti}]{Weekes1989}
Weekes TC, Cawley MF, Fegan DJ, Gibbs KG, Hillas AM, Kowk PW, Lamb RC, Lewis
  DA, Macomb D, Porter NA, Reynolds PT, Vacanti G (1989) {Observation of TeV
  gamma rays from the Crab nebula using the atmospheric Cerenkov imaging
  technique}. \apj 342:379--395, \doi{10.1086/167599}

\bibitem[{Hona(2021)}]{Hona2021}
Hona B (2021) {Matched Runs Method to Study Extended Regions of Gamma-ray
  Emission}. In: Proc.\ 37th Int.\ Cosmic Ray Conf.\ (ICRC2021), vol 395, p
  729, \doi{10.22323/1.395.0729}, \eprint{2108.07663}

\bibitem[{Mitchell et~al(2021)Mitchell, Caroff, Hinton, and
  Mohrmann}]{Mitchell2021}
Mitchell A, Caroff S, Hinton J, Mohrmann L (2021) {Detection of extended TeV
  emission around the Geminga pulsar with H.E.S.S.} In: Proc.\ 37th Int.\
  Cosmic Ray Conf.\ (ICRC2021), vol 395, p 780, \doi{10.22323/1.395.0780},
  \eprint{2108.02556}

\bibitem[{Berge et~al(2007)Berge, Funk, and Hinton}]{Berge2007}
Berge D, Funk S, Hinton J (2007) {Background modelling in very-high-energy
  $\gamma$-ray astronomy}. \aap 466:1219--1229,
  \doi{10.1051/0004-6361:20066674}, \eprint{astro-ph/0610959}

\bibitem[{Fomin et~al(1994)Fomin, Stepanian, Lamb, Lewis, Punch, and
  Weekes}]{Fomin1994}
Fomin VP, Stepanian AA, Lamb RC, Lewis DA, Punch M, Weekes TC (1994) {New
  methods of atmospheric Cherenkov imaging for gamma-ray astronomy. I. The
  false source method}. \aph 2:137--150, \doi{10.1016/0927-6505(94)90036-1}

\bibitem[{Rowell(2003)}]{Rowell2003}
Rowell GP (2003) {A new template background estimate for source searching in
  TeV $\gamma$-ray astronomy}. \aap 410:389--396,
  \doi{10.1051/0004-6361:20031194}, \eprint{astro-ph/0310025}

\bibitem[{Da~Vela et~al(2018)Da~Vela, Stamerra, Neronov, Prandini, Konno, and
  Sitarek}]{DaVela2018}
Da~Vela P, Stamerra A, Neronov A, Prandini E, Konno Y, Sitarek J (2018) {Study
  of the IACT angular acceptance and Point Spread Function}. \aph 98:1--8,
  \doi{10.1016/j.astropartphys.2018.01.002}

\bibitem[{Mohrmann et~al(2019)Mohrmann, Specovius, Tiziani, Funk, Malyshev,
  Nakashima, and van Eldik}]{Mohrmann2019}
Mohrmann L, Specovius A, Tiziani D, Funk S, Malyshev D, Nakashima K, van Eldik
  C (2019) {Validation of open-source science tools and background model
  construction in $\gamma$-ray astronomy}. \aap 632:A72,
  \doi{10.1051/0004-6361/201936452}, \eprint{1910.08088}

\bibitem[{Bernl\"ohr(2008)}]{Bernloehr2008}
Bernl\"ohr K (2008) {Simulation of imaging atmospheric Cherenkov telescopes
  with CORSIKA and sim\_telarray}. \aph 30:149--158,
  \doi{10.1016/j.astropartphys.2008.07.009}, \eprint{0808.2253}

\bibitem[{Kertzman and Sembroski(1994)}]{Kertzman1994}
Kertzman MP, Sembroski GH (1994) {Computer simulation methods for investigating
  the detection characteristics of TeV air Cherenkov telescopes}. \nima
  343:629--643, \doi{10.1016/0168-9002(94)90247-X}

\bibitem[{{Holler} et~al(2020){Holler}, {Lenain}, {de Naurois}, {Rauth}, and
  {Sanchez}}]{Holler2020}
{Holler} M, {Lenain} JP, {de Naurois} M, {Rauth} R, {Sanchez} DA (2020) {A
  run-wise simulation and analysis framework for Imaging Atmospheric Cherenkov
  Telescope arrays}. \aph 123:102491,
  \doi{10.1016/j.astropartphys.2020.102491}, \eprint{2007.01697}

\bibitem[{Hahn et~al(2014)Hahn, de~los Reyes, Bernl\"ohr, Kr\"{u}ger, Lo,
  Chadwick, Daniel, Deil, Gast, Kosack, and Marandon}]{Hahn2014}
Hahn J, de~los Reyes R, Bernl\"ohr K, Kr\"{u}ger P, Lo YTE, Chadwick PM, Daniel
  MK, Deil C, Gast H, Kosack K, Marandon V (2014) {Impact of aerosols and
  adverse atmospheric conditions on the data quality for spectral analysis of
  the H.E.S.S.\ telescopes}. \aph 54:25--32,
  \doi{10.1016/j.astropartphys.2013.10.003}, \eprint{1310.1639}

\bibitem[{Li and Ma(1983)}]{Li1983}
Li T, Ma Y (1983) {Analysis methods for results in gamma-ray astronomy}. \apj
  272:317--324, \doi{10.1086/161295}

\bibitem[{Piron et~al(2001)Piron, Djannati-Ata\"{i}, Punch, Tavernet, Barrau,
  Bazer-Bachi, Chounet, Debiais, Degrange, Dezalay, Espigat, Fabre, Fleury,
  Fontaine, Goret, Gouiffes, Khelifi, Malet, Masterson, Mohanty, Nuss, Renault,
  Rivoal, Rob, and Vorobiov}]{Piron2001}
Piron F, Djannati-Ata\"{i} A, Punch M, Tavernet JP, Barrau A, Bazer-Bachi R,
  Chounet LM, Debiais G, Degrange B, Dezalay JP, Espigat P, Fabre B, Fleury P,
  Fontaine G, Goret P, Gouiffes C, Khelifi B, Malet I, Masterson C, Mohanty G,
  Nuss E, Renault C, Rivoal M, Rob L, Vorobiov S (2001) {Temporal and spectral
  gamma-ray properties of Mkn 421 above 250 GeV from CAT observations between
  1996 and 2000}. \aap 374:895--906, \doi{10.1051/0004-6361:20010798},
  \eprint{astro-ph/0106196}

\bibitem[{{HEGRA Collaboration}(1999)}]{HEGRA1999}
{HEGRA Collaboration} (1999) {The temporal characteristicsof the TeV
  $\gamma$-emission from Mkn 501 in 1997: II. Results from HEGRA CT1 and CT2}.
  \aap 349:29--44, \eprint{astro-ph/9901284}

\bibitem[{{MAGIC Collaboration}(2007)}]{MAGIC_Unfolding_2007}
{MAGIC Collaboration} (2007) {Unfolding of differential energy spectra in the
  MAGIC experiment}. \nima 583:494--506, \doi{10.1016/j.nima.2007.09.048},
  \eprint{0707.2453}

\bibitem[{Milke et~al(2013)Milke, Doert, Klepser, Mazin, Blobel, and
  Rhode}]{Milke2013}
Milke N, Doert M, Klepser S, Mazin D, Blobel V, Rhode W (2013) {Solving inverse
  problems with the unfolding program TRUEE: Examples in astroparticle
  physics}. \nima 697:133--147, \doi{10.1016/j.nima.2012.08.105},
  \eprint{1209.3218}

\bibitem[{{H.E.S.S.\ Collaboration}(2021)}]{HESS_J1702_2021}
{HESS\ Collaboration} (2021) {Evidence of 100~TeV $\gamma$-ray emission from
  HESS~J1702$-$420: A new PeVatron candidate}. \aap 653:A152,
  \doi{10.1051/0004-6361/202140962}, \eprint{2106.06405}

\bibitem[{{H.E.S.S.\ Collaboration} et~al(2022){H.E.S.S.\ Collaboration},
  Blackwell, Braiding, Burton, Cubuk, Filipovi\'c, Tothill, and
  Wong}]{HESS_Wd1_2022}
{HESS\ Collaboration}, Blackwell R, Braiding C, Burton M, Cubuk K, Filipovi\'c
  M, Tothill N, Wong G (2022) {A deep spectromorphological study of the
  $\gamma$-ray emission surrounding the young massive stellar cluster
  Westerlund 1}. \aap 666:A124, \doi{10.1051/0004-6361/202244323},
  \eprint{2207.10921}

\bibitem[{{H.E.S.S.\ Collaboration}(2023)}]{HESS_J1809_2023}
{HESS\ Collaboration} (2023) {HESS J1809$-$193: a halo of escaped electrons
  around a pulsar wind nebula?} \aap, 672:A103,
  \doi{10.1051/0004-6361/202245459}, \eprint{2302.13663}

\bibitem[{Vovk et~al(2018)Vovk, Strzys, and Fruck}]{Vovk2018}
Vovk I, Strzys M, Fruck C (2018) {Spatial likelihood analysis for MAGIC
  telescope data: From instrument response modelling to spectral extraction}.
  \aap 619:A7, \doi{10.1051/0004-6361/201833139}, \eprint{1806.03167}

\bibitem[{Cardenzana(2017)}]{Cardenzana2017}
Cardenzana JV (2017) {A 3D maximum likelihood analysis for studying highly
  extended sources in VERITAS data}. PhD thesis, Iowa State University,
  \doi{10.31274/etd-180810-4900}

\bibitem[{Kn\"{o}dlseder et~al(2016)Kn\"{o}dlseder, Mayer, Deil, Cayrou, Owen,
  Kelley-Hoskins, Lu, Buehler, Forest, Louge, Siejkowski, Kosack, Gerard,
  Schulz, Martin, Sanchez, Ohm, Hassan, and Brau-Nogu\'{e}}]{Knoedlseder2016}
Kn\"{o}dlseder J, Mayer M, Deil C, Cayrou JB, Owen E, Kelley-Hoskins N, Lu CC,
  Buehler R, Forest F, Louge T, Siejkowski H, Kosack K, Gerard L, Schulz A,
  Martin P, Sanchez D, Ohm S, Hassan T, Brau-Nogu\'{e} S (2016) {GammaLib and
  ctools: A software framework for the analysis of astronomical gamma-ray
  data}. \aap 593:A1, \doi{10.1051/0004-6361/201628822}, \eprint{1606.00393}

\bibitem[{{Deil} et~al(2017){Deil}, {Zanin}, {Lefaucheur}, {Boisson},
  {Khelifi}, {Terrier}, {Wood}, {Mohrmann}, {Chakraborty}, {Watson},
  {Lopez-Coto}, {Klepser}, {Cerruti}, {Lenain}, {Acero}, {Djannati-Ata{\"\i}},
  {Pita}, {Bosnjak}, {Trichard}, {Vuillaume}, {Donath}, {King}, {Jouvin},
  {Owen}, {Sipocz}, {Lennarz}, {Voruganti}, {Spir-Jacob}, {Ruiz}, and
  {Arribas}}]{Deil_Gammapy_2017}
{Deil} C, {Zanin} R, {Lefaucheur} J, {Boisson} C, {Khelifi} B, {Terrier} R,
  {Wood} M, {Mohrmann} L, {Chakraborty} N, {Watson} J, {Lopez-Coto} R,
  {Klepser} S, {Cerruti} M, {Lenain} JP, {Acero} F, {Djannati-Ata{\"\i}} A,
  {Pita} S, {Bosnjak} Z, {Trichard} C, {Vuillaume} T, {Donath} A, {King} J,
  {Jouvin} L, {Owen} E, {Sipocz} B, {Lennarz} D, {Voruganti} A, {Spir-Jacob} M,
  {Ruiz} JE, {Arribas} MP (2017) {Gammapy - A prototype for the CTA science
  tools}. In: Proc.\ 35th Int.\ Cosmic Ray Conf.\ (ICRC2017), vol 301, p 766,
  \eprint{1709.01751}

\bibitem[{Kn\"odlseder et~al(2019)Kn\"odlseder, Tibaldo, Tiziani, Specovius,
  Cardenzana, Mayer, Kelley-Hoskins, Di~Venere, Bonnefoy, Ziegler, Eschbach,
  Martin, Louge, Brun, Haupt, and B\"uhler}]{Knoedlseder2019}
Kn\"odlseder J, Tibaldo L, Tiziani D, Specovius A, Cardenzana J, Mayer M,
  Kelley-Hoskins N, Di~Venere L, Bonnefoy S, Ziegler A, Eschbach S, Martin P,
  Louge T, Brun F, Haupt M, B\"uhler R (2019) {Analysis of the H.E.S.S.\ public
  data release with ctools}. \aap 632:A102, \doi{10.1051/0004-6361/201936010},
  \eprint{1910.09456}

\bibitem[{{HAWC Collaboration} et~al(2022){HAWC Collaboration}, Donath, and
  Funk}]{HAWC2022}
{HAWC Collaboration}, Donath A, Funk S (2022) {Validation of standardized data
  formats and tools for ground-level particle-based gamma-ray observatories}.
  \aap 667:A36, \doi{10.1051/0004-6361/202243527}, \eprint{2203.05937}

\bibitem[{{Zabalza}(2015)}]{Zabalza2015}
{Zabalza} V (2015) {Naima: a Python package for inference of particle
  distribution properties from nonthermal spectra}. In: Proc.\ 34th Int.\
  Cosmic Ray Conf.\ (ICRC2015), vol 236, p 922, \doi{10.22323/1.236.0922},
  \eprint{1509.03319}

\bibitem[{{HAWC Collaboration}(2019)}]{HAWC2019}
{HAWC Collaboration} (2019) {Measurement of the Crab Nebula Spectrum Past 100
  TeV with HAWC}. \apj 881:134, \doi{10.3847/1538-4357/ab2f7d},
  \eprint{1905.12518}

\bibitem[{{Joshi} et~al(2019){Joshi}, {Hinton}, {Schoorlemmer},
  {L{\'o}pez-Coto}, and {Parsons}}]{Joshi2019}
{Joshi} V, {Hinton} J, {Schoorlemmer} H, {L{\'o}pez-Coto} R, {Parsons} R (2019)
  {A template-based {\ensuremath{\gamma}}-ray reconstruction method for air
  shower arrays}. \jcap 2019(1):012, \doi{10.1088/1475-7516/2019/01/012},
  \eprint{1809.07227}

\bibitem[{{LHAASO Collaboration}(2021)}]{LHAASO2021a}
{LHAASO Collaboration} (2021) {Peta–electron volt gamma-ray emission from the
  Crab Nebula}. Science 373:425--430, \doi{10.1126/science.abg5137},
  \eprint{2111.06545}

\bibitem[{{Abdo} et~al(2012){Abdo}, {Allen}, {Atkins} et~al}]{MILAGRO2012}
{Abdo} AA, {Allen} BT, {Atkins} R, et~al (2012) {Observation and Spectral
  Measurements of the Crab Nebula with Milagro}. \apj 750:63,
  \doi{10.1088/0004-637X/750/1/63}, \eprint{1110.0409}

\bibitem[{Vianello et~al(2015)Vianello, Lauer, Younk, Tibaldo, Burgess,
  Ayala~Solares, Harding, Hui, Omodei, and Zhou}]{Vianello2015}
Vianello G, Lauer R, Younk P, Tibaldo L, Burgess JM, Ayala~Solares H, Harding
  JP, Hui CM, Omodei N, Zhou H (2015) {The Multi-Mission Maximum Likelihood
  framework}. In: Proc.\ 34th Int.\ Cosmic Ray Conf.\ (ICRC2015), vol 236, p
  1042, \doi{10.22323/1.236.1042}, \eprint{1507.08343}

\bibitem[{{Koldobskiy} et~al(2021){Koldobskiy}, {Kachelrie{\ss}}, {Lskavyan},
  {Neronov}, {Ostapchenko}, and {Semikoz}}]{koldobskiy21}
{Koldobskiy} S, {Kachelrie{\ss}} M, {Lskavyan} A, {Neronov} A, {Ostapchenko} S,
  {Semikoz} DV (2021) {Energy spectra of secondaries in proton-proton
  interactions}. \prd 104:123027, \doi{10.1103/PhysRevD.104.123027},
  \eprint{2110.00496}

\bibitem[{{Meyer} et~al(2010){Meyer}, {Horns}, and {Zechlin}}]{meyer10}
{Meyer} M, {Horns} D, {Zechlin} HS (2010) {The Crab Nebula as a standard candle
  in very high-energy astrophysics}. \aap 523:A2,
  \doi{10.1051/0004-6361/201014108}, \eprint{1008.4524}

\bibitem[{{Buehler} et~al(2012){Buehler}, {Scargle}, {Blandford}, {Baldini},
  {Baring}, {Belfiore}, {Charles}, {Chiang}, {D'Ammando}, {Dermer}, {Funk},
  {Grove}, {Harding}, {Hays}, {Kerr}, {Massaro}, {Mazziotta}, {Romani}, {Saz
  Parkinson}, {Tennant}, and {Weisskopf}}]{buhler12}
{Buehler} R, {Scargle} JD, {Blandford} RD, {Baldini} L, {Baring} MG, {Belfiore}
  A, {Charles} E, {Chiang} J, {D'Ammando} F, {Dermer} CD, {Funk} S, {Grove} JE,
  {Harding} AK, {Hays} E, {Kerr} M, {Massaro} F, {Mazziotta} MN, {Romani} RW,
  {Saz Parkinson} PM, {Tennant} AF, {Weisskopf} MC (2012) {Gamma-Ray Activity
  in the Crab Nebula: The Exceptional Flare of 2011 April}. \apj 749:26,
  \doi{10.1088/0004-637X/749/1/26}, \eprint{1112.1979}

\bibitem[{{Kuiper} et~al(2001){Kuiper}, {Hermsen}, {Cusumano}, {Diehl},
  {Sch{\"o}nfelder}, {Strong}, {Bennett}, and {McConnell}}]{kuiper01}
{Kuiper} L, {Hermsen} W, {Cusumano} G, {Diehl} R, {Sch{\"o}nfelder} V, {Strong}
  A, {Bennett} K, {McConnell} ML (2001) {The Crab pulsar in the 0.75-30 MeV
  range as seen by CGRO COMPTEL. A coherent high-energy picture from soft
  X-rays up to high-energy gamma-rays}. \aap 378:918--935,
  \doi{10.1051/0004-6361:20011256}, \eprint{astro-ph/0109200}

\bibitem[{{Sollerman} et~al(2000){Sollerman}, {Lundqvist}, {Lindler},
  {Chevalier}, {Fransson}, {Gull}, {Pun}, and {Sonneborn}}]{sollerman00}
{Sollerman} J, {Lundqvist} P, {Lindler} D, {Chevalier} RA, {Fransson} C, {Gull}
  TR, {Pun} CSJ, {Sonneborn} G (2000) {Observations of the Crab Nebula and Its
  Pulsar in the Far-Ultraviolet and in the Optical}. \apj 537:861--874,
  \doi{10.1086/309062}, \eprint{astro-ph/0002374}

\bibitem[{{Tziamtzis} et~al(2009){Tziamtzis}, {Lundqvist}, and
  {Djupvik}}]{tziamtis09}
{Tziamtzis} A, {Lundqvist} P, {Djupvik} AA (2009) {The Crab pulsar and its
  pulsar-wind nebula in the optical and infrared}. \aap 508:221--228,
  \doi{10.1051/0004-6361/200912031}, \eprint{0911.0608}

\bibitem[{{Aleksi{\'c}} et~al(2011){Aleksi{\'c}}, {Alvarez}, {Antonelli}
  et~al}]{aleksic11}
{Aleksi{\'c}} J, {Alvarez} EA, {Antonelli} LA, et~al (2011) {Observations of
  the Crab Pulsar between 25 and 100 GeV with the MAGIC I Telescope}. \apj
  742:43, \doi{10.1088/0004-637X/742/1/43}, \eprint{1108.5391}

\bibitem[{{Aliu} et~al(2011){Aliu}, {Arlen}, {Aune} et~al}]{aliu11}
{Aliu} E, {Arlen} T, {Aune} T, et~al (2011) {Detection of Pulsed Gamma Rays
  Above 100 GeV from the Crab Pulsar}. Science 334:69,
  \doi{10.1126/science.1208192}, \eprint{1108.3797}

\bibitem[{{Abdo} et~al(2013){Abdo}, {Ajello}, {Allafort} et~al}]{abdo13}
{Abdo} AA, {Ajello} M, {Allafort} A, et~al (2013) {The Second Fermi Large Area
  Telescope Catalog of Gamma-Ray Pulsars}. \apjs 208:17,
  \doi{10.1088/0067-0049/208/2/17}, \eprint{1305.4385}

\bibitem[{{Foreman-Mackey} et~al(2013){Foreman-Mackey}, {Hogg}, {Lang}, and
  {Goodman}}]{ForemanMackey2013}
{Foreman-Mackey} D, {Hogg} DW, {Lang} D, {Goodman} J (2013) {emcee: The MCMC
  Hammer}. \pasp 125:306, \doi{10.1086/670067}, \eprint{1202.3665}

\end{thebibliography}

\end{document}